\documentclass[a4paper,11pt]{article}
\pdfoutput=1 % if your are submitting a pdflatex (i.e. if you have images in pdf, png or jpg format)
\usepackage{jheppub}	% jheppub includes hyperref, color, natbib, amsmath, amssymb, epsfig, graphicx
\usepackage[utf8]{inputenc}
\usepackage[english]{babel}
\usepackage{makeidx}
\usepackage{amsfonts}
\usepackage{enumerate}
\usepackage{mathrsfs}
\usepackage{tensor}
\usepackage{xcolor,tikz,pgfplots}
\usepackage[autostyle]{csquotes}

\usepackage{subcaption}
\usepackage{caption}
\usepackage{sidecap}

\usepackage{braket}

\usetikzlibrary{matrix,calc,positioning,decorations.markings,decorations.pathmorphing,decorations.pathreplacing}%tikzmark,
\usetikzlibrary{arrows,cd}
\usepackage{bbding}
\usetikzlibrary{positioning}
\tikzset{>=stealth}

\usepackage{float}
\usepackage{booktabs}

\usepackage{bm}

\numberwithin{equation}{section}

\newcommand{\CC}{\mathbb{C}}
\newcommand{\FF}{\mathbb{F}}
\newcommand{\PP}{\mathbb{P}}
\newcommand{\RR}{\mathbb{R}}
\newcommand{\ZZ}{\mathbb{Z}}
\newcommand{\NN}{\mathbb{N}}
\newcommand{\SSS}{\mathbb{S}}
\newcommand{\TT}{\mathbb{T}}

\newcommand{\Cc}{\mathcal{C}}
\newcommand{\Nn}{\mathcal{N}}

\newcommand{\half}{\frac{1}{2}}

\newcommand{\Tr}{\text{Tr}}

\tikzstyle{vecArrow} = [thick, decoration={markings,mark=at position
   1 with {\arrow[semithick]{open triangle 60}}},
   double distance=3pt, shorten >= 4.5pt,
   preaction = {decorate},
   postaction = {draw,line width=3pt, white,shorten >= 4.5pt}]

  \title{Mass Deformations of Unoriented Quiver Theories
}
\author[a]{Massimo Bianchi,}
\author[b]{Davide Bufalini,}
\author[c]{Salvo Mancani,}
\author[d]{Fabio Riccioni}

\affiliation[a]{Dipartimento  di  Fisica,  Universit\`a  di  Roma  ``Tor  Vergata'', \\
	Sezione INFN Roma “Tor Vergata”, \\
	Via della Ricerca Scientifica 1, 00133, Roma, Italy }  
\affiliation[b]{Mathematical Sciences and STAG Research Centre, University of Southampton,\\
Highfield, Southampton, SO17 1BJ, UK}
\affiliation[c]{Dipartimento di Fisica, Università di Roma “La Sapienza”, \\
	Piazzale Aldo Moro 2, 00185 Roma, Italy}
\affiliation[d]{I.N.F.N.  Sezione  di  Roma, Dipartimento di Fisica,  Universit\`a di Roma “La Sapienza”, \\
	Piazzale Aldo Moro 2, 00185 Roma, Italy}
\emailAdd{massimo.bianchi@roma2.infn.it}
\emailAdd{D.Bufalini@soton.ac.uk}
\emailAdd{salvo.mancani@uniroma1.it}
\emailAdd{Fabio.Riccioni@roma1.infn.it}

\abstract{We study the interplay between mass deformations and unoriented projections of super-conformal quiver gauge theories resulting from D3-branes at (toric) Calabi-Yau singularities. We focus on simple orbifold cases ($\CC^3/\ZZ_3$ and $\CC^3/\ZZ_4$) and their non-orbifold descendants. This allows us to generalize the construction rules and clarify points that have been previously overlooked. In particular we spell out the conditions of anomaly cancellations as well as super-conformal invariance that typically require the introduction of flavour branes, which in turn may spoil toric symmetry. Finally, we discuss duality cascades in this context and the interplay between Seiberg/toric duality and unoriented projection with (or without) mass deformations.}


\preprint{PREPRINT}
 \clearpage

\begin{document}
\maketitle

%\tableofcontents

\section{Introduction and Motivations}

Open and unoriented strings, whose systematic construction was addressed long ago \cite{Bianchi:1990yu, Bianchi:1990tb}\footnote{For reviews and lists of original references, see for instance \cite{Bianchi:1997zs, Angelantonj:2002ct}.}, have proven to be an unprecedented tool in the exploration of gauge field dynamics after the introduction of D-branes \cite{Dai:1989ua,Polchinski:1995mt,Polchinski:1996na} and even more so after the advent of the holographic AdS/CFT correspondence \cite{Maldacena:1997re, Witten:1998qj, Gubser:1998bc, Aharony:1999ti}. A large class of exactly superconformal field theories (SCFT's) in $D=4$ can be realised on D3-branes at non-compact Calabi-Yau (CY) singularities \cite{Morrison:1998cs}, whose properties have been investigated from diverse perspectives \cite{Benvenuti:2004dy} and by means of different tools \cite{Forcella:2008ng}. A lot of attention has been devoted to orbifold singularities $\CC^3/\Gamma$ with $\Gamma$ an abelian \cite{Aharony:1996bi} or non-abelian \cite{Bruzzo:2017fwj} subgroup of $SU(3)$ acting on $\CC^3$. More recently toric singularities \cite{Martelli:2004wu, Franco:2005sm}, that admit global symmetries $G \supset U(1)^2$ in addition to $U(1)_R$ R-symmetry, have been studied, in particular those associated to `reflexive' polygons \cite{Hanany:2012hi, Hanany:2012vc, Cremonesi:2013aba}. Notwithstanding some important exceptions \cite{Franco:2007ii, GarciaEtxebarria:2012qx, Garcia-Etxebarria:2013tba, Garcia-Etxebarria:2015hua, Garcia-Etxebarria:2016bpb}, the investigation of unoriented singularities has been much less systematic. In this case an orientation reversing action $\sigma$ on the CY is combined with world-sheet parity $\Omega$ that entails an action $\gamma_\Omega$ on the Chan-Paton factors or, equivalently, on the D-branes. The geometric counterpart of the construction, a.k.a. orientifold, involve orientifold-planes or $\Omega$-planes for short. Even more so, the inclusion of `flavour' branes, i.e. branes with non-compact world-volumes or, equivalently, vanishingly small gauge couplings, has been addressed systematically only in the context of abelian orbifolds \cite{Bianchi:2013gka} or in some special instances of toric singularities \cite{Franco:2006es, Franco:2012mm, Franco:2013ana}. The conditions for superconformal symmetry and the effect of D-brane instantons have been spelled out only in a handful of cases \cite{Bianchi:2013gka,  Franco:2018vqd, Tenreiro:2017fon, Franco:2015kfa}, based on earlier work \cite{Billo:2002hm, Billo:2006jm, Bianchi:2007fx, Bianchi:2007wy, Bianchi:2009bg, Blumenhagen:2009qh, Bianchi:2012ud}. These and related issues deserve more attention for their triple role as models for (non-)perturbative gauge field dynamics (holographic superQCD and alike) \cite{Gursoy:2007er, Gursoy:2007cb}, for Standard Model embedding \cite{Angelantonj:1996uy, Uranga:2007zza, Blumenhagen:2006ci} and for cosmology (brane inflation) \cite{Berg:2004sj, Haack:2008yb}. 

The present analysis aims at clarifying some of the issues related to D3-branes, $\Omega$-planes and flavour branes. To this end, we will use various tools ranging from Quiver diagrams and, when toricity holds, Dimer and Toric diagrams. Unoriented toric singularities were already discussed in the Dimer literature and a number of anomaly-free models were found thanks to the identification of a set of rules for unoriented projections \cite{Franco:2007ii, GarciaEtxebarria:2012qx, Garcia-Etxebarria:2013tba, Garcia-Etxebarria:2015hua, Garcia-Etxebarria:2016bpb}. However, a Quiver description of these results was lacking and in the present work we are able to reproduce them by considering a generalization of the anomaly-cancellation equations previously used in \cite{Bianchi:2013gka}. Our analysis crucially relies on mass-deformations of orbifold models: we extend the validity of the consistency conditions to non-orbifold models when the latter can be obtained by mass-deformations  \cite{Bianchi:2014qma} and/or (Un-)Higgsing \cite{Franco:2012mm, Franco:2013ana} of orbifold theories. With these results, we recover a convenient Quiver description of unoriented singularities which, starting from simple orbifolds cases, can enlighten the relation between Orientifold charges, used in the context of Quiver diagrams, and T-parities, which are widely used in the context of Dimer models. Furthermore, Quiver diagrams can be used to easily compute beta functions and are suitable for the inclusion of flavour branes which are sometimes needed in order to obtain (conformal) non-anomalous theories in the perturbative regime \cite{Bianchi:2013gka}. However, it should be stressed that the anomalous dimensions can be determined once the superpotential is known, which in turn is determined from the dimer. In particular, when the theory is conformal, one can exploit the a-maximisation procedure \cite{Intriligator:2003jj}. Alas, several interesting non-perturbative unoriented models without flavour branes, found in \cite{Garcia-Etxebarria:2015hua, Garcia-Etxebarria:2016bpb}, are beyond the reach of our present approach.

It should be noted, however, that non-compact D7-branes can backreact on the local geometry and spoil the previously-existing toric symmetry. This fact raises some doubts about the validity of Toric and Dimer descriptions, whereby flavour branes are nicely represented  as open paths in the dimer or as paths connecting punctures in the mirror description \cite{Franco:2006es, Franco:2012mm, Franco:2013ana}. On the other hand, the Quiver description does not rely on toricity and a study of the above setup is still possible in the presence of flavour branes, even though the superpotential can be uniquely determined when toricity is present. In order to recognize the presence of $ \Omega3 $ and/or compact/non-compact $ \Omega7 $-plane in the resolved geometry, we combine the use of the Toric diagram \cite{Garcia-Etxebarria:2015hua, Garcia-Etxebarria:2016bpb} and a result from Algebraic Geometry, the Ito-Reid Theorem \cite{Ito:1994zx}. In case of orbifold models, the latter allows us to determine the number of compact two- and four-cycles in the resolved geometry by computing the \textit{age} of the elements of the orbifold group. For instance, compact four-cycles can be wrapped by $ \Omega $7-planes and D7-branes, making them `compact'. 

The presence of $\Omega$-planes alters the spectrum of the gauge theories and the conditions required for obtaining an anomaly-free super-conformal theory are no longer obvious. Furthermore, unoriented projections can be performed in different manners on Toric, Dimer and Quiver diagrams and we show that some of these unoriented projections cannot lead to consistent theories when (super)conformal invariance is additionally requested\footnote{Yet they can do so in the deep IR at strong coupling and possibly large $N$, as shown in  \cite{Garcia-Etxebarria:2015hua, Garcia-Etxebarria:2016bpb}. We thank I\~naki  Garcia-Etxebarria for clarifying this issue to us.}. One can discuss the above issue also in the presence of flavour branes and, as a simple example, we address this more delicate problem in the prototypical case of the unoriented $ \CC^3/\ZZ_3$ orbifold \cite{Bianchi:2007fx, Bianchi:2007wy}, that originally appeared in the context of the unoriented $\TT^6/\ZZ_3$, the first chiral Type I model, found in \cite{Angelantonj:1996uy}. 

We conclude our analysis by  using the previously described approach to discuss the unoriented projection in the context of Seiberg duality. We focus on a specific model, namely the chiral orbifold of the conifold, and we compare the theory that results from performing the unoriented projection at the beginning of a duality cascade to the theory in which the orientifold involution is performed at the end of the cascade. Duality cascades in unoriented quiver theories have already been studied in \cite{Franco:2015kfa,Argurio:2017upa}.

The plan of this paper is as follows. After briefly reviewing the D3-brane setup in Section \ref{Setup}, we pass to consider Orientifold projections of abelian orbifolds and more general toric singularities in Section \ref{Unorient} and, in particular, we describe how to make use of mass deformations to connect different (toric) theories. We devote Section \ref{Examples} to analyse interesting examples of unoriented quiver theories with a small number of nodes, starting from well-known cases such as $\CC^3$, $\CC^3/\ZZ_3$ and $\CC^3/\ZZ_4$, by exploiting mass-deformations of orbifold theories.  In Section \ref{Sec:Seiberg} we discuss duality cascades \cite{Klebanov:2000hb} of unoriented toric singularities. Finally, we draw some conclusions and outlook in Section \ref{Sec:Discussion}. We review the role of Higgsing in Appendix \ref{sec:higgs} and of Seiberg duality \cite{Seiberg:1994pq, Intriligator:1995au} in the above context in Appendix \ref{SeibergDual}.

\section{The Setup}\label{Setup}

For completeness, we describe the setup for our analysis. The reader familiar with the material can skip the present Section and Section \ref{Unorient} on unoriented projections and go directly to Section \ref{Examples}, which contains our original results.

We consider Type IIB Superstring Theory on a four-dimensional Minkowski space $ \RR^{1,3} $, transverse to a singular (toric \cite{Bouchard:2007ik}) non-compact Calabi-Yau (CY) three-fold, parametrized by three complex coordinates $ (X_1,X_2,X_3)$, $ I = 1,2,3$. The singularity is probed by fractional D3-branes on which, at low energy, a (non-)abelian supersymmetric gauge theory lives. The transverse CY three-fold can  either be an orbifold of $ \CC^3 $ of the form\footnote{Here $ \Gamma $ is a discrete abelian or non-abelian subgroup of $ SU(3) $.} $ \CC^3/\Gamma $ or a non-orbifold space. On top of this setup, we further consider the action of Orientifolds via $\Omega$-planes and, later on, flavour branes.

\subsection{D3-Branes at Toric Calabi-Yau Singularities}
\begin{figure}[H]
	\centerline{\begin{tikzpicture}[auto, scale= 0.5]
		%%%%%%%%%%%% Nodes %%%%%%%%%%
		\node [circle, draw=blue!50, fill=blue!20, inner sep=0pt, minimum size=5mm] (0) at (0,5) {$N_0$}; 
		\node [circle, draw=blue!50, fill=blue!20, inner sep=0pt, minimum size=5mm] (1) at (3,0) {$N_1$}; 			
		\node [circle, draw=blue!50, fill=blue!20, inner sep=0pt, minimum size=5mm] (2) at (-3,0) {$N_2$};
		%%%%%%%%%%% Lines %%%%%%%%%%%
		\draw (0) to node {} (1) [->>>, thick];
		\draw (1)  to node {} (2) [->>>, thick];
		\draw (2)  to node {} (0) [->>>, thick];
		\end{tikzpicture}}
	\caption{The quiver of the $\CC^3/\ZZ_3$ orbifold theory with $k_I=(1,1,1)$. The bifundamental fields are arrows connecting nodes representing $ U(N) $ gauge groups. The condition of anomaly cancellation gives $N_0= N_1 = N_2$ and, as a result, $ \beta_0 = \beta_1 = \beta_2 = 0 $. }		
	\label{fig:Quiver}
\end{figure}
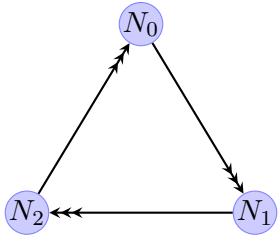

Consider the case of an abelian orbifold of the form $ \CC^3/\Gamma $ with $\Gamma = \ZZ_n$. On each complex coordinate $ (X_1,X_2,X_3) $ of the transverse space, the orbifold group acts as 
\begin{equation}
{{g}}: \quad X_I \mapsto \omega^{k_I} \, X_I, \qquad \omega = e^{\frac{2\pi i}{n}} \; ,
\end{equation}
where ${{g}}$ is the generator of $ \ZZ_n $ and $ k_I $ must satisfy the supersymmetry-preserving CY condition
\begin{equation}
\sum_{I=1}^3 k_I = 0 \; \textrm{mod }n \; ,
\end{equation}
where  $0 \leq k_I \leq (n-1)$. The quotient $ \CC^3/ \ZZ_n $ has only one fixed point at the origin, which is the singularity. There we place the fractional D3-branes, which we can think of as D5- and D7-branes wrapping respectively collapsed two- or four-cycles of the (resolved) CY three-fold. At low energy, the dynamics of open strings living on their world-volume is governed by a supersymmetric quiver gauge theory whose gauge group is a product of $U(N_a)$, $a=1,\ldots, n$ groups. The strings with endpoints connecting different fractional branes give rise to bifundamental fields $({{\bf{N}}}_a, \overline{{\bf{N}}}_b)$ denoted by ${X_{ab}}$, where ${\bf{N}}_a (\overline{{\bf{N}}}_a)$ is the fundamental (antifundamental) representation of the gauge group $U(N_a)$. A more detailed discussion of the spectrum in the orbifold case is presented in \cite{Bianchi:2013gka}. The quiver for the simple example of the $\CC^3/\ZZ_3$ orbifold is shown in Fig.\,\eqref{fig:Quiver}. Note that the gauge theories arising from orbifold may be chiral (as $k_I=(1,1,n-2)$) or non-chiral (as $k_I=(1,n-1,0)$). While the latter are non-anomalous by construction, the former need the RR tadpoles to vanish for the theory to be anomaly-free, thus giving constraints on the ranks of the gauge groups. Furthermore, even if for $\CC^3/\ZZ_n$ the sum of all the beta functions is zero, i.e. $\sum_a \beta_a=0$, each (non-abelian) gauge group may have a non-zero $\beta_a$. 

The above example is only one of infinitely many possibilities where fractional D3-branes are placed at CY singularities. In the present work we focus on the case of toric CY spaces, i.e. admitting at least a $U(1) \times U(1)$ isometry in addition to the $U(1)_R$ R-symmetry. We recall that the non-compact CY need not be necessarily an abelian discrete orbifold of $ \CC^3 $ but can be a general real cone over a five dimensional Sasaki-Einstein space or, even better if it is the case, a complex cone over a K\"ahler-Einstein base.  

Strings on the fractional D3-branes do not see the singular space $\widehat{Y}$, but they effectively live on a smooth resolved space $Y$, related to the former by a blow-down morphism $Y \to \widehat{Y}$. For $\widehat{Y}= \CC^m/\Gamma$ orbifold theories, one can determine some useful geometrical properties of the resolved space by introducing the concept of \textit{age grading} \cite{Ito:1994zx, Bruzzo:2017fwj, Bruzzo:2019asa}. Consider an element $g \in \Gamma$, that is such that $g^n = 1$. The \textit{age} of the element ${{g}}$ is defined as
\begin{equation}
\mathrm{age}({{g}}) = \frac{1}{n} \sum_{I=1}^m k_I \; .
\end{equation} 
From the age, one can organize the elements of $\Gamma$ in various conjugacy classes: null or baby classes have age $=0$, junior classes have age $=1$ and senior classes with age $=2$. According to the Ito-Reid Theorem \cite{Ito:1994zx}, each conjugacy class is associated with the dimension of de Rham cohomology groups of the crepant\footnote{We recall that a \textit{crepant resolution} is a resolution that preserves the Calabi-Yau condition, namely the first Chern class of the tangent bundle of $ Y $ vanishes.} resolution, according to
\begin{equation}
\mathrm{dim }H^{2k}(Y) = \text{number of age $ k $ conjugacy classes of $ \Gamma $} 
\end{equation}
while all odd cohomology groups are trivial\footnote{This implies that the singularity, as most of the toric CY singularities, is `isolated' in that it does not admit complex (marginal) deformations. We will later on discuss relevant mass deformations which trigger RG-flows.}. The conjugacy classes of age $k$ are then related to the existence of $2k$-cycles in the smooth space $Y$. In particular, a non-trivial senior class implies the existence of a compact 4-cycle. A more general and detailed discussion is presented, for instance, in \cite{Bruzzo:2017fwj}. As an example, the crepant resolution of $ \CC^3/\ZZ_4 $ is addressed in \cite{Bruzzo:2019asa}. This classification helps identifying the cycles wrapped by a D-brane or an $\Omega$-plane.

\subsection{The Dimer}

The above general configuration is T-dual to D5-branes and smeared NS5-branes, where the former wrap a 2-torus $ \TT^2 $ along the (real) directions $ 5 $ and $ 7 $, as shown in Tab.\,\eqref{tab:D5NS5NS5'}. Indeed, after performing two T-dualities along the two one-cycles of $ \TT^2 $, each D5-brane turns into a D3-brane and the smeared NS5-branes turn into geometry \cite{Franco_2006,Yamazaki_2008}. The complex surface $ \Sigma $ extends along the (real) directions $ 4,5,6,7 $. The directions 8 and 9 give an $ SO(2) \simeq U(1)$ isometry, which corresponds to the R-symmetry of the ${\cal N}=1$ SCFT.

\begin{center}
	\begin{tabular*}{0.5\textwidth}{@{\extracolsep{\fill}}cc|cccccc}
		\toprule
		& 0 1 2 3 & 4 5 & 6 7 & 8 9 \\
		\midrule
		D5 & - - - - & $\cdot$ - & $\cdot$ - & $\cdot$ $\cdot$ \\
		NS5 & - - - - & $\Sigma$ & $\Sigma$ & $\cdot$ $\cdot$ \\  
		\bottomrule
	\end{tabular*}
	\captionof{table}{D5-branes and smeared NS5-branes are T-dual to D3-branes at CY threefold singularity.}
	\label{tab:D5NS5NS5'}
\end{center} 

\begin{figure}[H]
\centerline{\includegraphics[scale=0.25, trim={3cm 3cm 1.5cm 10cm}, clip]{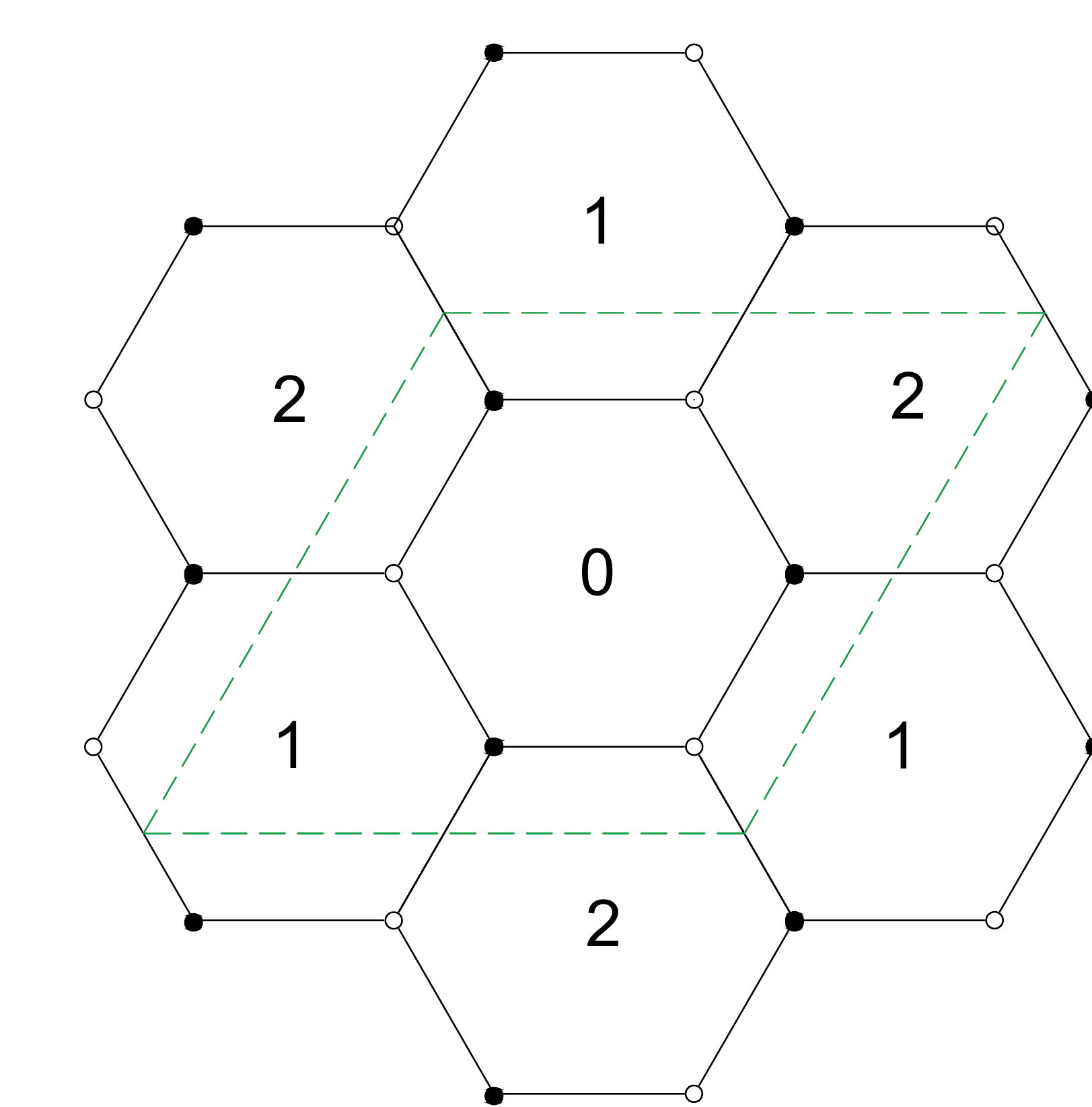}}
\caption{The dimer of the $\CC^3/\ZZ_3$ orbifold theory. The green dashed line defines the fundamental cell.}\label{fig:Dimer}
\end{figure}

In this setup, the NS5-branes intersect the D5-branes in such a way that the two-torus $ \TT^2 $ gets tiled. The tiling of the torus forms a bipartite graph called the \emph{dimer} \cite{Franco_2006}, made by even-sided polygons. A graph is called bipartite when its nodes are colored in white and black in such a way that the edges connect a white node to a black one (or viceversa). Each face represents a gauge group $U(N_a)$ and two adjacent polygons $a$ and $b$ are connected by bifundamental fields ${X_{ab}}$. The tiling for $ \CC^3/\ZZ_3 $ is drawn in Fig.\,\eqref{fig:Dimer}. 

From the dimer one can easily read off the superpotential of the gauge theory. Going clockwise (counter-clockwise) around a white (black) node and forming products of fields associated to edges, each white (black) node give rise to a  monomial term in the superpotential with positive (negative) sign. Each monomial is a mesonic operator, i.e. a trace operator of the form\footnote{We will not consider baryonic operators which characterize the baryonic branch of the moduli space.}
\begin{equation}
(X_{ab})_{i_a}^{\phantom{i_a}i_b} (X_{bc})_{i_b}^{\phantom{i_b}i_c} \ldots (X_{fa})_{i_f}^{\phantom{i_f}i_a} \; ,
\end{equation}
where lower indices $i_a=1, \ldots, \dim N_a$ belong to the fundamental representation of the gauge group $U(N_a)$, while upper indices $i_a$ to the anti-fundamental representation. Adding up all the monomials (with their signs) in the fundamental cell of the torus we obtain the superpotential $W$ of the theory. The bipartite nature of the dimer is such that $W$ contains each bi-fundamental chiral field $X_{ab}$ twice, once with positive sign and once with negative sign: this is called the \textit{toric condition}.

\subsection{The Toric Diagram}

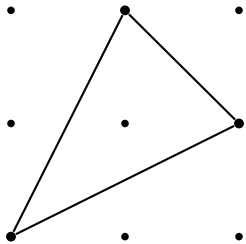
\begin{figure}[H]
	\centerline{\begin{tikzpicture}[auto, scale= 0.5]
		%%%%%%%%%%%% Nodes %%%%%%%%%%
		\node [circle, fill=black, inner sep=0pt, minimum size=1mm] (1) at (-3,3) {};		
		\node [circle, fill=black, inner sep=0pt, minimum size=1.3mm] (2) at (0,3) {}; 
		\node [circle, fill=black, inner sep=0pt, minimum size=1mm] (3) at (3,3) {};
		\node [circle, fill=black, inner sep=0pt, minimum size=1mm] (4) at (-3,0) {};
		\node [circle, fill=black, inner sep=0pt, minimum size=1mm] (5) at (0,0) {}; 
		\node [circle, fill=black, inner sep=0pt, minimum size=1.3mm] (6) at (3,0) {}; 			
		\node [circle, fill=black, inner sep=0pt, minimum size=1.3mm] (7) at (-3,-3) {};
		\node [circle, fill=black, inner sep=0pt, minimum size=1mm] (8) at (0,-3) {}; 
		\node [circle, fill=black, inner sep=0pt, minimum size=1mm] (9) at (3,-3) {};
		%%%%%%%%%%% Lines %%%%%%%%%%%
		\draw (2) edge (6) [thick];
		\draw (6) edge (7) [thick];
		\draw (7) edge (2) [thick];
		\end{tikzpicture}}
	\caption{The toric diagram of the $\CC^3/\ZZ_3$ orbifold.} \label{fig:Toric}
\end{figure}

The metric of the non-compact toric CY threefold is of the form
\begin{equation}
ds^2 = dr^2 + r^2 ds^2_{SE} \; ,
\end{equation}   
where the five-dimensional \textit{horizon} \cite{Morrison:1998cs} is a Sasaki-Einstein space and serves as the base of the real cone, whose radial coordinate is denoted by $r$. An infinite family of such spaces is well known \cite{Gauntlett:2004hh,Martelli:2004wu,Cvetic:2005ft}. This horizon can be described as a $\TT^3$ fibration over a polygon called \textit{toric diagram}. On the edges of this polygon the $\TT^3$ degenerates to a $\TT^2$ and on the vertices it further degenerates to an $\SSS^1$ \cite{Feng:2005gw}. The toric diagram is specified by a set of vectors, all lying on the same two-dimensional lattice, that represent the vertices of the polygon. As an example, the toric diagram of the $\CC^3/\ZZ_3$ orbifold is shown in Fig.\,\eqref{fig:Toric}.

The above discussion highlights the fact that there are different tools that allow the study of gauge theories arising as low energy theories on fractional D3-branes probing toric CY manifolds. Each diagram has its own peculiarities and advantages and they can be related to one another. This is still valid when we consider the addition of $\Omega$-planes as long as toricity is preserved, a problem that we will address in the forthcoming sections.

\section{Orientifolds of Orbifold and Non-orbifold Toric Singularities}\label{Unorient}

We consider quotients of gauge theories by unoriented involutions, which involve the action of the world-sheet parity operator $\Omega$, general spacetime reflections $\sigma$ and a suitable $ \ZZ_2 $ symmetry such as $ (-1)^F $. The original theory will be called the \textit{parent theory}. Labelling by $ a $ the gauge group, its image under the orientifold action is denoted by $  a' \, $. If $ a = a' $ the $ U(N_a) $ group is projected to $ SO(N_a) $ or $ Sp(N_a) $ (with $N_a$ even). When the strings connect a node $ a $ and its image $ a' $, the bifundamental fields give rise to symmetric or antisymmetric representations of the gauge group. We now briefly recall how to perform an unoriented projection on the diagrams discussed above.  
 
\subsection{Orientifolding the Quiver Diagram}

The orientifold projection is performed by placing a (system of) $\Omega$-plane(s) on the D3-branes configuration, hence one can represent the orientifold action on the quiver diagram. The case of unoriented projections of abelian orbifolds of the form $ \CC^3/\ZZ_n  $ or $ \CC^3/(\ZZ_n \times \ZZ_m) $ was systematically adressed in \cite{Bianchi:2013gka} where it was represented by a reflection line denoted by $\Omega$, identifying specular nodes and fields with opposite orientation.

For orbifold theories, we can arrange the information of the orientifold action into four signs with a clear physical interpretation: ${\epsilon}_0$ is the sign of the RR-charge of the orientifold plane, the other three ${\epsilon}_I$, $I=1,2,3$, determine the $\ZZ_2$ action of the $\Omega$-plane on the complex directions $X_I$ of the CY transverse to the D3-branes. Thus, $({\epsilon}_0, {\epsilon}_1, {\epsilon}_2, {\epsilon}_3)$ is our set of orientifold signs describing the $\Omega$ plane. For instance, we can have 
\begin{align}
\Omega3^{\pm}: \qquad (\pm, -, -, - ) \; ,  \nonumber \\
\Omega7^{\pm}: \qquad (\mp, +, +, - ) \; .
\end{align}
Note that the $\Omega7$ has $\epsilon_I$ with different signs, while the $\Omega3$ has only identical $\epsilon_I$. 

When a node is on top of the orientifold plane $\Omega$, the corresponding gauge group is projected down to $Sp$ or $SO$ if $\epsilon_0 = +1$ or $\epsilon_0 = - 1$, respectively. In case of odd $n$, there is always at least one node on top of the orientifold line, that for convenience and without loss of generality we denote by 0,  while nodes $a$ and $n{-}a$ are reflected into each other. For even $n$, there are also various configurations with all nodes away from the orientifold line is allowed, which we denote as $\widehat{\Omega}$, in such a way that nodes $a$ and $n{-}a{-}1$ are reflected into each other (up to renumbering of the nodes). Thus
\begin{align}\label{Mirror}
\Omega:& \qquad {\bf{N}}_a \leftrightarrow \overline{{\bf{N}}}_{n-a} \; , \nonumber \\
\widehat{\Omega}:& \qquad {\bf{N}}_a \leftrightarrow \overline{{\bf{N}}}_{n-a-1} \; , 
\end{align}
consequently fields are identified as 
\begin{equation}
{\left( X_{ab} \right)}_{i_a}^{i_b} = ({\bf{N}}_a, \overline{{\bf{N}}}_{b})_{i_a}^{i_b} \leftrightarrow (\overline{{\bf{N}}}_{a'}, {\bf{N}}_{b'})_{i_{b}}^{i_{a}} = \left[ {\left( X_{ab} \right)}_{i_a}^{i_b} \right]^T \; ,
\end{equation} 
where nodes with prime $a'$, $b'$ are the image of $a$, $b$. This means that in the superpotential we may replace the bifundamental field $ X_{a' \, b'} $ with $ (X_{a \,b})^T $. Clearly, an orientifold $\Omega$ yields a theory with at least one gauge group $Sp/SO$, while $\widehat{\Omega}$ only gives $U(N)$ groups only. When a field $X_{aa'}^I$ connects a node and its image $a'$, it gives rise to the symmetric representation if $\epsilon_0 \epsilon_I = +1$ or to the antisymmetric representation if $\epsilon_0 \epsilon_I = -1$. In fact, the orientifold charge $\epsilon_I$ acts on the fields $X_{ab}^I$ if $b = a' =  a + k_I$ and
\begin{equation}
{\left( X_{aa'} \right) }_{i_a}^{i_{a'}} = ({\bf{N}}_a, \overline{{\bf{N}}}_{a'})_{i_a}^{i_{a'}} = ({\bf{N}}_{a}, {\bf{N}}_{a})_{i_a i_{a'}} \; .
\end{equation}   

There are also quiver diagrams associated to non-orbifold singularities, which may be related to orbifold ones via Higgsing/un-Higgsing or mass deformations, as we will see later. 

\subsection{Orientifolding the Dimer}

An orientifold projection on the dimer is performed by a $\ZZ_2$ involution of the fundamental cell of the torus, as explained in \cite{Franco:2007ii}. We briefly recall here the set of rules derived there.

The $\ZZ_2$ involution may act on one of the coordinate of the torus or both, with different physical interpretation. Let us discuss them schematically:
\begin{itemize}
\item When both coordinates of the $ \TT^2 $ are projected, there are four fixed points in the fundamental cell, independently of its shape. Each fixed point corresponds to an $\Omega$5-plane with positive or negative signs ${\tau}_i$, with $i=1,2,3,4$, called \textit{T-parities}. This type of involution maps black nodes on the dimer to white nodes and {\it vice versa}, consequently the number of terms in the superpotential, $N_W$, is halved. The overall product of the T-parities satisfies
\begin{equation}\label{eq:dimersgn}
\prod_{i=1}^4 {\tau}_i = (-1)^{N_W/2} \; .
\end{equation}
When the fixed point lays at the center of a face, the gauge group is projected down to $SO$ $(Sp)$ if the sign is $ +\,(-) $. If a fixed point lies on an edge, adjacent groups are identified with one another and the field connecting face $a$ and its image $a'$, transforms in the (anti)symmetric representation if ${\tau}_i = +$ $(-)$. In general, fields $X_{ab}$ are identified with $X_{b'a'}$ as for the quiver, which entails  that the orientation is reversed. The involution with fixed points preserves at least the mesonic symmetry $U(1)^2$, hence toricity may still hold.
\item Projecting only one coordinate of the torus, we obtain fixed lines depending on the shape of the fundamental cell: if it is a rhombus there may be only a fixed line, if it is a square also a second line is allowed. Fixed lines must map black nodes to black nodes and white nodes to white nodes. A fixed line is an $\Omega$7-plane crossing the torus, with ${\tau}=\pm $. The effect on the tiling is to project a gauge group down to $SO$ $(Sp)$ if an $\Omega^+$ ($\Omega^-$) passes through a face. Adjacent gauge groups are identified if the $\Omega^{\pm}$ lays on an edge, with the corresponding bi-fundamental fields giving rise to the symmetric (antisymmetric) representation of $U(N_a)$ if ${\tau} = +$ $(-)$. This involution breaks the mesonic symmetries $U(1)^2$, thus toricity is broken \cite{Garcia-Etxebarria:2016bpb}.
\end{itemize}

It is important to stress that the orientifold charges $(\epsilon_0, \epsilon_I)$ for orbifold theories (which have a more direct geometric interpretation) and the T-parities defined on the dimer are not always the same \cite{Imamura:2008fd}, and in fact they also act in different ways. The mesonic moduli space is spanned by gauge-invariant mesonic operators and a subclass of them can be regarded as the transverse coordinates to the D3 branes. On the dimer, they are constructed with fields lying on closed oriented paths (see \cite{Franco:2007ii}) and the product of T-parities they intersect gives the orientifold action on the mesonic operators. For the unoriented projection of $\CC^3$ and $\CC^3/\ZZ_3$ the product of (pairs of) T-parities, i.e. the orientifold action on mesonic operators, is exactly the same as $\epsilon_0 \epsilon_I$.

\subsection{Orientifolding the Toric Diagram}

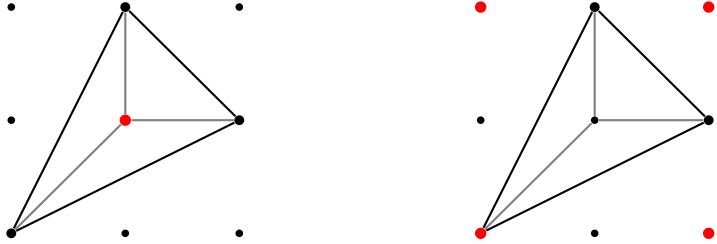
\begin{figure}[H]
\centering
\begin{subfigure}{0.4\textwidth}
	\centerline{\begin{tikzpicture}[auto, scale= 0.5]
		%%%%%%%%%%%% Nodes %%%%%%%%%%
		\node [circle, fill=black, inner sep=0pt, minimum size=1mm] (1) at (-3,3) {};		
		\node [circle, fill=black, inner sep=0pt, minimum size=1.3mm] (2) at (0,3) {}; 
		\node [circle, fill=black, inner sep=0pt, minimum size=1mm] (3) at (3,3) {};
		\node [circle, fill=black, inner sep=0pt, minimum size=1mm] (4) at (-3,0) {};
		\node [circle, fill=red, inner sep=0pt, minimum size=1.5mm] (5) at (0,0) {}; 
		\node [circle, fill=black, inner sep=0pt, minimum size=1.3mm] (6) at (3,0) {}; 			
		\node [circle, fill=black, inner sep=0pt, minimum size=1.3mm] (7) at (-3,-3) {};
		\node [circle, fill=black, inner sep=0pt, minimum size=1mm] (8) at (0,-3) {}; 
		\node [circle, fill=black, inner sep=0pt, minimum size=1mm] (9) at (3,-3) {};
		%%%%%%%%%%% Lines %%%%%%%%%%%
		\draw (2) edge (6) [thick];
		\draw (6) edge (7) [thick];
		\draw (7) edge (2) [thick];
		\draw (2) edge (5) [thick, gray];
		\draw (5) edge (6) [thick, gray];
		\draw (7) edge (5) [thick, gray];
		\end{tikzpicture}}
\end{subfigure}
\begin{subfigure}{0.4\textwidth}
\centerline{\begin{tikzpicture}[auto, scale= 0.5]
		%%%%%%%%%%%% Nodes %%%%%%%%%%
		\node [circle, fill=red, inner sep=0pt, minimum size=1.5mm] (1) at (-3,3) {};		
		\node [circle, fill=black, inner sep=0pt, minimum size=1.3mm] (2) at (0,3) {}; 
		\node [circle, fill=red, inner sep=0pt, minimum size=1.5mm] (3) at (3,3) {};
		\node [circle, fill=black, inner sep=0pt, minimum size=1mm] (4) at (-3,0) {};
		\node [circle, fill=black, inner sep=0pt, minimum size=1mm] (5) at (0,0) {}; 
		\node [circle, fill=black, inner sep=0pt, minimum size=1.3mm] (6) at (3,0) {}; 			
		\node [circle, fill=red, inner sep=0pt, minimum size=1.5mm] (7) at (-3,-3) {};
		\node [circle, fill=black, inner sep=0pt, minimum size=1mm] (8) at (0,-3) {}; 
		\node [circle, fill=red, inner sep=0pt, minimum size=1.5mm] (9) at (3,-3) {};
		%%%%%%%%%%% Lines %%%%%%%%%%%
		\draw (2) edge (6) [thick];
		\draw (6) edge (7) [thick];
		\draw (7) edge (2) [thick];
		\draw (2) edge (5) [thick, gray];
		\draw (5) edge (6) [thick, gray];
		\draw (7) edge (5) [thick, gray];
		\end{tikzpicture}}
\end{subfigure}
	\caption{The toric involution on the toric diagram of the resolved $\CC^3/\ZZ_3$ orbifold theory, with the fixed points displayed in red. On the left the resolved orientifolded theory with a compact $\Omega$7, while on the right it is showed the resolved unoriented theory with a non-compact $\Omega$7 and a $\Omega$3.} \label{fig:ToricRes}
\end{figure}

As we have seen, projections with four fixed points on the dimer preserve the mesonic symmetries while projections with fixed lines break them. In a similar fashion, we can represent the toric involution on the toric diagram by fixed points on an even sublattice and non-toric involution by a fixed line \cite{Garcia-Etxebarria:2016bpb}. In performing the orientifold projection, there are various choices for the even sub-lattice and the position of the fixed point on the toric diagram. The complete resolution of the singular space is described by triangles of minimal area on the toric diagram. If one corner lies on an extremal point, the triangle corresponds to a non-compact divisor in the geometry, otherwise it is a compact one. If one of the corners is an extremal fixed point, the minimal triangle represents a point in a non-compact $\Omega$7-plane, whereas with an internal fixed point the triangle is a point in a compact $\Omega$7. In the remaining case, all fixed points are external to the diagram and then the triangles corresponds to $\Omega$3-planes. Given a particular toric diagram, for example for $\CC^3/\ZZ_3$, we have different orientifold configurations as displayed in Fig.\,\eqref{fig:ToricRes}. The $\Omega$3 and/or $\Omega$7-planes emerge \emph{after} the complete resolution of the singular space, i.e. they live in the smooth resolved space. Thus, the cycles they wrap correspond to the cycles identified by the conjugacy classes we introduced previously. For instance, the presence of a senior class makes it possible for an $\Omega$7 to wrap a compact 4-cycle.

%Relation with the tiling;
We should stress that from the dimer it is easier to find all possible orientifolds and related projections of fields and nodes and the superpotential, from the toric diagram we easily read which $\Omega$-planes emerge after the resolution and finally from the quiver we easily read the matter content of the unoriented theory and the anomaly-cancellation conditions. Furthermore, the quiver diagram allows us to easily introduce additional flavour branes in the theory and read off the new RR tadpole cancellation conditions. As already mentioned we will not explore the construction of non-perturbative orientifolds along the lines of \cite{Garcia-Etxebarria:2015hua, Garcia-Etxebarria:2016bpb} because they require the inclusion of strongly coupled sectors in the IR. 

%%%%

\subsection{Adding Flavour Branes}

In the setup with D3-branes transverse to a singular toric CY and $\Omega$-planes we may add stacks of $M$ non-compact D7-branes which act as flavour branes. Their presence generates a non-dynamical D7-D7 open string sector, and a dynamical D3-D7 sector which adds new matter fields to the spectrum and to the superpotential. The group $U(M)$ is seen as a global symmetry by strings living on the D3. Under an orbifold quotient by $\ZZ_n$, the flavour groups becomes $U(M_{\alpha})$, $\alpha=0,\ldots,n{-}1$. We denote the new matter fields as $q_{a\alpha} = ({\bf{N}}_a, \overline{\mathbf{M}}_\alpha)$ for fields transforming in the fundamental of $U(N_a)$ of the $ a $-th fractional D3-branes and in the anti-fundamental of $U(M_\alpha)$ of the $ \alpha $-th flavour branes (i.e. starting from the D3-brane and ending on the D7-brane) while $\tilde{q}_{\alpha a} = (\mathbf{M}_\alpha, \overline{{\bf{N}}}_a)$ for open strings with the opposite orientation. They enter in the super-potential with terms of the form
\begin{equation}
W_{37} = (\tilde{q}_{\alpha a})_{i_{\alpha}}^{i_a} (M_{ab})_{i_a}^{i_b} (q_{b \alpha})_{i_b}^{i_{\alpha}} \; , 
\end{equation}
where $M_{ab} = ({X_{ac}})_{i_a}^{i_c} ({X_{cd}})_{i_c}^{i_d} \ldots ({X_{fb}})_{i_f}^{i_b}$ is a composite operator and the indices $i_a$ run over the $N_a$ colours of the $a$-th gauge group, while the indices $i_{\alpha}$ run over the $M_{\alpha}$ flavours of the $\alpha$-th flavour groups. 

When the D7's wrap non-compact cycles, they are much heavier than the D3's and thus backreact on the geometry, in which case the toricity may not be preserved. This can be further seen from the super-potential where the new terms coming from the D3-D7 sector break the toric condition. Notwithstanding the possibility of representing flavour branes as open paths in the dimer \cite{Franco:2006es}, one cannot define a tiling on a torus, but on other geometries \cite{Franco:2013ana}. In this case the machinery of toric geometry cannot be fully exploited beyond this point. However, the quiver description still holds, as described in \cite{Bianchi:2013gka}. %Hence, in order to maintain toricity the flavour branes must be taken as probes such that they are not much heavier than the D3 branes. As a result the D7 branes must wrap compact cycles.\\

\begin{figure}[H]
\centering
    \begin{subfigure}{0.35\textwidth}
     \centerline{\begin{tikzpicture}[auto, scale= 0.5]
        \node [circle, draw=blue!50, fill=blue!20, inner sep=0pt, minimum size=5mm] (a) at (0,0) {$N_a$};
        \node [circle, draw=blue!50, fill=blue!20, inner sep=0pt, minimum size=5mm] (b) at (4,0) {$N_b$};
        \draw (a) to node {$X_{ab}$} (b) [->, thick];
     \end{tikzpicture}}
    \end{subfigure}
    \begin{subfigure}{0.35\textwidth}
     \centerline{\begin{tikzpicture}[auto, scale= 0.5]
        \node [circle, draw=blue!50, fill=blue!20, inner sep=0pt, minimum size=5mm] (a) at (0,0) {$N_a$};
        \node [circle, draw=blue!50, fill=blue!20, inner sep=0pt, minimum size=5mm] (b) at (4,0) {$N_b$};
        \draw (a) to node {$X_{ab}^{p=1,2}$} (b) [->>, thick];
     \end{tikzpicture}}
    \end{subfigure}\\[15pt]
    \begin{subfigure}{0.35\textwidth}
     \centerline{\begin{tikzpicture}[auto, scale= 0.5]
        \node [circle, draw=blue!50, fill=blue!20, inner sep=0pt, minimum size=5mm] (a) at (0,0) {$N_a$};
        \node [rectangle, draw=red!50, fill=red!20, inner sep=0pt, minimum size=5mm] (b) at (4,0) {$M_{\alpha}$};
        \draw (a) to node [red] {$q_{a \alpha}$} (b) [->, thick, red];
     \end{tikzpicture}}
    \end{subfigure}
    \begin{subfigure}{0.35\textwidth}
     \centerline{\begin{tikzpicture}[auto, scale= 0.5]
        \node [rectangle, draw=red!50, fill=red!20, inner sep=0pt, minimum size=5mm] (a) at (0,0) {$M_{\alpha}$};
        \node [circle, draw=blue!50, fill=blue!20, inner sep=0pt, minimum size=5mm] (b) at (4,0) {$N_a$};
        \draw (a) to node [red] {${\tilde{q}}_{ \alpha a}$} (b) [->, thick, red];
     \end{tikzpicture}}
    \end{subfigure}
\caption{The various fields appearing in this work and their denomination.}\label{fig:legend}
\end{figure}
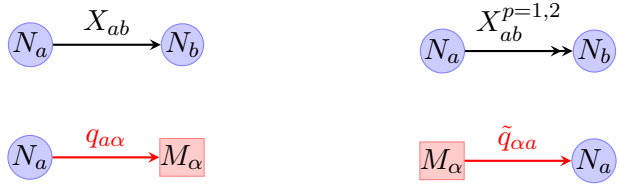

We can add flavour branes to the quiver as new nodes and links for the global symmetry groups and 3-7 fields, see Fig.\,\eqref{fig:legend}. The D3-D7 open strings sector contains chiral multiplets $\mathbf{C}^{\dot{a}}$ which transform in $Spin(4)$ with weights $\pm(\half, \half)$. Under the orbifold projection by $\ZZ_n$, $\mathbf{C}^{\dot{a}}$ transform as
\begin{equation}
\mathbf{C}^{\dot{a}} \mapsto \omega^{\pm \half (k_I + k_J)} \mathbf{C}^{\dot{a}} = \omega^{\pm s} \mathbf{C}^{\dot{a}} \; ,
\end{equation}
where $\omega = e^{i2\pi/n}$, $I \neq J \neq K = 1,2,3$, we have used the supersymmetry preserving condition $k_1 + k_2 = - k_3 \; \textrm{mod}\,n$ and we have defined  $s=(k_I + k_J)/2$. For a supersymmetric embedding of the D7 branes $\left( \omega^{\pm s}\right)^n = 1$, thus $k_K$ must be even. Moreover, $s$ determines the connection between gauge and global groups, as a colored node $a$ is connected to a flavour node $\alpha{+}s$. Note that we can embed different D7-branes on the same divisor $X_K=0$ but with different Chan-Paton factors \cite{Franco:2006es}. There are various choices, but here we show only one of them for $\CC^3/\ZZ_3$. Under an orientifold projection, flavour groups are projected to $SO/Sp(M_{\alpha})$ while D3-D7 strings are identified as in Eq.\,(\ref{Mirror}), namely 
\begin{align}\label{eq:FlavMirror}
\Omega & \, : \qquad \mathbf{M}_{\alpha} \leftrightarrow \overline{\textbf{M}}_{n{-}\alpha}\; , \nonumber \\ 
\widehat{\Omega} & \, : \qquad \mathbf{M}_{\alpha} \leftrightarrow \overline{\textbf{M}}_{n{-}\alpha{-}1}\; , \nonumber \\ 
q_{a \alpha} &= ({\bf{N}}_a, \overline{\mathbf{M}}_\alpha) \leftrightarrow (\mathbf{M}_{\alpha'}, \overline{{\bf{N}}}_{a'}) = \tilde{q}_{\alpha' a'} \; .
\end{align}

\subsection{Anomaly Cancellation Conditions}\label{Sec:Anomaly}
The low-energy gauge theories we have so far considered involve chiral fields, hence they can be potentially anomalous. In the toric and dimer literature it is sometimes pointed out that, if the original theory is non-anomalous and the ranks of the gauge groups are fixed, after the orientifold projection one needs to add flavour branes in order to cancel the potential anomaly. The reason is that, in the T-dual picture with 5-branes, the D5 charge flows along the NS5 branes and it can be transferred along cycles on the torus. One can set a system in which this flow is absorbed by four flavour D5' branes parallel to the $\Omega$5-planes, and the network of NS5 is such that the flow of D5 charge is indeed absorbed \cite{Imamura:2008fd}. Since D5 charge conservation leads to anomaly cancellation, the resulting theory is anomaly-free.

The method developed in \cite{Bianchi:2013gka} for unoriented $\CC^3/\ZZ_n$ orbifolds is different, since they put conditions on the rank of gauge groups after the inclusion of orientifold planes and flavour branes, not in the parent theory. It works as follows. Consider a node $a$ in the quiver and count how many fields transform under the gauge group. Those which go out from the node, i.e. transform in the fundamental representation, are counted with a positive sign, while those which enter in the node, i.e. transform in the anti-fundamental representation, take a negative sign. Also chiral fields with flavour indices must be counted, if flavour branes are present. When the $\Omega$-plane crosses fields which connect the node $a$ to its image $a'$, these are projected to their symmetric or antisymmetric representation and thus their contribution to the anomaly is $(N_a{+}4)$ and $(N_a{-}4)$ respectively, for each field. On the dimer, this corresponds to a $\Omega 5^\pm$ plane, i.e. a fixed point, on top of an edge. For each node $ a $, the anomaly cancellation condition for the orbifold theory reads
\begin{equation}\label{eq:anomalyGen} 
\sum_{b=1}^{n-1}\left( I_{ab} N_b + J_{ab} M_b \right) + 4 \sum_I \left(\epsilon_{aa'}^{(I)} -   \epsilon_{a'a}^{(I)} \right) = 0 \; ,
\end{equation}
where $I_{ab}$ and $J_{ab}$ count with orientation how many fields start from (or end on) the node $a$,
\begin{align}
I_{ab} =  \sum_{I=1}^3 \left( \delta_{a, \,b - k_I} - \delta_{a, \,b + k_I} \right) \; , \qquad
J_{ab} =  \, \delta_{a, \,b - s} - \delta_{a, \,b + s} \; , 
\end{align} 
and ${\epsilon}_{aa'}^{(I)}, {\epsilon}_{a'a}^{(I)} = \pm$ account for symmetric $ (+) $ or antisymmetric $ (-) $ fields, connecting nodes $a$ and $a' = a + k_I$ (or the opposite orientation). It is important to note their relative sign in the anomaly cancellation, due to their different orientation in the spectrum\footnote{Actually, in \cite{Bianchi:2013gka} the anomaly cancellation condition was different, because it was assumed that fields along the direction $I=1,2,3$ have always the same orientation. In the present work the general scenario is allowed, generalizing previous results. See below the example of $\CC^3/\ZZ_4$ $(1,1,2)$.}. In simple cases where fields along the direction $I$ have the same orientation, $\epsilon_{aa'}^{(I)} = - \epsilon_0 \epsilon_I$ and $\epsilon_{a'a}^{(I)} = \epsilon_0 \epsilon_I$. The above Eq.\,\eqref{eq:anomalyGen} is a generalization of the anomaly cancellation condition used in \cite{Bianchi:2013gka}, where it is was derived from the partition function of orbifold theories along the lines of \cite{Bianchi:2000de, Aldazabal:1999nu}. We will use it also for non-orbifold theories, since these can be related via mass deformations (see Sec.\,\eqref{MassDefo}) or Higgsing (see App.\,\eqref{sec:higgs}) to orbifold theories. In this case we still use the notation $\epsilon_{aa'}^{(I)}$, where $I$ stands for the fields in a multiplet rather than the orbifold directions. The above generalization allow us to exactly reproduce known results in the literature and to gain further physical intuition about unoriented gauge theories at general Calabi-Yau singularities, as we will see in the forthcoming sections.

\subsection{Conformal Invariance}

The presence of both $\Omega$-planes and flavour branes can alter the conformal properties of the original parent theory. It is thus an interesting question to ask which unoriented projections allow for a theory which is simultaneously anomaly-free and conformal in the perturbative regime, i.e. without the inclusion of strongly coupled sectors in the IR as done in \cite{Garcia-Etxebarria:2015hua, Garcia-Etxebarria:2016bpb}. We will show that not all the orientifold projections we consider fulfill this request and we will give a physical interpretation of the results in terms of $ \Omega $3 and compact/non-compact $ \Omega $7 planes.\\

Recall that the $\beta$-function of an  $\Nn=1$ gauge theory is
\begin{equation}
\beta \propto \frac{1}{2} \left( 3 {\ell}(\mathbf{Adj}) - \sum_I (1- \gamma_{_I}) {\ell}(\bf{R}_I) \right) \; ,
\end{equation}
where ${\ell}$ denotes the Dynkin index of the representation $ \textbf{R} $, $\gamma_{_I}$ is the anomalous dimension of the (bi-fundamental) chiral fields\footnote{Note that we define the anomalous dimension as $\gamma= - \partial Z / \partial \log \mu$, with $Z$ the field renormalisation $q \rightarrow \sqrt{Z} q$, then $\Delta = 1 + \frac{1}{2}\gamma$.} and the sum runs over all chiral fields transforming under the gauge group. When there are several gauge groups, each of them has its own $\beta$-function. This is the case for quiver gauge theories. To determine the anomalous dimensions of the chiral fields $X_{ab}$, one may use the properties of the superpotential and the fact that at the conformal point $\Delta = \frac{3}{2} R$, where $\Delta$ is the scaling dimension and $R$ is the $R$-charge, that has to be positive for chiral fields in a unitary theory. However, we should note that this relation holds for gauge-invariant operators, then we deduce $\Delta$ of the chiral fields from the composite mesonic operators. Further constraints that usually allow to determine the $R$-charges arise via $a$-maximization \cite{Intriligator:2003jj}.

%In~\cite{Bianchi:2013gka} it is argued that the anomalous dimensions vanish for fields in $\Nn=1$ unoriented theories $\CC^3/(\ZZ_n \times \ZZ_m)$, since they are obtained from $\Nn=2$ hypermultiplets. For example, this is the case for the unoriented projection of $\CC^3/\ZZ_3$ and $\CC^3/\ZZ_4$. On the other hand, fields in conifold $\Cc$ and its orbifold, e.g. $\Cc/\ZZ_2'$, have non-zero anomalous dimensions. As well known, this is because the conifold theory is obtained from the $\Nn=2$ theory $\CC^2/\ZZ_2 \times \CC$ with a mass deformation for the adjoint fields and in the IR the fields have different scaling dimension due to the presence of deformation scale parameter $m$ \cite{KlebanovWittenSuperconformalFieldTheoryonThreebranes1998}. 
  
The anomalous dimensions of fields in the parent theory and in the unoriented one differ by terms of order $1/N$, which are suppressed in the large $N$ limit. Moreover, fields transforming in the symmetric or antisymmetric representation are usually present in unoriented theories and their contribution to the beta function differs from the original fields in the parent theory. For these reasons, we should {\it a priori} allow for non-zero anomalous dimensions even in the simple case of an unoriented $\CC^3/\ZZ_n$ model.

Starting with a super-conformal quiver gauge theory, the orientifold projection may break conformal invariance. In case of an $\Omega$3-plane, which acts projecting fields $X_{aa'}^{I}$ onto different symmetric or antisymmetric representations for different $I$, conformal invariance may hold. On the other hand, an $\Omega$7-plane couples to the dilaton and then its presence usually breaks conformal invariance. In some cases, in order to obtain a conformal theory, we may add a suitable number of flavour branes, as shown in \cite{Bianchi:2013gka}.
For quiver gauge theories with flavour branes and $\Omega$-planes, the general $\beta$-function $\beta_a$ at the node $a$ reads
\begin{align}\label{eq:beta}
	\beta_a^{SU} &=
	3 N_a -  \sum_I \left( \epsilon^{(I)}_{aa'} + \epsilon^{(I)}_{a'a} \right) \left(1-\gamma^{(I)}_{aa'}\right) - \frac{1}{2} \left(I_{ab}^+ N_b + J_{ab}^+ M_b  \right)  \; , \nonumber \\[7pt]
	\beta_a^{SO/Sp} &= \frac{3}{2} N_a + 3 {\epsilon_a} - \frac{1}{2} \sum_I \left( \epsilon^{(I)}_{aa'} + \epsilon^{(I)}_{a'a} \right) \left(1-\gamma^{(I)}_{aa'}\right) - \frac{1}{4}  \left(I_{ab}^+ N_b + J_{ab}^+ M_b  \right) \; ,
	\end{align}
where $ \gamma_{ab} $ are the anomalous dimensions of the fields starting from the node $a$ and ending on the node $b$ and $\gamma_{aa'}=\gamma_{a'a}$. For (oriented) abelian orbifolds $\CC^3/\ZZ_n$, $\gamma_{ab}=0$. In (\ref{eq:beta}) we have used  
\begin{align}
I_{ab}^{+} &= \sum_{I=1}^3 \left( \delta_{a, b-k_I} + \delta_{a, b+k_I} \right)\left(1-\gamma_{ab}^{(I)}\right) \; , \qquad
J_{ab}^{+} = \delta_{a, b - s} + \delta_{a, b + s} \; ,
\end{align}
and $\epsilon_a=\pm$ projects a gauge group to $Sp$ $ (+) $ or $SO$ $ (-) $. The second term on the right hand side of both expressions in Eq.\,(\ref{eq:beta})  only exists if in the quiver there are fields that are  cut  by the orientifold action and start from the node $ a $. The above equations are a generalization of the action of a unique orientifold charge $\epsilon_0$ i.e. we allow different projections for gauge groups in the same theory: the same orientifold may project a gauge group to $SO(Sp)$ while another group is projected to $ Sp(SO) $. This generalization works also for the anomaly cancellation equation and is perfectly consistent with previous results obtained in the literature, as we will see in the examples below.

Furthermore, for $\CC^3/\ZZ_n$ theories the sum of all the beta functions is zero, i.e. $\sum_a \beta_a = 0$ since far away from the singularity the theory is conformal, but each $\beta_a$ may be non-zero. Notice that, as observed in \cite{Bianchi:2013gka}, $U(1)$ and $O(2)$ groups are free in the IR and so decouple from the dynamics at large distances. The vanishing of their $\beta$ functions cannot be achieved and should be relaxed. Similarly $O(1) = \ZZ_2$ has no proper
$\beta$ function. Needless to say, the beta function of an empty node should not be considered.

\subsection{Mass Deformation}\label{MassDefo}

The chiral superfields of the parent theory $X_{ab}$ transform in the bi-fundamentals under the gauge groups $({\bf{N}}_a, \overline{{\bf{N}}}_b)$ if $a\neq b$ or in the adjoint if $a\neq b$. If  \emph{a pair} of vector-like fields $X_{ab}$ and $X_{ba}$ or adjoint fields $X_{aa}=\phi_a$ are present one can deform the superpotential $W$ with relevant mass terms as
\begin{align}
\Delta W_X &=  m (X_{ab}X_{ba} - X_{cd}X_{dc}) \; , \nonumber \\
\Delta W_{\phi} &=  \frac{m}{2} (\phi_a^2 - \phi_b^2) \; , 
\end{align}
which explicitly break the toricity condition. One can then integrate out the fields by solving the corresponding F-term equations of the deformed superpotential $W_{\mathrm{def}} = W + \Delta W$ and plugging the solutions back in $W_{\mathrm{def}}$. The final theory is an IR theory, defined at energies below the scale parameter $m$ \cite{Klebanov:1998hh}. In some simple cases the resulting theory is still toric, but in general we need to redefine properly the remaining fields as
\begin{align}
X_{ab}' &= X_{ab} + \frac{1}{m} \sum_k c_k^{(ab)}X_{ak}X_{kb} \; , \nonumber \\
\phi_a' &= \phi_a + \frac{1}{m} \sum_k c_k^{a}X_{ak}X_{ka} \; , 
\end{align}
where $c_k^{(ab)}$ and $c_k^{a}$ are some coefficients that ensure the restoration of toric symmetry. A more detailed discussion is presented in \cite{Bianchi:2014qma}.

\begin{figure}[H]
	\centerline{\includegraphics[scale=0.35]{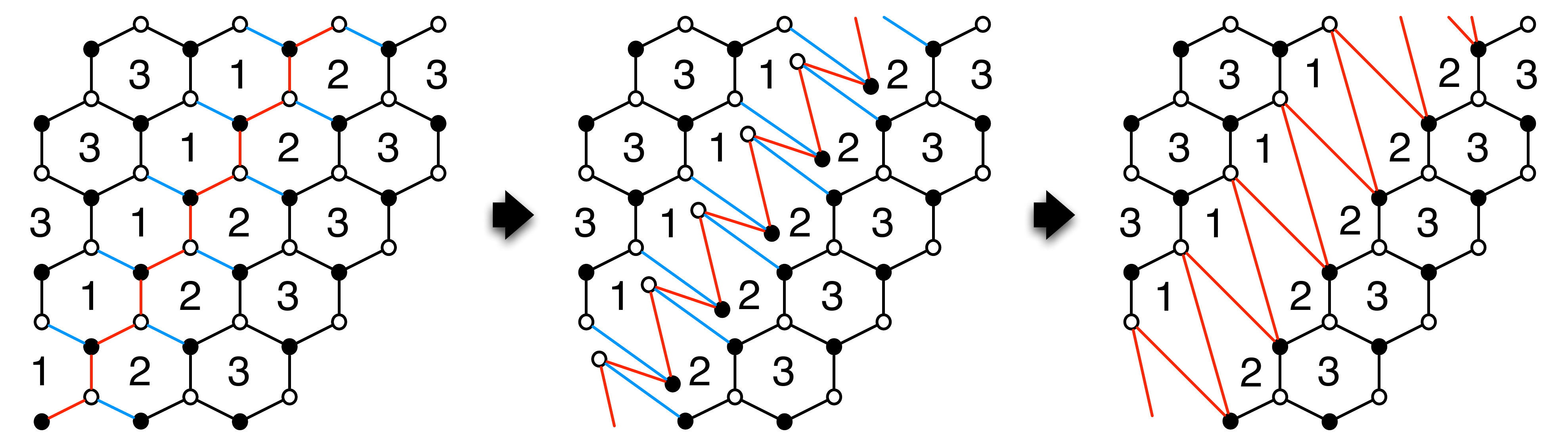}}
	\caption{The mass deformation of $\CC^2/\ZZ_3 \times \CC$ on the dimer. Image from~\cite{Bianchi:2014qma}.}\label{fig:massdef}
\end{figure}

The mass deformation has effects on quivers, dimers and toric diagrams. In quivers, the arrows corresponding to fields which are integrated out are eliminated. On the dimer, the tiling is deformed taking away the integrated fields, reducing the number of edges in the corresponding faces/gauge groups. An example is shown in Fig.\,\eqref{fig:massdef}. On the toric diagram, external points are moved around the perimeter, but their number is preserved. Since the toric diagram changes, the mesonic moduli space changes as well.

This procedure connects different toric theories, such as $\CC^3/\ZZ_4$ in Sec.~\eqref{Sec:C3Z4} and the chiral orbifold of the Conifold $\Cc/{\ZZ}'_2$ in Sec.~\eqref{Sec:CZ2}, whose unoriented projections can  thus  be related.

\section{Unoriented Toric Singularities and their Mass Deformation} \label{Examples}
In this Section, we will present several examples with small number of nodes. We will revisit by-now prototypical examples, like $\CC^3$ and $\CC^3/\ZZ_3$, as well as new ones. We will try and emphasize the connection between orbifold and non-orbifold singularities via mass-deformations or Higgsing that preserve toricity, where possible. In the next section we will discuss RG flows and Seiberg duality.

The super-potential $W$ consists of mesonic operators as $(X_{ab})_{i_a}^{i_b} (X_{bc})_{i_b}^{i^c} \ldots (X_{fa})_{i_f}^{i_a}$. For convenience of notation, we just write $X_{ab} X_{bc} \ldots X_{fa}$. The same holds when 3-7 fields $q_{a \alpha}$ and $\tilde{q}_{\alpha a}$ are involved. In this way, we use upper indices $I=1,2,3$ or $p=1,2$ to denote the different fields in a multiplet as $X_ {ab}^{I\,(p)}$, which enter with different combinations in mesonic operators. We use the extended notation only when needed for clarity.

\subsection{Unoriented Projections of D3-branes on $\CC^3$}

The first prototypical example is the unoriented projection of $\CC^3$ \cite{Franco:2007ii, Garcia-Etxebarria:2016bpb}. The parent gauge theory is $\Nn=4$ Super Yang-Mills (SYM) theory with super-potential
\begin{equation}
W = \Tr{\;\Phi_1 [\Phi_2, \Phi_3]} \; ,
\end{equation}
where $\Phi_I$ are adjoint fields, which parametrize the three complex directions transverse to the stack of D3-branes. The toric diagram of this theory is a triangle with no internal point and consequently there are two different involutions given by an $\Omega3$ or a non-compact $\Omega7$.

\begin{figure}[H]
\centering
	 \begin{subfigure}{0.3 \textwidth}
		\centerline{\begin{tikzpicture}[auto, scale= 0.9]
		%%%%%%%%%%%% Nodes %%%%%%%%%%
		\node [circle, draw=blue!50, fill=blue!20, inner sep=0pt, minimum size=5mm] (0) at (0,0) {$N$}; 
		%%%%%%%%%%% Lines %%%%%%%%%%%
        \draw (0) to [out=120, in=60, looseness=10, thick] (0) ;
		\draw (0) to [out=130, in=50, looseness=13] (0) ;
		\draw (0) to [out=140, in=40, looseness=16] (0) ;
		\draw [thick, dashed, gray] (0, -1) to node [pos=0.01]{$\Omega$}(0,2);
		\end{tikzpicture}}
	\end{subfigure}
	\begin{subfigure}{0.3\textwidth}
		\centerline{\includegraphics[scale=0.3]{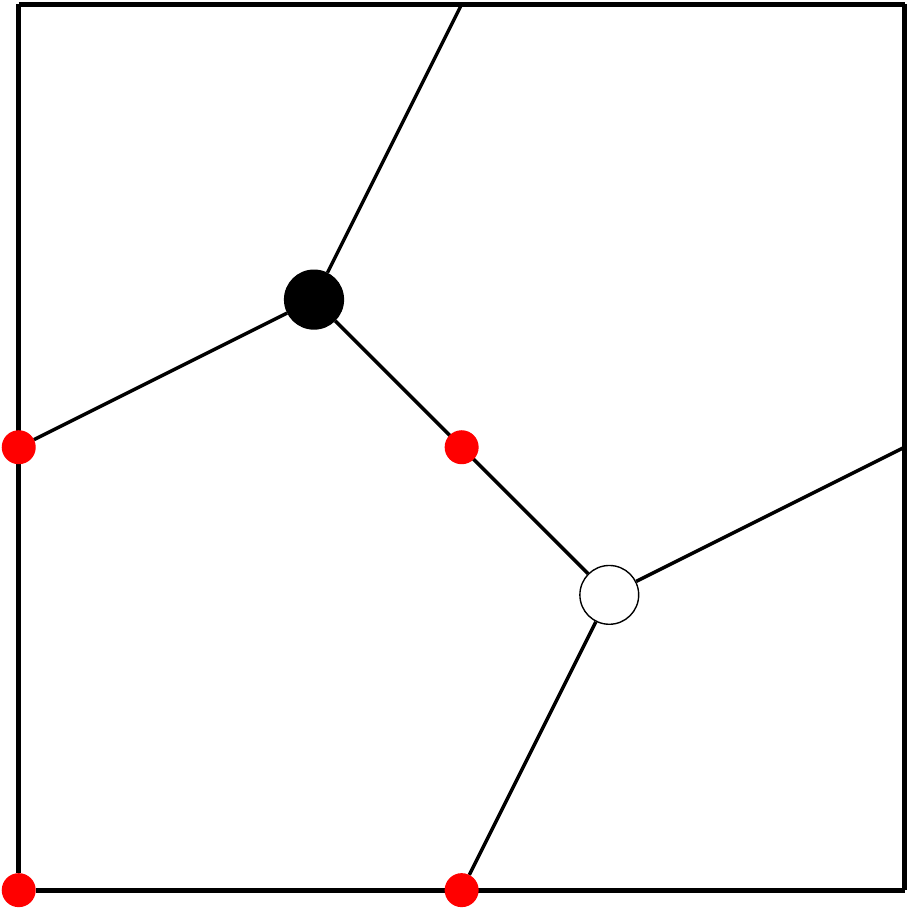}}
	\end{subfigure}
\centering
\begin{subfigure}{0.3\textwidth}
		\centerline{\begin{tikzpicture}[auto, scale= 0.4]
		%%%%%%%%%%%% Nodes %%%%%%%%%%
        \node [circle, fill=red, inner sep=0pt, minimum size=1.5mm] (1) at (-3,3) {};		
		\node [circle, fill=black, inner sep=0pt, minimum size=1.3mm] (2) at (0,3) {}; 
		\node [circle, fill=red, inner sep=0pt, minimum size=1.5mm] (3) at (3,3) {};
		\node [circle, fill=black, inner sep=0pt, minimum size=1mm] (4) at (-3,0) {};
		\node [circle, fill=black, inner sep=0pt, minimum size=1.3mm] (5) at (0,0) {}; 
		\node [circle, fill=black, inner sep=0pt, minimum size=1.3mm] (6) at (3,0) {}; 			
		\node [circle, fill=red, inner sep=0pt, minimum size=1.5mm] (7) at (-3,-3) {};
		\node [circle, fill=black, inner sep=0pt, minimum size=1mm] (8) at (0,-3) {}; 
		\node [circle, fill=red, inner sep=0pt, minimum size=1.5mm] (9) at (3,-3) {};
		\node [circle, fill=black, inner sep=0pt, minimum size=1mm] (10) at (6,3) {};
		\node [circle, fill=black, inner sep=0pt, minimum size=1.3mm] (11) at (6,0) {};
		\node [circle, fill=black, inner sep=0pt, minimum size=1.3mm] (12) at (6,-3) {};
		%%%%%%%%%%% Lines %%%%%%%%%%%
        \draw (2) edge (6) [thick];
		\draw (6) edge (5) [thick];
		\draw (5) edge (2) [thick];
		\draw (11) edge (12) [thick];
		\draw (12) edge (9) [thick];
		\draw (9) edge (11) [thick];
		\end{tikzpicture}}
	\end{subfigure} \\[20pt]
    \caption{The unoriented quiver (left) and the dimer (center) of $\CC^3$, while the unoriented toric diagram (right) with the two orientifolds $\Omega3$ and $\Omega7$.}\label{fig:C3}
\end{figure}
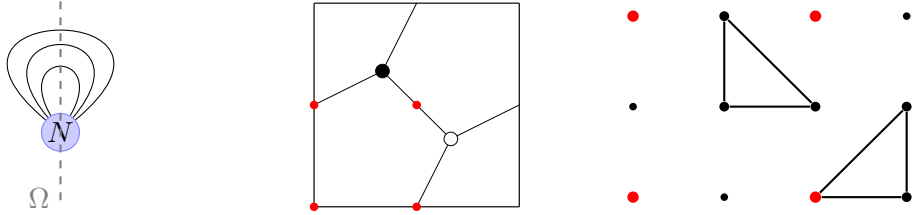

Following the quiver description presented in \cite{Bianchi:2013gka}, the orientifold charges for an $\Omega3^{\pm}$, $(\pm, - ,- ,-)$ give $\Nn=4$ SYM with $Sp (SO)(N)$ gauge group and the three fields transforming in the (anti)symmetric representation; the orientifold charges for a $\Omega7^{\pm}$, $(\mp,+,+,-)$ yield a $\Nn=2$ gauge theory with either $Sp(N)$ and two antisymmetric fields and one symmetric, or $SO(N)$ with two symmetric fields and one antisymmetric. The super-potential of the parent theory is easily read from the dimer and, after the unoriented projection, the product of the T-parities on the dimer must be negative due to the constraint given by Eq.\,\eqref{eq:dimersgn}. The first of various choices\footnote{The first T-parity lays on the face, the other three follow in clockwise order on the dimer.} is $(\tau_0, \tau_1, \tau_2, \tau_3) = (\mp, \pm, \pm ,\pm)$, which gives $\Nn=4$ SYM. In fact, the three mesonic operators on the dimer flip their signs under $\Omega3^{\pm}$. For this model the relation with the orientifold charges is $\epsilon_0 = - \tau_0 = \pm$, $\epsilon_1 = \tau _0 \tau_1=-$, $\epsilon_2 = \tau _0 \tau_2=-$, $\epsilon_3 = \tau_0 \tau_3=-$, and corresponds to an $\Omega3^{\pm}$. On the other hand, the action of an $\Omega7$ is encoded in such choices of T-parities as $(\pm, \pm, \pm, \mp)$ or permutation of the last three signs. The three diagrams are drawn in Fig.\,\eqref{fig:C3}. Finally, note that $\Omega3$ yields a conformally invariant theory with canonical dimensions and R-charges for the three adjoint chiral fields. 

The next simplest case is a stack of regular D3-branes placed at the singularity $\CC^3/\ZZ_n$. This gives a quiver theory with $n$ nodes, representing fractional branes wrapping internal cycles. Away from the origin, the space looks like $\CC^3$, thus, moving the stack of D3 branes away of the singularity, the theory goes back to the conformal $\Nn=4$ SYM and the same reasoning holds with the unoriented projection $\Omega3$ of an orbifold theory $\CC^3/\ZZ_n$. This imposes that the beta functions $\beta_a$ add up to zero
\begin{equation}\label{eq:BetaRule}
\sum_a \beta_a = 0 \; ,
\end{equation} 
which corresponds to a constant complex dilaton $S$. In fact, for $\CC^3/\ZZ_n$ the gauge kinetic function at each node, $f_a$, sum up to $S$. Note that each node represent a stack of different kind of fractional D-branes: point-like D3's, D5's wrapped around 2-cycles and `compact' D7's wrapped around 4-cycles.

\subsection{Orientifold of $\Nn=1$ Orbifold $\CC^3/\ZZ_3$, $(1,1,1)$}

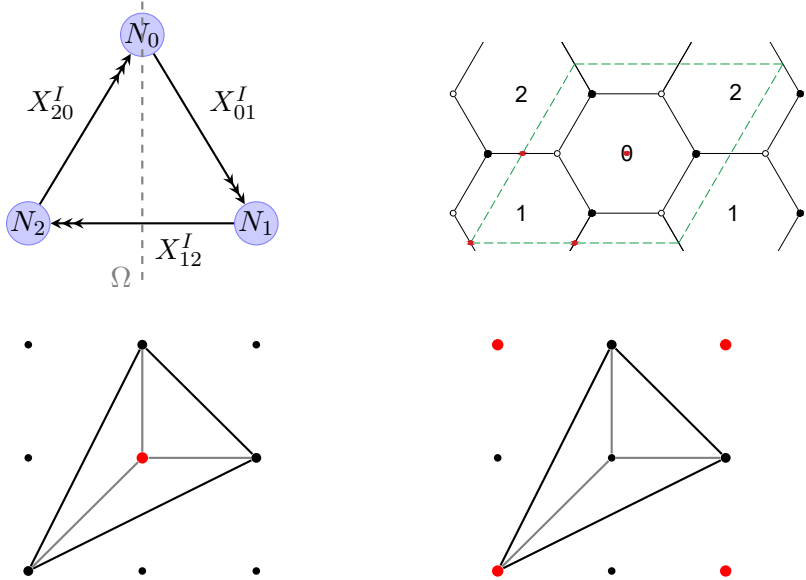
\begin{figure}[h]
\centering
	 \begin{subfigure}{0.4 \textwidth}
		\centerline{\begin{tikzpicture}[auto, scale= 0.5]
		%%%%%%%%%%%% Nodes %%%%%%%%%%
		\node [circle, draw=blue!50, fill=blue!20, inner sep=0pt, minimum size=5mm] (0) at (0,5) {$N_0$}; 
		\node [circle, draw=blue!50, fill=blue!20, inner sep=0pt, minimum size=5mm] (1) at (3,0) {$N_1$}; 			
		\node [circle, draw=blue!50, fill=blue!20, inner sep=0pt, minimum size=5mm] (2) at (-3,0) {$N_2$};
		%%%%%%%%%%% Lines %%%%%%%%%%%
        \draw (0) to node {$X_{01}^{I}$} (1) [->>>, thick];
		\draw (1) to node [pos=0.3] {$X_{12}^{I}$} (2) [->>>, thick];
		\draw (2) to node {$X_{20}^{I}$} (0) [->>>, thick];
		\draw [thick, dashed, gray] (0, -1.5) to node [pos=0.01]{$\Omega$}(0,6);
		\end{tikzpicture}}
	\end{subfigure}
	\begin{subfigure}{0.4\textwidth}
		\centerline{\includegraphics[scale=0.3, trim={2cm 5cm 2cm 15.5cm}, clip]{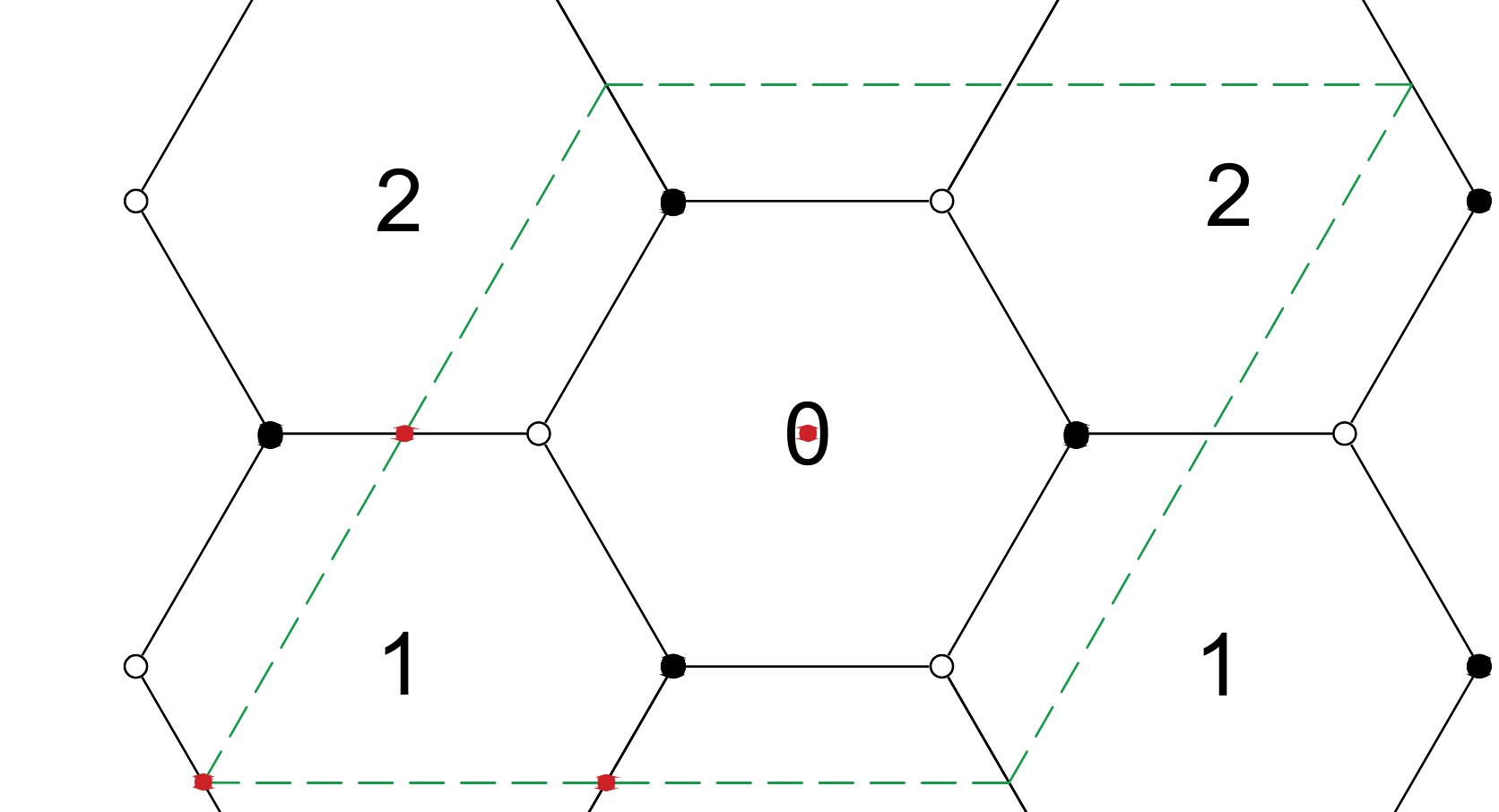}}
	\end{subfigure}
	 \\[15pt]
	\begin{subfigure}{0.4\textwidth}
		\centerline{\begin{tikzpicture}[auto, scale= 0.5]
		%%%%%%%%%%%% Nodes %%%%%%%%%%
        \node [circle, fill=black, inner sep=0pt, minimum size=1mm] (1) at (-3,3) {};		
		\node [circle, fill=black, inner sep=0pt, minimum size=1.3mm] (2) at (0,3) {}; 
		\node [circle, fill=black, inner sep=0pt, minimum size=1mm] (3) at (3,3) {};
		\node [circle, fill=black, inner sep=0pt, minimum size=1mm] (4) at (-3,0) {};
		\node [circle, fill=red, inner sep=0pt, minimum size=1.5mm] (5) at (0,0) {}; 
		\node [circle, fill=black, inner sep=0pt, minimum size=1.3mm] (6) at (3,0) {}; 			
		\node [circle, fill=black, inner sep=0pt, minimum size=1.3mm] (7) at (-3,-3) {};
		\node [circle, fill=black, inner sep=0pt, minimum size=1mm] (8) at (0,-3) {}; 
		\node [circle, fill=black, inner sep=0pt, minimum size=1mm] (9) at (3,-3) {};
		%%%%%%%%%%% Lines %%%%%%%%%%%
        \draw (2) edge (6) [thick];
		\draw (6) edge (7) [thick];
		\draw (7) edge (2) [thick];
		\draw (2) edge (5) [thick, gray];
		\draw (5) edge (6) [thick, gray];
		\draw (7) edge (5) [thick, gray];
		\end{tikzpicture}}
	\end{subfigure}
\begin{subfigure}{0.4\textwidth}
\centerline{\begin{tikzpicture}[auto, scale= 0.5]
		%%%%%%%%%%%% Nodes %%%%%%%%%%
		\node [circle, fill=red, inner sep=0pt, minimum size=1.5mm] (1) at (-3,3) {};		
		\node [circle, fill=black, inner sep=0pt, minimum size=1.3mm] (2) at (0,3) {}; 
		\node [circle, fill=red, inner sep=0pt, minimum size=1.5mm] (3) at (3,3) {};
		\node [circle, fill=black, inner sep=0pt, minimum size=1mm] (4) at (-3,0) {};
		\node [circle, fill=black, inner sep=0pt, minimum size=1mm] (5) at (0,0) {}; 
		\node [circle, fill=black, inner sep=0pt, minimum size=1.3mm] (6) at (3,0) {}; 			
		\node [circle, fill=red, inner sep=0pt, minimum size=1.5mm] (7) at (-3,-3) {};
		\node [circle, fill=black, inner sep=0pt, minimum size=1mm] (8) at (0,-3) {}; 
		\node [circle, fill=red, inner sep=0pt, minimum size=1.5mm] (9) at (3,-3) {};
		%%%%%%%%%%% Lines %%%%%%%%%%%
		\draw (2) edge (6) [thick];
		\draw (6) edge (7) [thick];
		\draw (7) edge (2) [thick];
		\draw (2) edge (5) [thick, gray];
		\draw (5) edge (6) [thick, gray];
		\draw (7) edge (5) [thick, gray];
		\end{tikzpicture}}
\end{subfigure}	
    \caption{The various unoriented descriptions of $ \CC^3/\ZZ_3 $. The upper row shows the quiver (left) and the $\Omega$-line, whereas on the right side the dimer, with the four fixed points in red. In the lower row: the toric diagram and the toric involution with a compact $\Omega$7 (left) and a non-compact $\Omega7$ (right).}\label{fig:C3Z3}
\end{figure}

In this section we analyze the second prototypical example, the unoriented projection of the chiral orbifold $\CC^3/\ZZ_3$ \cite{Feng_2001, Franco:2007ii, Garcia-Etxebarria:2016bpb}, whose different descriptions are drawn in Fig.\,\eqref{fig:C3Z3}. We study the $\Nn=1$ theory with $k_I=(1,1,1)$, whose quiver is the same as the theory with $k_I=(2,2,2)$, up to relabelling the nodes. Among the conjugacy classes, summarized in Tab.~\eqref{tab:C3Z3classes}, the single senior class signals the presence of a single compact 4-cycle and this information will be useful later.

\begin{center}
\begin{tabular*}{0.55\textwidth}{@{\extracolsep{\fill}}ccc}
\toprule
$(k_1,\, k_2,\, k_3)$ & Age$=\frac{1}{3} \sum_I k_I$ & Conjugacy class\\
\midrule
$(0,\, 0,\, 0)$ & 0 & Baby \\
$(1,\, 1,\, 1)$ & 1 & Junior \\
$(2,\, 2,\, 2)$ & 2 & Senior \\
\bottomrule
\end{tabular*}
\end{center}  
\captionof{table}{The conjugacy classes of the orbifold model $\CC^3/\ZZ_3$.}\label{tab:C3Z3classes}
\vspace{15pt}

The super-potential is 
\begin{equation}
W = {\epsilon}_{_{IJK}} X_{01}^{I}X_{12}^{J}X_{20}^{K} \; ,
\end{equation}
which enjoys the mesonic symmetry $SU(3)$. In fact, the theory has symmetry $SU(3) \times U(1)_R$, where $U(1)_R$ is the R-symmetry of the $\Nn=1$ SCFT. 

Let us perform the unoriented projection with a compact $\Omega$7-plane in the resolved space, whose toric diagram is the left one showed in Fig.\,\eqref{fig:C3Z3}. Since $n=3$ is odd, there are only three equivalent projections with the $\Omega$-plane on top of a node. Here, without loss of generality, we consider only an $\Omega$-plane through the node 0. This orientifold involution acts on the orbifold as 
\begin{align}
\overline{{\bf{N}}}_2 = {\bf{N}}_1 \; , \qquad
U(N_0)  \rightarrow Sp/SO (N_0) \; 
\end{align}
and the super-potential becomes
\begin{equation}
W' = {\epsilon}_{_{IJK}} X_{01}^{I}X_{11'}^{J}X_{1'0}^{K} \; .
\end{equation}

The anomaly cancellation condition Eq.~\eqref{eq:anomalyGen} reads
\begin{equation}\label{eq:Z3Anomaly}
N_0 = N_1 + \frac{4}{3} \sum_{I=1}^3 \epsilon^{(I)}_{11'} = N_1 \pm 4 \; ,
\end{equation}
which is indeed what we would have obtained with an $\Omega3^{\pm}$ placed at the origin of the singular space. According to the toric diagram, the orientifold plane is a compact $\Omega7$ in the smooth resolved space, which wraps the compact 4-cycle (whose presence is signaled by the senior conjugacy class) that in the singular space corresponds to an $ \Omega3^{\pm} $-plane. The results are summarized in Tab.~\eqref{tab:C3Z3}.
From the dimer, we can reproduce this unoriented projection with four fixed points ${\tau}_i=(\pm, \mp, \mp, \mp)$ as displayed in Fig.\,\eqref{fig:C3Z3} and negative overall product of the T-parities ${\tau}_i$. Again in this case, the relation between T-parities and orientifold charges is given by $\epsilon_0 = - \tau_0$ and the relative sign of the mesonic operators under the orientifold involution $\epsilon_I = \tau_0 \tau_I$.

\begin{center}
\begin{tabular*}{0.95\textwidth}{@{\extracolsep{\fill}}cccc}
\toprule
Orientifold & Gauge groups & Anomaly condition & $(X_{11'}^{1}, X_{11'}^{2}, X_{11'}^{3})$\\
\midrule
$\Omega3^+$ & $Sp(N_0)\times U(N_1) $ & $N_0 = N_1 + 4$ & $ (S,S,S) $ \\
$\Omega3^-$ & $SO(N_0)\times U(N_1) $ & $N_0 = N_1 - 4$ & $ (A,A,A) $ \\
\bottomrule
\end{tabular*}
\end{center}  
\captionof{table}{The unoriented projection $\Omega3^{\pm}$ (on the singular space) of the orbifold model $\CC^3/\ZZ_3$ which, in the resolved space, is a compact $\Omega7$-plane. The fields $X_{11'}^{I}$ are projected onto symmetric or antisymmetric representation, where “A” stands for “Antisymmetric representation”, while “S” for “Symmetric representation”.\\[10pt]}\label{tab:C3Z3}

Let us compute the $\beta$-functions. From Eq.~\eqref{eq:beta} and using Eq.~\eqref{eq:Z3Anomaly} we obtain
\begin{align}
\beta_0^{SO/Sp} &= \frac{3}{2}N_1\gamma_{01} + 2 \sum_{I=1}^3 \epsilon_{11'}^{(I)} + 3 \epsilon_0 \; , \nonumber \\
\beta_1^{SU} &= \frac{3}{2} N_1 \left( \gamma_{01} + \gamma_{11'} \right) + \sum_{I=1}^3 \epsilon_{11'}^{(I)} \left( -3+ \gamma_{11'} + 2 \gamma_{01} \right) \; .
\end{align}
Assuming there is a conformal point, where $1-\gamma_{ab}=3(1-R_{ab})$, we have for the R-charges
\begin{align}
&R_{01} \frac{9}{2}N_1 = 3 N_1 - 3 \epsilon_0 - 2 \sum_{I=1}^3 \epsilon_{11'}^{(I)} \; , \nonumber \\
&R_{01} \left( \frac{9}{2}N_1 + 6 \sum_{I=1}^3\epsilon_{11'}^{(I)} \right) + R_{11'} \left( \frac{9}{2}N_1 + 3 \sum_{I=1}^3 \epsilon_{11'}^{(I)} \right)= 3 \left( 2N_1 + 3 \sum_{I=1}^3 \epsilon_{11'}^{(I)} \right) \; , \nonumber \\
&2 R_{01} + R_{11'} = 2 \; ,  
\end{align}
where the last equation comes from the super-potential. The solution of the system is
\begin{align}
R_{01} = \frac{2}{3} \frac{N_1-\sum_{I=1}^3 \epsilon_{11'}^{(I)}}{N_1} \; , \nonumber \\
R_{11'} = \frac{2}{3} \frac{N_1 + 2\sum_{I=1}^3 \epsilon_{11'}^{(I)}}{N_1} \; , 
\end{align}
from which $\gamma_{01} = - 2  \frac{ \sum_{I=1}^3\epsilon_{11'}^{(I)}}{N_1} $, $\gamma_{11'} = + 2 \frac{ \sum_{I=1}^3 \epsilon_{11'}^{(I)}}{N_1}$. Note that in the large $N$ limit we retrieve back the anomalous dimensions of the parent theory, namely, $\gamma_{11'}=\gamma_{01}=0$ and the sum of the beta functions vanishes if $3\epsilon_0 = \sum_{I=1}^3 \epsilon_{11'}^{(I)}$. 

Note also that in the resolved space the compact $\Omega$7 acts in such a way that all the three fields $X_{11'}^I$ are projected in the same way, as we can see from the anomaly cancellation condition Eq.\,\eqref{eq:Z3Anomaly}.

\begin{figure}[H]
\centerline{\begin{tikzpicture}[auto,scale= 0.4]
		%%%%%%%%%%%% Nodes %%%%%%%%%%
		\node [circle, draw=blue!50, fill=blue!20, inner sep=0pt, minimum size=5mm] (0) at (0,5) {$N_0$}; 
		\node [circle, draw=blue!50, fill=blue!20, inner sep=0pt, minimum size=5mm] (1) at (3,0) {$N_1$}; 			
		\node [circle, draw=blue!50, fill=blue!20, inner sep=0pt, minimum size=5mm] (2) at (-3,0) {$N_2$};
		%%Flavour%% 
		\node [rectangle, draw=red!50, fill=red!20, inner sep=0pt, minimum size=5mm] (a) at (0,-5) {$M_0$};
		\node [rectangle, draw=red!50, fill=red!20, inner sep=0pt, minimum size=5mm] (b) at (6,5) {$M_1$};
		\node [rectangle, draw=red!50, fill=red!20, inner sep=0pt, minimum size=5mm] (c) at (-6,5) {$M_2$};
		%%%%%%%%%%% Lines %%%%%%%%%%%
        \draw (0) to (1) [->>>, thick];
		\draw (1) to (2) [->>>, thick];
		\draw (2) to (0) [->>>, thick];
		\draw [->, red, thick] (0) to node [swap] {${q_0}_2$} (c);
		\draw [->, red, thick] (c)  to node [swap] {$\tilde{q}_2{}_2$} (2); 
		\draw [->, red, thick] (2) to node [swap] {${q_2}_0$} (a); 
		\draw [->, red, thick] (a) to node [swap] {$\tilde{q}_0{}_1$} (1); 
		\draw [->, red, thick] (1) to node [swap] {${q_1}_1$} (b);
		\draw [->, red, thick] (b) to node [swap] {$\tilde{q}_1{}_0$} (0);
		\draw [thick, dashed, gray] (0, -5.5) to node {$\Omega$}(0,6.5);
		\end{tikzpicture}}
\caption{The quiver of $\CC^3/\ZZ_3$ with the addition of flavour branes.}\label{fig:C3Z3Flav}
\end{figure}
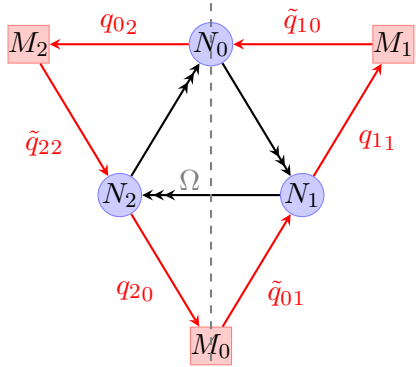

In \cite{Bianchi:2013gka} it is been argued that one may add non-compact flavour branes in order to recover conformal invariance as shown in Fig.\,\eqref{fig:C3Z3Flav}. We discuss the presence of flavour branes only for this model. Since the D7-branes yield additional chiral fields, the new anomaly equations read
\begin{equation}\label{eq:Z3FlavAnomaly}
M_0 - M_1 = 3 (N_1 - N_0) + 4 \sum_{I = 1}^3 \epsilon^{(I)}_{11'} \; ,
\end{equation}
that can be solved even with the presence of $\Omega7$ in the singular space and a judicious choice of $M_1$ and $M_0$. These branes enter in the super-potential with new terms such as
\begin{equation}
W = {\epsilon}_{_{IJK}} X_{01}^{I}X_{12}^{J}X_{20}^{K} + {\tilde{q}}_{\alpha a} X_{ab}^3 q_{b \alpha} \; 
\end{equation}
wrapping the flavour branes along the divisor $X_3 = 0$. The orientifold action on the flavour groups and super-potential is
\begin{align}
\bf{\overline{M}}_3 &= \bf{M}_2 \; , \nonumber \\
U(M_0) & \rightarrow Sp/SO (M_0) \; , \nonumber \\
W' = {\epsilon}_{_{IJK}} X_{02'}^{I}X_{2'2}^{J}X_{20}^{K} + {\tilde{q}}_{0 2'} &X_{2'2}^3 q_{2  \, 0} + {\tilde{q}}_{1 0} X_{01}^3 q_{1  \, 1} + {\tilde{q}}_{1' 2} X_{20}^3 q_{0  \, 1'}\; .
\end{align}

The presence of non-compact flavour branes breaks the toric condition and we cannot use the orientifold rules from the dimer. The tiling would not be defined on a torus and the orientifold involution gives rise to different geometries, as already mentioned. The beta-functions, together with the anomaly-free condition, in this case read
\begin{align}
2 \beta_0 &= 3 N_1 \gamma_{01} + 6 \epsilon_0 + 4 \sum_{I = 1}^3 \epsilon_{11'}^{(I)} - M_0 \; , \nonumber \\
2 \beta_1 &= N_1 \left( 3 \gamma_{01} + \sum_{I=1}^3 \gamma_{11'}^{(I)}   \right)   + 2 \sum_{I=1}^3 \epsilon_{11'}^{(I)} \left(-3 + 2 \gamma_{01} + \gamma_{11'}^{(I)}\right) + M_0 \left(-2 + \gamma_{01}\right) - M_0 \gamma_{01} \; . 
\end{align}
It is important to note that now, thanks to the presence of flavour branes, the beta functions can vanish separately if a judicious choice of the signs and number of flavour branes is made. Imposing the vanishing of the sum of the beta functions when the anomalous dimensions are trivial, we get
\begin{equation} 
3 \epsilon_{0} - \sum_{I=1}^3 \epsilon_{11'}^{(I)} = \half (M_0 + 2 M_1) \; ,
\end{equation}
which is solved, for instance, for $\epsilon_{0}=+1, \sum_{I=1}^3 \epsilon_{11'}^{(I)} = -3$ and $M_0 = M_1 = 4$. This scenario corresponds to the presence of a $ \Omega 3 $-plane or, better, a compact $ \Omega 7 $-plane in the resolved space wrapped on a 4-cycle, whose existence is guaranteed by the Ito-Reid theorem. Note that in this case the beta functions do no vanish separately. A second choice corresponds to considering $ \epsilon_{0} = +1 $ and $ \sum_{I=1}^3 \epsilon_{11'}^{(I)} = \pm 1 $ which is related to the presence of a non-compact $ \Omega 7 $-plane.

\subsection{Orientifold of the First del Pezzo Surface ($ dP_1 $)}\label{Sec:dP1}

\begin{figure}[h]
\centering
	\begin{subfigure}{0.25\textwidth}
			\centerline{\begin{tikzpicture}[auto, scale= 0.5]
		%%%%%%%%%%%% Nodes %%%%%%%%%
		\node [circle, draw=blue!50, fill=blue!20, inner sep=0pt, minimum size=5mm] (0) at (3,3) {$N_0$}; 
		\node [circle, draw=blue!50, fill=blue!20, inner sep=0pt, minimum size=5mm] (1) at (3,-3) {$N_1$}; 			
		\node [circle, draw=blue!50, fill=blue!20, inner sep=0pt, minimum size=5mm] (2) at (-3,-3) {$N_2$};
		\node [circle, draw=blue!50, fill=blue!20, inner sep=0pt, minimum size=5mm] (3) at (-3,3) {$N_3$}; 
		%%%%%%%%%%% Lines %%%%%%%%%%%
		\draw (0)  to node [pos=0.6] {$X_{01}^{p}$} (1) [->>, thick];
		\draw (1)  to node [swap] {$X_{12}^{p}$} (2) [->>, thick];
		\draw (1)  [bend left=70] to node [pos=0.5, swap] {$X_{12}^3$} (2) [->, thick];
		\draw (2)  to node [pos=0.4] {$X_{23}^{p}$} (3) [->>, thick];
		\draw (3)  to node {$X_{30}$} (0) [->, thick];
        \draw (3)  to node [pos=0.3, swap] {$X_{31}$} (1) [->, thick];
        \draw (2)  to node [pos=0.7, swap] {$X_{20}$} (0) [->, thick];
		\end{tikzpicture}}
	\end{subfigure}
	\hfill
	\begin{subfigure}{0.3\textwidth}
		\centerline{\includegraphics[scale=0.3, trim={3cm 5cm 1.5cm 13cm}, clip]{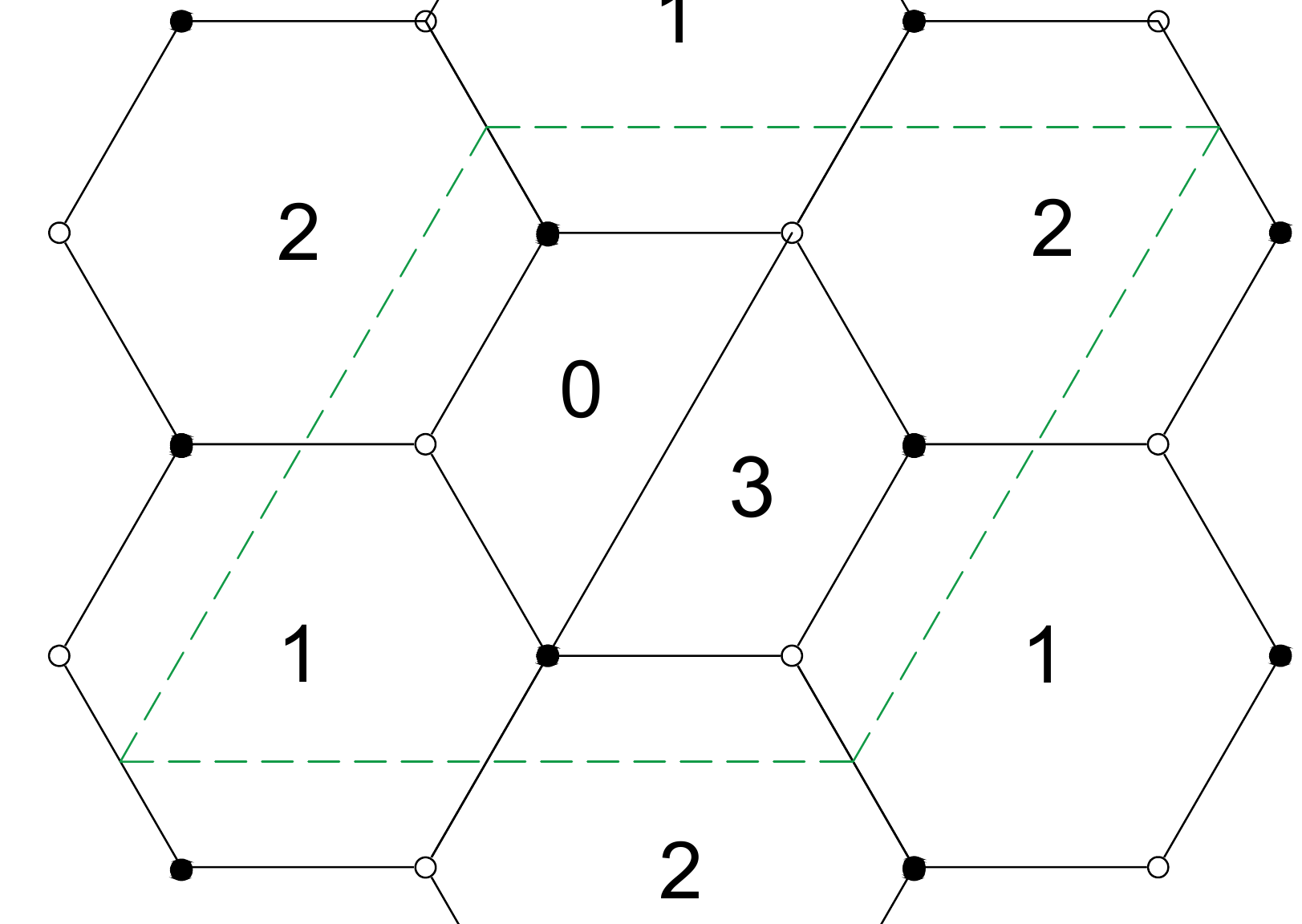}}
	\end{subfigure}
	\hfill
\begin{subfigure}{0.3\textwidth}
        \centerline{\begin{tikzpicture}[auto, scale= 0.5]
		%%%%%%%%%%%% Nodes %%%%%%%%%%
        \node [circle, fill=black, inner sep=0pt, minimum size=1mm] (1) at (-3,3) {};		
		\node [circle, fill=black, inner sep=0pt, minimum size=1mm] (2) at (0,3) {}; 
		\node [circle, fill=black, inner sep=0pt, minimum size=1mm] (3) at (3,3) {};
		\node [circle, fill=black, inner sep=0pt, minimum size=1mm] (4) at (-3,0) {};
		\node [circle, fill=black, inner sep=0pt, minimum size=1mm] (5) at (0,0) {}; 
		\node [circle, fill=black, inner sep=0pt, minimum size=1mm] (6) at (3,0) {}; 			
		\node [circle, fill=black, inner sep=0pt, minimum size=1mm] (7) at (-3,-3) {};
		\node [circle, fill=black, inner sep=0pt, minimum size=1mm] (8) at (0,-3) {}; 
		\node [circle, fill=black, inner sep=0pt, minimum size=1mm] (9) at (3,-3) {};
		%%%%%%%%%%% Lines %%%%%%%%%%%
        \draw (2) edge (6) [thick];
		\draw (6) edge (7) [thick];
		\draw (7) edge (4) [thick];
		\draw (4) edge (2) [thick];
		\end{tikzpicture}}
\end{subfigure}
    \caption{The quiver (left), the dimer (center) and the toric diagram (right) of $dP_1$.}\label{fig:dP1}
\end{figure}
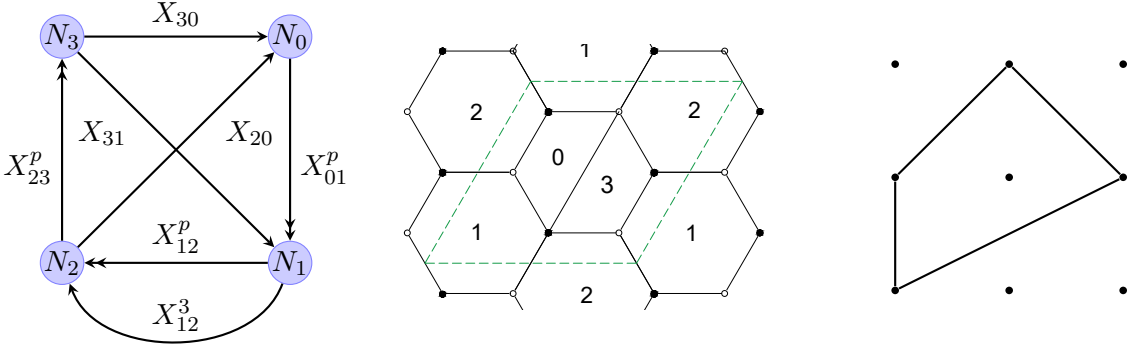

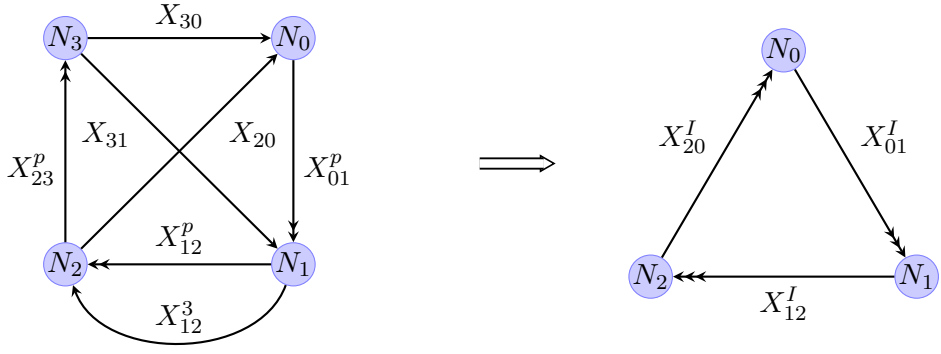
\begin{figure}
\centering
\begin{subfigure}{0.3\textwidth}
			\centerline{\begin{tikzpicture}[auto, scale= 0.5]
		%%%%%%%%%%%% Nodes %%%%%%%%%
		\node [circle, draw=blue!50, fill=blue!20, inner sep=0pt, minimum size=5mm] (0) at (3,3) {$N_0$}; 
		\node [circle, draw=blue!50, fill=blue!20, inner sep=0pt, minimum size=5mm] (1) at (3,-3) {$N_1$}; 			
		\node [circle, draw=blue!50, fill=blue!20, inner sep=0pt, minimum size=5mm] (2) at (-3,-3) {$N_2$};
		\node [circle, draw=blue!50, fill=blue!20, inner sep=0pt, minimum size=5mm] (3) at (-3,3) {$N_3$}; 
		%%%%%%%%%%% Lines %%%%%%%%%%%
		\draw (0)  to node [pos=0.6] {$X_{01}^{p}$} (1) [->>, thick];
		\draw (1)  to node [swap]{$X_{12}^{p}$} (2) [->>, thick];
		\draw (1)  [bend left=70] to node [pos=0.5, swap] {$X_{12}^3$} (2) [->, thick];
		\draw (2)  to node [pos=0.4] {$X_{23}^{p}$} (3) [->>, thick];
		\draw (3)  to node {$X_{30}$} (0) [->, thick];
        \draw (3)  to node [pos=0.3, swap] {$X_{31}$} (1) [->, thick];
        \draw (2)  to node [pos=0.7, swap] {$X_{20}$} (0) [->, thick];
		\end{tikzpicture}}
	\end{subfigure}
	\begin{subfigure}{0.6\textwidth}
			\centerline{\begin{tikzpicture}[auto, scale= 0.5]
		%%%%%%%%%%%% Nodes %%%%%%%%%
		\node [circle, draw=blue!50, fill=blue!20, inner sep=0pt, minimum size=5mm] (0) at (0,6) {$N_0$}; 
		\node [circle, draw=blue!50, fill=blue!20, inner sep=0pt, minimum size=5mm] (1) at (3.5,0) {$N_1$}; 			
		\node [circle, draw=blue!50, fill=blue!20, inner sep=0pt, minimum size=5mm] (2) at (-3.5,0) {$N_2$};
		%%%%%%%%%%% Lines %%%%%%%%%%%
        \draw (0) to node {$X_{01}^{I}$} (1) [->>>, thick];
		\draw (1) to node {$X_{12}^{I}$} (2) [->>>, thick];
		\draw (2) to node {$X_{20}^{I}$} (0) [->>>, thick];		
		\draw[vecArrow] (-8,3) to (-6,3);
		\end{tikzpicture}}
	\end{subfigure}
\caption{The Higgsing of nodes 0 and 3 on the quiver of $dP_1$ gives the quiver of $\CC^3/\ZZ_3$.\label{fig:HigssdP1}}
\end{figure}

\begin{figure}
\centering
\centerline{\begin{tikzpicture}[auto, scale= 0.5]
		%%%%%%%%%%%% Nodes dP1 %%%%%%%%%%
        \node [circle, fill=black, inner sep=0pt, minimum size=1mm] (1) at (-13,3) {};		
		\node [circle, fill=black, inner sep=0pt, minimum size=1mm] (2) at (-10,3) {}; 
		\node [circle, fill=black, inner sep=0pt, minimum size=1mm] (3) at (-7,3) {};
		\node [circle, fill=black, inner sep=0pt, minimum size=1mm] (4) at (-13,0) {};
		\node [circle, fill=black, inner sep=0pt, minimum size=1mm] (5) at (-10,0) {}; 
		\node [circle, fill=black, inner sep=0pt, minimum size=1mm] (6) at (-7,0) {}; 			
		\node [circle, fill=black, inner sep=0pt, minimum size=1mm] (7) at (-13,-3) {};
		\node [circle, fill=black, inner sep=0pt, minimum size=1mm] (8) at (-10,-3) {}; 
		\node [circle, fill=black, inner sep=0pt, minimum size=1mm] (9) at (-7,-3) {};
		%%%%%%%%%%% Lines dP1%%%%%%%%%%%
        \draw (2) edge (6) [thick];
		\draw (6) edge (7) [thick];
		\draw (7) edge (4) [gray];
		\draw (4) edge (2) [gray];
		\draw (2) edge (7) [gray, dashed];		
		%%%%%%%%%%% Nodes C3Z3 %%%%%%%%%%%
        \node [circle, fill=black, inner sep=0pt, minimum size=1mm] (11) at (5,3) {};		
		\node [circle, fill=black, inner sep=0pt, minimum size=1mm] (22) at (8,3) {}; 
		\node [circle, fill=black, inner sep=0pt, minimum size=1mm] (33) at (11,3) {};
		\node [circle, fill=black, inner sep=0pt, minimum size=1mm] (44) at (5,0) {};
		\node [circle, fill=black, inner sep=0pt, minimum size=1mm] (55) at (8,0) {}; 
		\node [circle, fill=black, inner sep=0pt, minimum size=1mm] (66) at (11,0) {}; 			
		\node [circle, fill=black, inner sep=0pt, minimum size=1mm] (77) at (5,-3) {};
		\node [circle, fill=black, inner sep=0pt, minimum size=1mm] (88) at (8,-3) {}; 
		\node [circle, fill=black, inner sep=0pt, minimum size=1mm] (99) at (11,-3) {};
		%%%%%%%%%%% Lines C3Z3 %%%%%%%%%%%
        \draw (22) edge (66) [thick];
		\draw (66) edge (77) [thick];
		\draw (77) edge (22) [thick];		
        \draw[vecArrow] (-2,0) to (0,0);
		\end{tikzpicture}}
		\caption{Higgsing $dP_1$ (left) takes out an external node from the toric diagram, resulting in $\CC^3/\ZZ_3$ (right).}\label{fig:ToricHiggsing}
\end{figure}
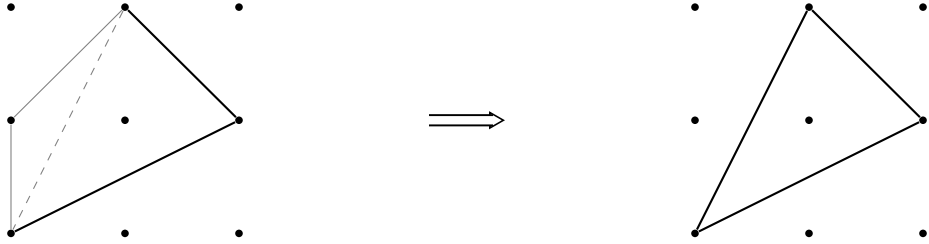

We begin the study of the unoriented projections of some non-orbifold theories with the complex cone over the first del Pezzo surface $dP_1$ \cite{Feng_2001, Franco_2006,Franco:2007ii, Garcia-Etxebarria:2016bpb}, whose different diagrams are drawn in Fig.\,\eqref{fig:dP1}. The anomaly cancellation condition Eq.\,\eqref{eq:anomalyGen} is derived from partition functions of orbifold theories, then in principle we are not allowed to use it for these cases. However, the $dP_1$ theory is related to the orbifold model $\CC^3/\ZZ_3$ by Higgsing two gauge groups \cite{Feng:2002fv}. From the quiver and the dimer it is easy to see that merging nodes/faces 0 and 3 into one node/face produce the diagrams of $\CC^3/\ZZ_3$, see for example Fig.\,\eqref{fig:HigssdP1}. On the toric diagram, the Higgsing procedure takes out an external node, as displayed in Fig.\,\eqref{fig:ToricHiggsing}. We will see that also the super-potential matches. Let us begin with the super-potential of $dP_1$, which reads 
\begin{equation}
W = {\epsilon}_{pq} \left[ X_{12}^q \left( X_{20} X_{01}^p - X_{23}^p X_{31} \right) + X_{12}^3 X_{23}^q X_{30} X_{01}^p \right] \; .
\end{equation}
with $p,q=1,2$. As explained in Appendix \eqref{sec:higgs}, we give VEV to $\langle X_{30} \rangle=1$ and the super-potential becomes
\begin{equation}
W = {\epsilon}_{pq} \left[ X_{12}^q \left( X_{20} X_{01}^p - X_{20}^p X_{01} \right) + X_{12}^3 X_{20}^q X_{01}^p \right] \; .
\end{equation}
Re-defining the fields as  $X_{20} \rightarrow X_{20}^3$, $X_{01} \rightarrow X_{01}^3$ we end up with the super-potential of the $\CC^3/\ZZ_3$ theory
\begin{equation}
W = {\epsilon}_{_{IJK}} \left( X_{12}^I X_{20}^J X_{01}^K \right) \; .
\end{equation}

The idea is to use the argument the other way around, namely from the unoriented $\CC^3/\ZZ_3$ we un-Higgs the group at node $ 0 $ and obtain the unoriented $dP_1$. The anomaly cancellation condition is thus inherited from the orbifold theory.

\vspace{10pt}

\subsubsection*{Orientifold $\widehat{\Omega}$ of $dP_1$}

\begin{figure}[H]
\centering
   \begin{subfigure}{0.4\textwidth}
	\centerline{\begin{tikzpicture}[auto, scale= 0.5]
				%%%%%%%%%%%% Nodes %%%%%%%%%
		\node [circle, draw=blue!50, fill=blue!20, inner sep=0pt, minimum size=5mm] (0) at (3,3) {$N_0$}; 
		\node [circle, draw=blue!50, fill=blue!20, inner sep=0pt, minimum size=5mm] (1) at (3,-3) {$N_1$}; 			
		\node [circle, draw=blue!50, fill=blue!20, inner sep=0pt, minimum size=5mm] (2) at (-3,-3) {$N_2$};
		\node [circle, draw=blue!50, fill=blue!20, inner sep=0pt, minimum size=5mm] (3) at (-3,3) {$N_3$}; 
		%%%%%%%%%%% Lines %%%%%%%%%%%
		\draw (0)  to node [pos=0.6] {$X_{01}^{p}$} (1) [->>, thick];
		\draw (1)  to node [pos=0.2]{$X_{12}^{p}$} (2) [->>, thick];
		\draw (1)  [bend left=70] to node [pos=0.4] {$X_{12}^3$} (2) [->, thick];
		\draw (2)  to node [pos=0.4] {$X_{23}^{p}$} (3) [->>, thick];
		\draw (3)  to node [pos=0.8] {$X_{30}$} (0) [->, thick];
        \draw (3)  to node [pos=0.3, swap] {$X_{31}$} (1) [->, thick];
        \draw (2)  to node [pos=0.7, swap] {$X_{20}$} (0) [->, thick];
		\draw [thick, dashed, gray] (0, -6) to node [pos=0.01] {$\widehat{\Omega}$} (0,4) ;
		\end{tikzpicture}}
		\end{subfigure}
		\begin{subfigure}{0.4\textwidth}
		\centerline{\includegraphics[scale=0.3, trim={3cm 5cm 1.5cm 13cm}, clip]{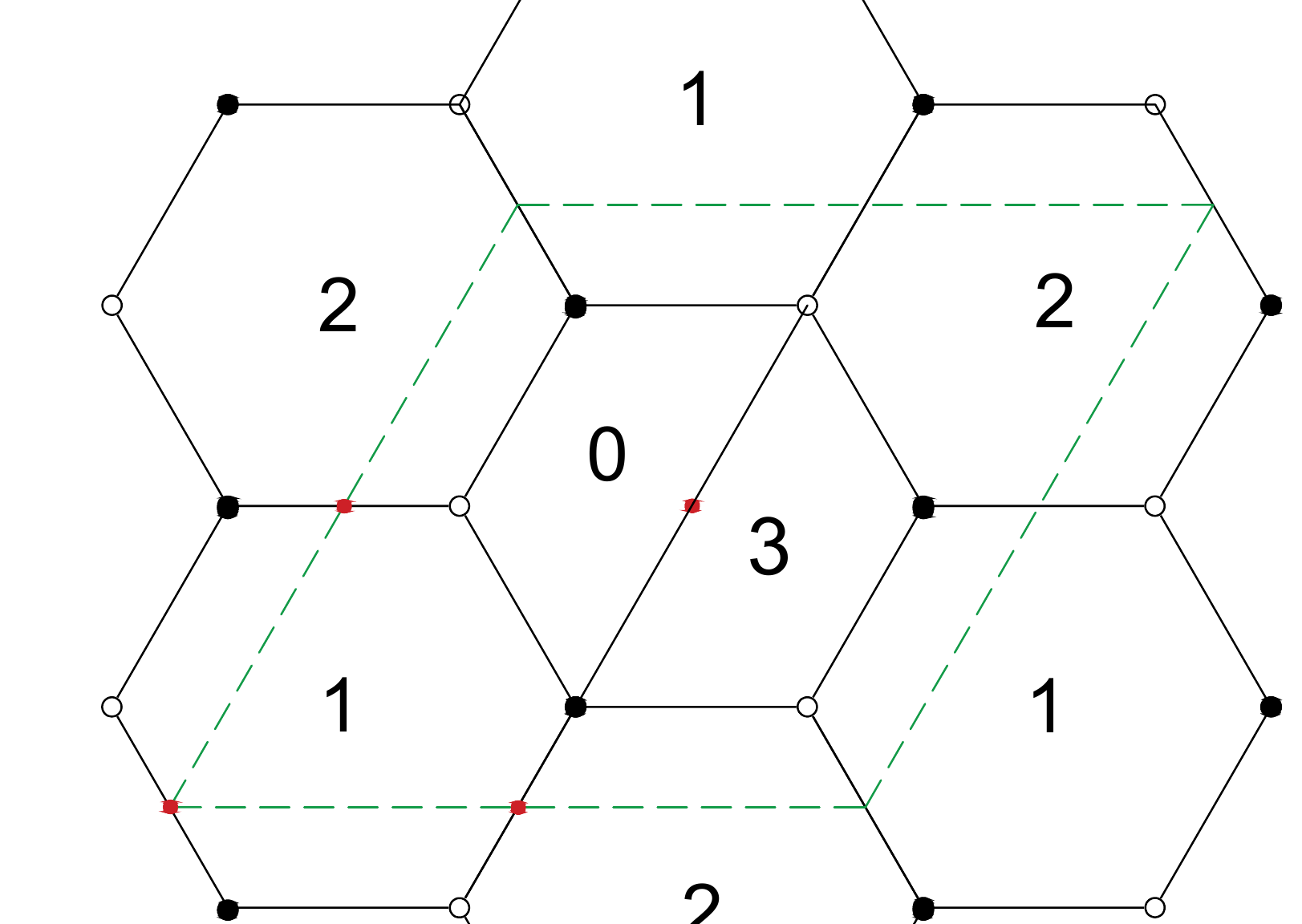}}
		\end{subfigure}
\caption{The orientifold projection $\widehat{\Omega}$ of $dP_1$, whose quiver is on the left and dimer on the right.}\label{fig:dP1OmegaHat}
\end{figure}
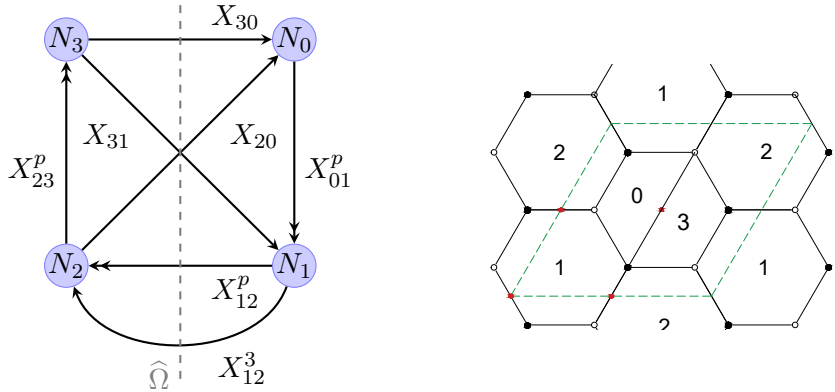

\begin{figure}[h]
\centering
  \begin{subfigure}{0.3\textwidth}
    \centerline{\begin{tikzpicture}[auto, scale= 0.5]
		%%%%%%%%%%%% Nodes %%%%%%%%%%
        \node [circle, fill=red, inner sep=0pt, minimum size=1.5mm] (1) at (-3,3) {};		
		\node [circle, fill=black, inner sep=0pt, minimum size=1.3mm] (2) at (0,3) {}; 
		\node [circle, fill=red, inner sep=0pt, minimum size=1.5mm] (3) at (3,3) {};
		\node [circle, fill=black, inner sep=0pt, minimum size=1.3mm] (4) at (-3,0) {};
		\node [circle, fill=black, inner sep=0pt, minimum size=1mm] (5) at (0,0) {}; 
		\node [circle, fill=black, inner sep=0pt, minimum size=1.3mm] (6) at (3,0) {}; 			
		\node [circle, fill=red, inner sep=0pt, minimum size=1.5mm] (7) at (-3,-3) {};
		\node [circle, fill=black, inner sep=0pt, minimum size=1mm] (8) at (0,-3) {}; 
		\node [circle, fill=red, inner sep=0pt, minimum size=1.5mm] (9) at (3,-3) {};
		%%%%%%%%%%% Lines %%%%%%%%%%%
        \draw (2) edge (6) [thick];
		\draw (6) edge (7) [thick];
		\draw (7) edge (4) [thick];
		\draw (4) edge (2) [thick];
		\end{tikzpicture}}
  \end{subfigure} 
  \begin{subfigure}{0.3\textwidth}
  \centerline{\begin{tikzpicture}[auto, scale= 0.5]
		%%%%%%%%%%%% Nodes %%%%%%%%%%
        \node [circle, fill=black, inner sep=0pt, minimum size=1mm] (1) at (-3,3) {};		
		\node [circle, fill=black, inner sep=0pt, minimum size=1.3mm] (2) at (0,3) {}; 
		\node [circle, fill=black, inner sep=0pt, minimum size=1mm] (3) at (3,3) {};
		\node [circle, fill=red, inner sep=0pt, minimum size=1.5mm] (4) at (-3,0) {};
		\node [circle, fill=black, inner sep=0pt, minimum size=1mm] (5) at (0,0) {}; 
		\node [circle, fill=red, inner sep=0pt, minimum size=1.5mm] (6) at (3,0) {}; 			
		\node [circle, fill=black, inner sep=0pt, minimum size=1.3mm] (7) at (-3,-3) {};
		\node [circle, fill=black, inner sep=0pt, minimum size=1mm] (8) at (0,-3) {}; 
		\node [circle, fill=black, inner sep=0pt, minimum size=1mm] (9) at (3,-3) {};
		%%%%%%%%%%% Lines %%%%%%%%%%%
        \draw (2) edge (6) [thick];
		\draw (6) edge (7) [thick];
		\draw (7) edge (4) [thick];
		\draw (4) edge (2) [thick];
		\end{tikzpicture}}
  \end{subfigure}
  \begin{subfigure}{0.3\textwidth}
  \centerline{\begin{tikzpicture}[auto, scale= 0.5]
		%%%%%%%%%%%% Nodes %%%%%%%%%%
        \node [circle, fill=black, inner sep=0pt, minimum size=1mm] (1) at (-3,3) {};		
		\node [circle, fill=black, inner sep=0pt, minimum size=1.3mm] (2) at (0,3) {}; 
		\node [circle, fill=black, inner sep=0pt, minimum size=1mm] (3) at (3,3) {};
		\node [circle, fill=black, inner sep=0pt, minimum size=1.3mm] (4) at (-3,0) {};
		\node [circle, fill=red, inner sep=0pt, minimum size=1.5mm] (5) at (0,0) {}; 
		\node [circle, fill=black, inner sep=0pt, minimum size=1.3mm] (6) at (3,0) {}; 			
		\node [circle, fill=black, inner sep=0pt, minimum size=1.3mm] (7) at (-3,-3) {};
		\node [circle, fill=black, inner sep=0pt, minimum size=1mm] (8) at (0,-3) {}; 
		\node [circle, fill=black, inner sep=0pt, minimum size=1mm] (9) at (3,-3) {};
		%%%%%%%%%%% Lines %%%%%%%%%%%
        \draw (2) edge (6) [thick];
		\draw (6) edge (7) [thick];
		\draw (7) edge (4) [thick];
		\draw (4) edge (2) [thick];
		\end{tikzpicture}}
  \end{subfigure}
  \caption{The various orientifold projections of toric diagram of $dP_1$. The left figure shows the orientifold with a non-compact $\Omega$7 and $\Omega$3, in the center a non-compact $\Omega$7 and on the right a compact $\Omega$7.}\label{fig:ToricdP1}
\end{figure}
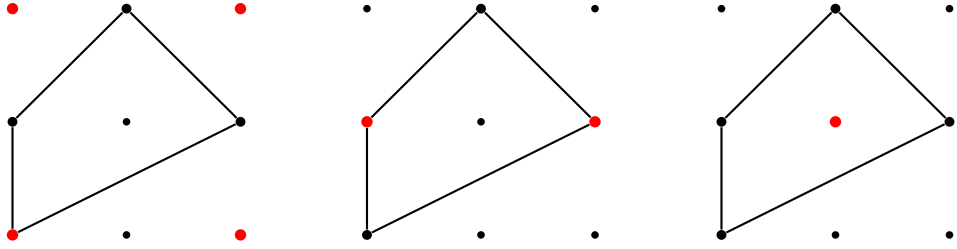

Considering the toric diagram of this model, there are three orientifold involutions allowed, one with a compact $\Omega$7, one with a non-compact $\Omega$7 and an $\Omega$3, and one with a non-compact $\Omega$7, see Fig.\eqref{fig:ToricdP1}. The unoriented projections from quiver and dimer are shown in Fig.\eqref{fig:dP1OmegaHat}. Identifications are
\begin{align}
{\bf{\overline{N}}_3} = {\bf{N}_0} \; , \qquad
{\bf{\overline{N}}_2} = {\bf{N}_1} \; ,
\end{align}
the super-potential reads
\begin{equation}
W = {\epsilon}_{pq} \left[ X_{11'}^q \left( X_{1'0} X_{01}^p - X_{1'0'}^p X_{0'1} \right) + X_{11'}^3 X_{1'0'}^q X_{0'0} X_{01}^p \right] \; 
\end{equation}
and the anomaly-cancellation equations are
\begin{align}
{\epsilon}_{0'0} &= - \frac{1}{3} \sum_{I=1}^3 {\epsilon}_{11'}^{(I)} \; , \nonumber \\
N_1 &= N_0 - \frac{4}{3} \sum_{I=1}^3 {\epsilon}_{11'}^{(I)} \; ,
\end{align}
which requires $\left( {\epsilon}_{11'}^{(1)}, {\epsilon}_{11'}^{(I)}, {\epsilon}_{11'}^{(I)} \right) = (\pm, \pm, \pm)$ and $\epsilon_{0'0}=\mp$. This agrees with Eq. (\ref{eq:dimersgn}), from which the overall product of the signs must be negative. The possible choices are summarized in Tab.\,\eqref{tab:dP1OmegaHat}.

\begin{center}
	\begin{tabular*}{0.7\textwidth}{@{\extracolsep{\fill}}cccc}
		\toprule
		Orientifold & Anomaly condition & $X_{0'0}$ & $\left( X_{11'}^1,X_{11'}^2,X_{11'}^3 \right)$ \\
		\midrule
		$\widehat{\Omega}^+$ & $N_1 = N_0 + 4$ & $ S $ &  $ (A,A,A) $ \\
		$\widehat{\Omega}^-$ & $N_1 = N_0 - 4$ & $ A $ &  $ (S, S, S) $ \\
		\bottomrule
	\end{tabular*}
\end{center} 
\captionof{table}{The orientifold involutions $\widehat{\Omega}$ of $dP_1$ with gauge groups $U(N_0) \times U(N_1)$. “A” stands for “Antisymmetric representation”, while “S” for “Symmetric representation”}\label{tab:dP1OmegaHat}
\vspace{15pt}

The beta functions of this model read
\begin{align}
2 \beta_0 &= N_0 \left( 2 + 2 \gamma_{01} + \gamma_{0'0}  + \gamma_{1'0}  \right) + 2 {\epsilon}_{0'0} \left( -7+ \gamma_{0'0}+ 2 \gamma_{1'0} + 4\gamma_{01}  \right) \; , \nonumber \\
2 \beta_1 &= N_0 \left( \gamma_{0'1} + 2 \gamma_{01} + 2\gamma_{11'} + \overline{\gamma}_{11'}  \right) +  2 {\epsilon}_{0'0} \left( 9 + \gamma_{11'} + 2 \overline{\gamma}_{11'} \right) + 2 \epsilon_{11'}^3 \left(  \overline{\gamma}_{11'}-\gamma_{11'}  \right) \; ,
\end{align}
where $\gamma_{ab}$ are the anomalous dimensions of $X_{ab}$ and $\overline{\gamma}_{11'}$ is the anomalous dimension the third field $X_{11'}^3$. Note that the parent theory is conformal \cite{Bertolini:2004xf, Berenstein:2005xa,Bertolini:2005di} if $N_0 = N_1 = N_2 = N_3$, but the unoriented theory can be anomalous. 

\subsection{Orientifold of the Chiral Orbifold ${\ZZ}'_2$ of the Conifold  $ \Cc$ ($\Cc/{\ZZ}'_2$)} \label{Sec:CZ2}

\begin{figure}[H]
\centering
        \centerline{\begin{tikzpicture}[auto, scale= 0.5]
		%%%%%%%%%%%% Nodes %%%%%%%%%%
        \node [circle, fill=black, inner sep=0pt, minimum size=1mm] (1) at (-3,3) {};		
		\node [circle, fill=black, inner sep=0pt, minimum size=1.3mm] (2) at (0,3) {}; 
		\node [circle, fill=black, inner sep=0pt, minimum size=1mm] (3) at (3,3) {};
		\node [circle, fill=black, inner sep=0pt, minimum size=1.3mm] (4) at (-3,0) {};
		\node [circle, fill=black, inner sep=0pt, minimum size=1mm] (5) at (0,0) {}; 
		\node [circle, fill=black, inner sep=0pt, minimum size=1.3mm] (6) at (3,0) {}; 			
		\node [circle, fill=black, inner sep=0pt, minimum size=1mm] (7) at (-3,-3) {};
		\node [circle, fill=black, inner sep=0pt, minimum size=1.3mm] (8) at (0,-3) {}; 
		\node [circle, fill=black, inner sep=0pt, minimum size=1mm] (9) at (3,-3) {};
		%%%%%%%%%%% Lines %%%%%%%%%%%
        \draw (2) edge (6) [thick];
		\draw (6) edge (8) [thick];
		\draw (8) edge (4) [thick];
		\draw (4) edge (2) [thick];
		\end{tikzpicture}}
\caption{The toric diagram of $\Cc/\ZZ'_2$.}
\label{fig:ToricCZ2}
\end{figure}
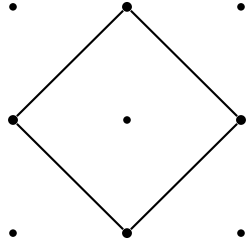

We now pass to study the chiral orbifold of the Conifold, denoted by $\Cc/\ZZ'_2$ \cite{Feng_2001, Franco_2006,Franco:2007ii, Garcia-Etxebarria:2016bpb}, and its orientifold. The theory has two dual phases, called ``electric'' and ``magnetic'', with the same toric diagram drawn in Fig.\,\eqref{fig:ToricCZ2}. Later, we will compare our results with those in $\CC^3/\ZZ_4$ and its mass deformation, with and without the orientifold.

\subsubsection{Electric Phase of $ \Cc/\ZZ'_2$}

\begin{figure}[H]
\centering
	\begin{subfigure}{0.4\textwidth}
		\centerline{\begin{tikzpicture}[auto, scale= 0.5]
		%%%%%%%%%%%% Nodes %%%%%%%%%
		\node [circle, draw=blue!50, fill=blue!20, inner sep=0pt, minimum size=5mm] (0) at (3,3) {$N_0$}; 
		\node [circle, draw=blue!50, fill=blue!20, inner sep=0pt, minimum size=5mm] (1) at (3,-3) {$N_1$}; 			
		\node [circle, draw=blue!50, fill=blue!20, inner sep=0pt, minimum size=5mm] (2) at (-3,-3) {$N_2$};
		\node [circle, draw=blue!50, fill=blue!20, inner sep=0pt, minimum size=5mm] (3) at (-3,3) {$N_3$}; 
		%%%%%%%%%%% Lines %%%%%%%%%%%
		\draw (0)  to node {$X_{01}^{p}$} (1) [->>, thick];
		\draw (1)  to node {$X_{12}^{p'}$} (2) [->>, thick];
		\draw (2)  to node {$X_{23}^ {p}$} (3) [->>, thick];
		\draw (3)  to node {$X_{30}^{p'}$} (0) [->>, thick];
		\end{tikzpicture}}
	\end{subfigure}
	\begin{subfigure}{0.4\textwidth}
		\centerline{\includegraphics[scale=0.35, trim={2cm 2cm 6.1cm 15.1cm}, clip]{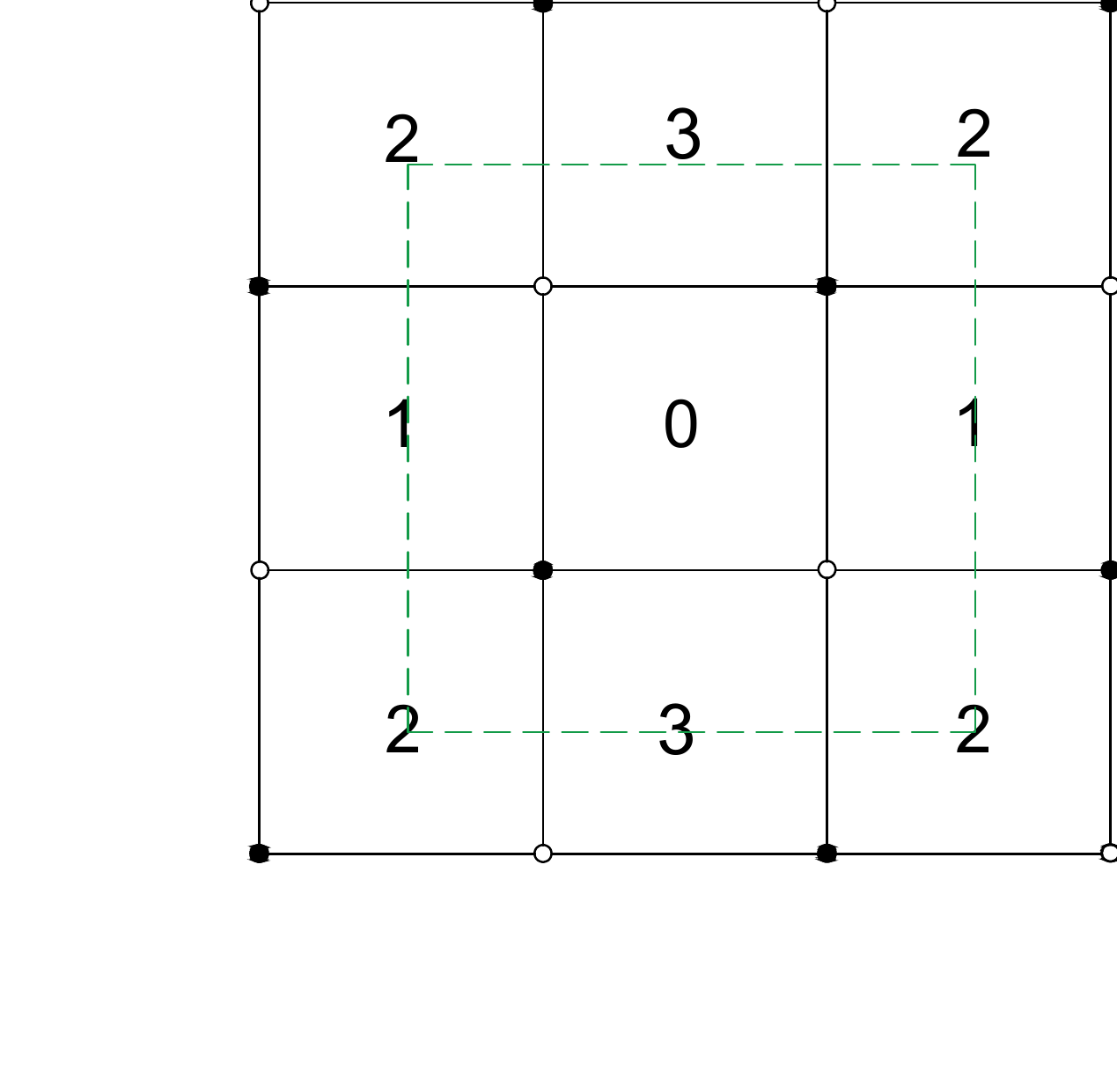}}
	\end{subfigure}
    \caption{The quiver of the eletric phase of $ \Cc/\ZZ'_2$ is shown on the left, while the corresponding dimer is on the right.}\label{fig:CZ2El}
\end{figure}
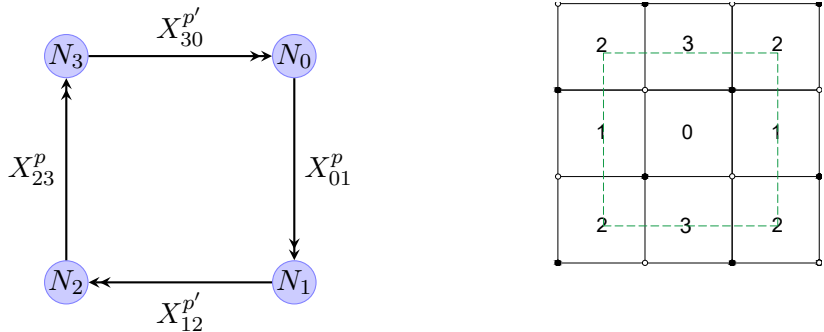

The super-potential reads
\begin{align}
W = {\epsilon}_{pq}{\epsilon}_{p'q'}X_{01}^p X_{12}^{p'} X_{23}^q X_{30}^{q'} \; ,
\end{align}
with $p,q=1,2$ and $p',q'=1,2$ indices of $SU(2) \times SU(2)'$, the group of mesonic symmetry enjoyed by the model. By looking at the quiver present in Fig.\,\eqref{fig:CZ2El} we see that, up to equivalence, we have two possible unoriented projections: the first denoted by $\Omega$ which passes through the nodes 0 and 2, while the second is denoted by $\widehat{\Omega}$ and passes only through fields.

\subsubsection*{Orientifold $\Omega$ of the Electric Phase of $ \Cc/\ZZ'_2$}

\begin{figure}[H]
\centering
   \begin{subfigure}{0.3\textwidth} 
	\centerline{\begin{tikzpicture}[auto, scale= 0.5]
		%%%%%%%%%%%% Nodes %%%%%%%%%
		\node [circle, draw=blue!50, fill=blue!20, inner sep=0pt, minimum size=5mm] (0) at (3,3) {$N_0$}; 
		\node [circle, draw=blue!50, fill=blue!20, inner sep=0pt, minimum size=5mm] (1) at (3,-3) {$N_1$}; 			
		\node [circle, draw=blue!50, fill=blue!20, inner sep=0pt, minimum size=5mm] (2) at (-3,-3) {$N_2$};
		\node [circle, draw=blue!50, fill=blue!20, inner sep=0pt, minimum size=5mm] (3) at (-3,3) {$N_3$}; 
		%%%%%%%%%%% Lines %%%%%%%%%%%
		\draw (0)  to node {$X_{01}^{p}$} (1) [->>, thick];
		\draw (1)  to node {$X_{12}^{p}$} (2) [->>, thick];
		\draw (2)  to node {$X_{23}^{p}$} (3) [->>, thick];
		\draw (3)  to node {$X_{30}^{p}$} (0) [->>, thick];
		\draw [thick, dashed, gray] (4, 4) to node {$\Omega$} (-4,-4) ;
		\end{tikzpicture}}
		\end{subfigure}
	\begin{subfigure}{0.3\textwidth}
	\centerline{\includegraphics[scale=0.35, trim={2cm 2cm 6.1cm 16.3cm}, clip]{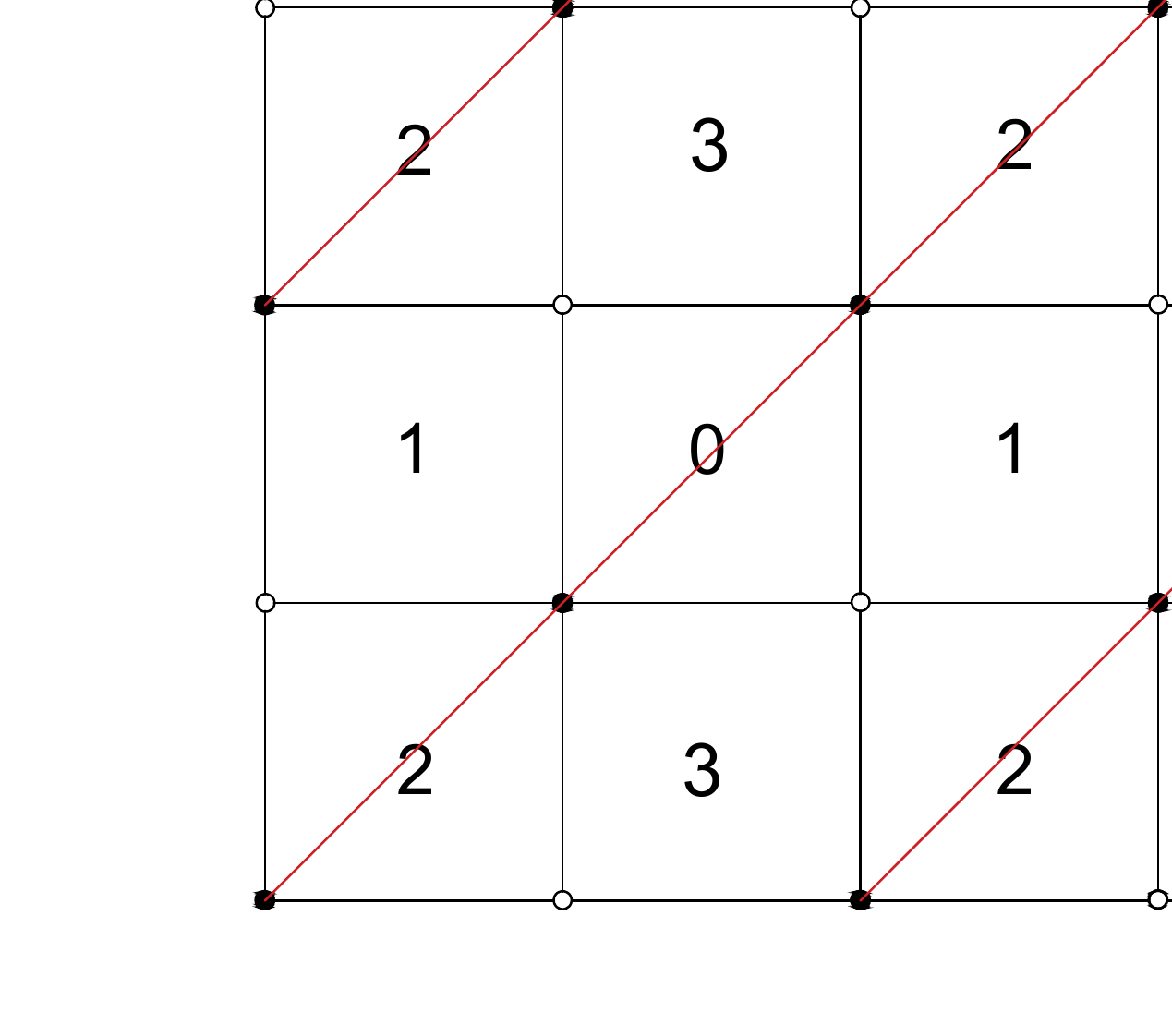}}
	\end{subfigure}
   \begin{subfigure}{0.3\textwidth}
   \centering
        \centerline{\begin{tikzpicture}[auto, scale= 0.5]
		%%%%%%%%%%%% Nodes %%%%%%%%%%
        \node [circle, fill=black, inner sep=0pt, minimum size=1mm] (1) at (-3,3) {};		
		\node [circle, fill=black, inner sep=0pt, minimum size=1.3mm] (2) at (0,3) {}; 
		\node [circle, fill=black, inner sep=0pt, minimum size=1mm] (3) at (3,3) {};
		\node [circle, fill=black, inner sep=0pt, minimum size=1.3mm] (4) at (-3,0) {};
		\node [circle, fill=black, inner sep=0pt, minimum size=1mm] (5) at (0,0) {}; 
		\node [circle, fill=black, inner sep=0pt, minimum size=1.3mm] (6) at (3,0) {}; 			
		\node [circle, fill=black, inner sep=0pt, minimum size=1mm] (7) at (-3,-3) {};
		\node [circle, fill=black, inner sep=0pt, minimum size=1.3mm] (8) at (0,-3) {}; 
		\node [circle, fill=black, inner sep=0pt, minimum size=1mm] (9) at (3,-3) {};
		%%%%%%%%%%% Lines %%%%%%%%%%%
        \draw (2) edge (6) [thick];
		\draw (6) edge (8) [thick];
		\draw (8) edge (4) [thick];
		\draw (4) edge (2) [thick];
		\draw (2) edge (8) [thick, red];
		\end{tikzpicture}}
   \end{subfigure}
\caption{The orientifold involution $\Omega$ of $ \Cc/\ZZ'_2$, whose quiver is drawn on the left and the dimer on the center, while the toric diagram is shown on the right.}		
\label{fig:CZ2b}
\end{figure}
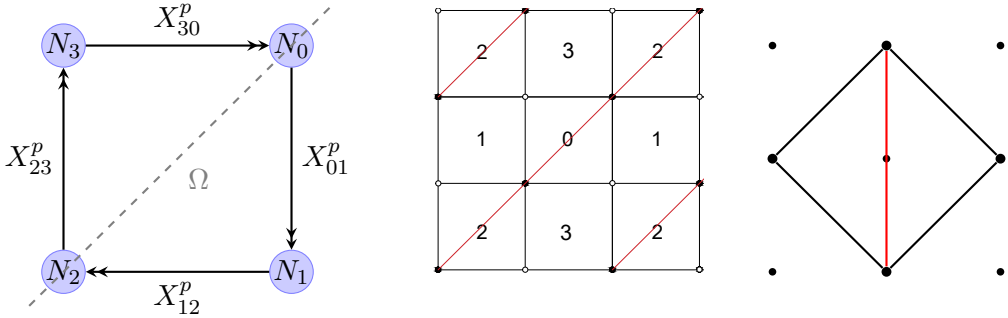

This orientifold acts as
\begin{align}
\overline{{\bf{N}}}_3 = {\bf{N}}_1 \; , \qquad
U(N_0) \rightarrow Sp/SO(N_0) \; , \qquad
U(N_2) \rightarrow Sp/SO(N_2) \; \nonumber \\ 
\end{align}
and the super-potential becomes
\begin{align} \label{eq:CZ2W}
W' = {\epsilon}_{pq}{\epsilon}_{rs}X_{01}^p X_{12}^{r} X_{21'}^q X_{1'0}^{s} \; .
\end{align}
where the $ SU(2) $ indices refer now to the diagonal subgroup to which the mesonic symmetry $ SU(2)\times SU(2)' $ is broken.
Indeed, this unoriented projection is obtained by a fixed line in the dimer as shown in Fig.\,\eqref{fig:CZ2b} and it breaks the mesonic symmetries and thus it is a non-toric involution. In the super-potential there are only three terms, since two are identified being the transpose of each other. The sign of the fixed line determines the projection to $SO$ or $Sp$ gauge groups, and we must have $\epsilon_0 = \epsilon_2$. We denote the two involutions as $\Omega^{+}$ for $Sp$ and $\Omega^{-}$ for $SO$, following the sign convention for the quiver. The theory is anomaly-free if 
\begin{equation}
N_0 = N_2 \; .
\end{equation}

In this theory, the chiral fields acquire anomalous dimensions, that appear after the mass deformation from $\CC^3/\ZZ_4$ (see next subsection). With the anomaly-free condition, the beta functions read
\begin{align}
2 \beta_0 &= 3 N_0 + 6 {\epsilon_0} - 2 N_1\left( 1 - \gamma_{01} \right) \; , \nonumber \\
\beta_1 &= 3 N_1 -  N_0 (2-\gamma_{01}- \gamma_{12}) , \nonumber \\
2\beta_2 &= 3 N_0 + 6 {\epsilon_2} - 2 N_1\left( 1-\gamma_{12} \right) \; ,
\end{align}
From the super-potential before the orientifold, each field has R-charge $1/2$ and the parent theory is conformal if all $\gamma  = - 1/2$. The same reasoning yields again all $\gamma=-1/2$ for the involution $\Omega$ of $\Cc/{\ZZ}'_2$. With these values for the anomalous dimensions, we can impose the beta functions to vanish simultaneously: from the second beta function we need that $N_0=N_1$. Then, if we start with a conformal parent theory, after the unoriented projection we obtain
\begin{align}
\beta_0 & = 3 {\epsilon_0} \; , \nonumber \\
\beta_1 & = 0 \; , \nonumber \\
\beta_2 & = 3 {\epsilon_0} \;  
\end{align}
Where we have used the fact that $ \epsilon_0 = \epsilon_2 $, since by construction the orientifold involution corresponds to a fixed line on the dimer. We conclude that the unoriented projection $\Omega$ breaks conformal invariance. We can still verify if the sum of the beta function vanishes. Summing the three equations above with all $ \gamma = - 1/2 $ we get
\begin{equation} 
\epsilon_{0} = - \epsilon_2
\end{equation}
which is in contrast with the original condition $ \epsilon_{0} = \epsilon_2 $. We conclude that this unoriented projection spoils conformal invariance, it does not allow the fractional branes to recombine into a single bulk brane and it also breaks toricity.\\

The results for the anomaly-free (but non toric and non conformal) theories are summarized in the following Tab.\,\eqref{tab:CZ2El}.  

\begin{center}
\begin{tabular*}{0.7\textwidth}{@{\extracolsep{\fill}}ccc}
\toprule
Orientifold & Gauge groups & Anomaly condition \\
\midrule
$\Omega^+$ & $Sp(N_0)\times U(N_1) \times Sp(N_2)$ & $N_0 = N_2$ \\
$\Omega^-$ & $SO(N_0)\times U(N_1) \times SO(N_2)$ & $N_0 = N_2$ \\
\bottomrule
\end{tabular*}
\end{center}  
\captionof{table}{The unoriented $ \Cc/\ZZ'_2$.\\[10pt]}\label{tab:CZ2El}
\vspace{10pt}

\subsubsection*{Orientifold $\widehat{\Omega}$ of the Electric Phase of $ \Cc/\ZZ'_2$}

\begin{figure}[H]
\centering
   \begin{subfigure}{0.4\textwidth} 
	\centerline{\begin{tikzpicture}[auto, scale= 0.5]
		%%%%%%%%%%%% Nodes %%%%%%%%%
		\node [circle, draw=blue!50, fill=blue!20, inner sep=0pt, minimum size=5mm] (0) at (3,3) {$N_0$}; 
		\node [circle, draw=blue!50, fill=blue!20, inner sep=0pt, minimum size=5mm] (1) at (3,-3) {$N_1$}; 			
		\node [circle, draw=blue!50, fill=blue!20, inner sep=0pt, minimum size=5mm] (2) at (-3,-3) {$N_2$};
		\node [circle, draw=blue!50, fill=blue!20, inner sep=0pt, minimum size=5mm] (3) at (-3,3) {$N_3$}; 
		%%%%%%%%%%% Lines %%%%%%%%%%%
		\draw (0)  to node {$X_{01}^{p}$} (1) [->>, thick];
		\draw (1)  to node [pos=0.3] {$X_{12}^{p'}$} (2) [->>, thick];
		\draw (2)  to node {$X_{23}^{p}$} (3) [->>, thick];
		\draw (3)  to node [pos=0.7] {$X_{30}^{p'}$} (0) [->>, thick];
		\draw [thick, dashed, gray] (0, -5) to node [pos=0.01]{$\widehat{\Omega}$} (0,5) ;
		\end{tikzpicture}}
		\end{subfigure}
	\begin{subfigure}{0.4\textwidth}
	\centerline{\includegraphics[scale=0.35, trim={2cm 2cm 5.8cm 16.4cm}, clip]{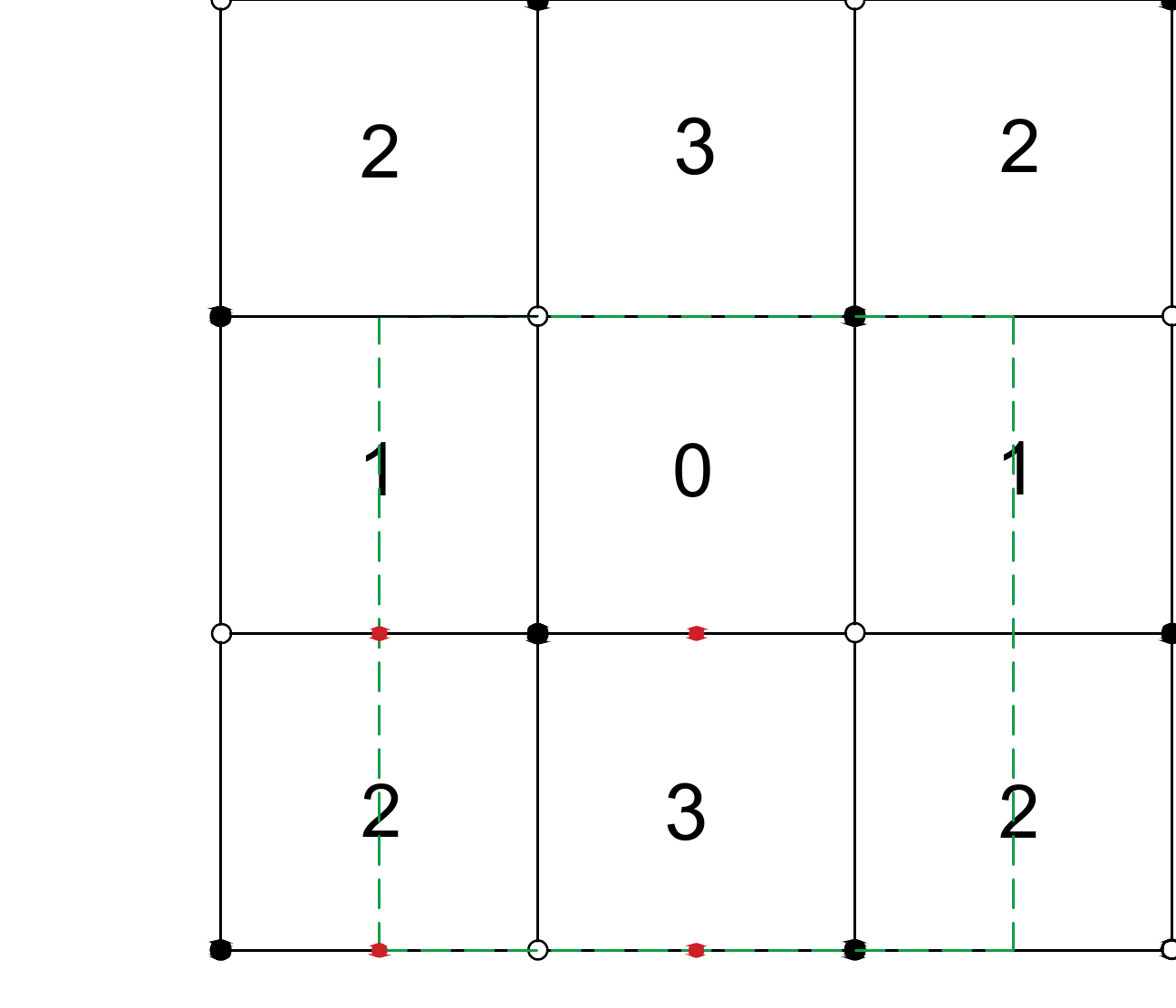}}
	\end{subfigure}
\caption{The orientifold projection $\widehat{\Omega}$ of $\Cc/\ZZ'_2$, whose quiver is drawn on the left and the dimer on the right.}		
\label{fig:CZ2a}
\end{figure}

This orientifold acts as 
\begin{align}
\bf{\overline{N}_3} = \bf{N_0} \; , \qquad
\bf{\overline{N}}_2 = \bf{N_1} \; 
\end{align}
and the super-potential reads
\begin{align}
W' = {\epsilon}_{pq}{\epsilon}_{p'q'}X_{01}^p X_{11'}^{p'} X_{1'0'}^q X_{0'0}^{q'} \; .
\end{align}
This unoriented projection is obtained by four fixed points in the dimer as in Fig.\,\eqref{fig:CZ2a} and it preserves the mesonic symmetries. The product of the four T-parities must be positive.

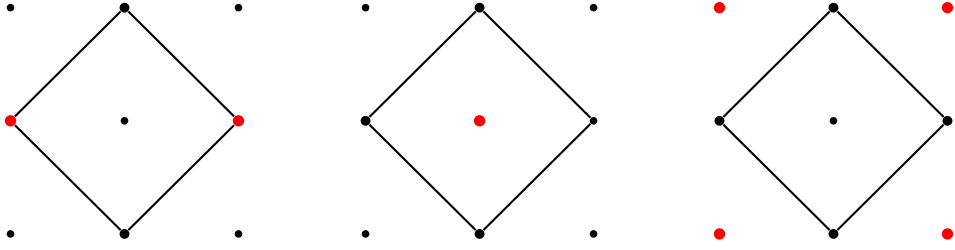
\begin{figure}[H]
\centering
   \begin{subfigure}{0.3\textwidth}
   \centering
        \centerline{\begin{tikzpicture}[auto, scale= 0.5]
		%%%%%%%%%%%% Nodes %%%%%%%%%%
        \node [circle, fill=black, inner sep=0pt, minimum size=1mm] (1) at (-3,3) {};		
		\node [circle, fill=black, inner sep=0pt, minimum size=1.3mm] (2) at (0,3) {}; 
		\node [circle, fill=black, inner sep=0pt, minimum size=1mm] (3) at (3,3) {};
		\node [circle, fill=red, inner sep=0pt, minimum size=1.5mm] (4) at (-3,0) {};
		\node [circle, fill=black, inner sep=0pt, minimum size=1mm] (5) at (0,0) {}; 
		\node [circle, fill=red, inner sep=0pt, minimum size=1.5mm] (6) at (3,0) {}; 			
		\node [circle, fill=black, inner sep=0pt, minimum size=1mm] (7) at (-3,-3) {};
		\node [circle, fill=black, inner sep=0pt, minimum size=1.3mm] (8) at (0,-3) {}; 
		\node [circle, fill=black, inner sep=0pt, minimum size=1mm] (9) at (3,-3) {};
		%%%%%%%%%%% Lines %%%%%%%%%%%
        \draw (2) edge (6) [thick];
		\draw (6) edge (8) [thick];
		\draw (8) edge (4) [thick];
		\draw (4) edge (2) [thick];
		\end{tikzpicture}}
   \end{subfigure}
\begin{subfigure}{0.3\textwidth}
   \centering
        \centerline{\begin{tikzpicture}[auto, scale= 0.5]
		%%%%%%%%%%%% Nodes %%%%%%%%%%
        \node [circle, fill=black, inner sep=0pt, minimum size=1mm] (1) at (-3,3) {};		
		\node [circle, fill=black, inner sep=0pt, minimum size=1.3mm] (2) at (0,3) {}; 
		\node [circle, fill=black, inner sep=0pt, minimum size=1mm] (3) at (3,3) {};
		\node [circle, fill=black, inner sep=0pt, minimum size=1.3mm] (4) at (-3,0) {};
		\node [circle, fill=red, inner sep=0pt, minimum size=1.5mm] (5) at (0,0) {}; 
		\node [circle, fill=black, inner sep=0pt, minimum size=1mm] (6) at (3,0) {}; 			
		\node [circle, fill=black, inner sep=0pt, minimum size=1mm] (7) at (-3,-3) {};
		\node [circle, fill=black, inner sep=0pt, minimum size=1.3mm] (8) at (0,-3) {}; 
		\node [circle, fill=black, inner sep=0pt, minimum size=1mm] (9) at (3,-3) {};
		%%%%%%%%%%% Lines %%%%%%%%%%%
        \draw (2) edge (6) [thick];
		\draw (6) edge (8) [thick];
		\draw (8) edge (4) [thick];
		\draw (4) edge (2) [thick];
		\end{tikzpicture}}
   \end{subfigure}
\begin{subfigure}{0.3\textwidth}
   \centering
        \centerline{\begin{tikzpicture}[auto, scale= 0.5]
		%%%%%%%%%%%% Nodes %%%%%%%%%%
        \node [circle, fill=red, inner sep=0pt, minimum size=1.5mm] (1) at (-3,3) {};		
		\node [circle, fill=black, inner sep=0pt, minimum size=1.3mm] (2) at (0,3) {}; 
		\node [circle, fill=red, inner sep=0pt, minimum size=1.5mm] (3) at (3,3) {};
		\node [circle, fill=black, inner sep=0pt, minimum size=1.3mm] (4) at (-3,0) {};
		\node [circle, fill=black, inner sep=0pt, minimum size=1mm] (5) at (0,0) {}; 
		\node [circle, fill=black, inner sep=0pt, minimum size=1.3mm] (6) at (3,0) {}; 			
		\node [circle, fill=red, inner sep=0pt, minimum size=1.5mm] (7) at (-3,-3) {};
		\node [circle, fill=black, inner sep=0pt, minimum size=1.3mm] (8) at (0,-3) {}; 
		\node [circle, fill=red, inner sep=0pt, minimum size=1.5mm] (9) at (3,-3) {};
		%%%%%%%%%%% Lines %%%%%%%%%%%
        \draw (2) edge (6) [thick];
		\draw (6) edge (8) [thick];
		\draw (8) edge (4) [thick];
		\draw (4) edge (2) [thick] ;
		\end{tikzpicture}}
   \end{subfigure}
   \caption{The various toric involutions $\widehat{\Omega}$ of $\Cc/\ZZ'_2$. The right figure shows the toric involution with a non-compact $\Omega$7, in the center a compact $\Omega$7 and on the right a $\Omega$3.}\label{ToricCZ2OmegaHat}
\end{figure}  

The anomaly cancellation conditions read
\begin{align}
N_0 = N_1 + 2 &\left( {\epsilon}_{11'}^{(1)} + {\epsilon}_{11'}^{(2)} \right) \; , \nonumber \\
\left( {\epsilon}_{0'0}^{(1)} + {\epsilon}_{0'0}^{(2)} \right) = - &\left( {\epsilon}_{11'}^{(1)} + {\epsilon}_{11'}^{(2)} \right)
\end{align}
This means that we have various unoriented theories with gauge groups $U(N_0) \times U(N_1)$, with $N_1 = N_0$ or $N_1 = N_0 \pm 4$, summarized in Tab.\,\eqref{tab:CZ2OmegaHat}.

\begin{center}
	\begin{tabular*}{0.8\textwidth}{@{\extracolsep{\fill}}ccc}
\toprule
Anomaly condition & $(X_{11'}^{1},X_{11'}^{2})$ & $(X_{0'0}^{1}, X_{0'0}^{2})$\\
\midrule
$N_0 = N_1 + 4$ & $ (S,S) $ & $ (A,A) $ \\
$N_0 = N_1 - 4$ & $ (A,A) $ & $ (S,S) $\\
$N_0 = N_1$ & $ (S,A) $ or $ (A,S) $ & $ (S,A) $ or $ (A,S) $\\
$N_0 = N_1$ & $ (S,A) $ or $ (A,S) $ & $ (A,S) $ or $ (S,A) $\\
\bottomrule
	\end{tabular*}
\end{center}
\captionof{table}{The various unoriented projections $\widehat{\Omega}$ of $\Cc/\ZZ'_2$, all of them with gauge groups $U(N_0) \times U(N_1)$.“A” stands for “Antisymmetric representation”, while “S” for “Symmetric representation”\\[10pt]}\label{tab:CZ2OmegaHat}
\vspace{10pt}

Plugging in the anomaly cancellation condition, the beta functions of this model read
\begin{align}
\beta_0 &= N_1 \left(1 + \gamma_{01}+\gamma_{0'0}\right) + \left( \epsilon_{11'}^{(1)} + \epsilon_{11'}^{(2)} \right) \left( 5 + \gamma_{0'0}\right) \; , \nonumber \\
\beta_1 &= N_1 \left(1 + \gamma_{01} +\gamma_{11'} \right) - \left( \epsilon_{11'}^{(1)} + \epsilon_{11'}^{(2)} \right) \left( 3 - \gamma_{11'} - 2 \gamma_{01}\right) \; , 
\end{align}
The anomalous dimensions of this unoriented theory are different from the ones of the parent theory and still non-zero.

\subsubsection{Magnetic Phase of $ \Cc/\ZZ'_2$}

\begin{figure}[H]
\centering
	\begin{subfigure}{0.4\textwidth}
		\centerline{\begin{tikzpicture}[auto, scale= 0.5]
		%%%%%%%%%%%% Nodes %%%%%%%%%
		\node [circle, draw=blue!50, fill=blue!20, inner sep=0pt, minimum size=5mm] (0) at (3,3) {$N_4$}; 
		\node [circle, draw=blue!50, fill=blue!20, inner sep=0pt, minimum size=5mm] (1) at (3,-3) {$N_1$}; 			
		\node [circle, draw=blue!50, fill=blue!20, inner sep=0pt, minimum size=5mm] (2) at (-3,-3) {$N_2$};
		\node [circle, draw=blue!50, fill=blue!20, inner sep=0pt, minimum size=5mm] (3) at (-3,3) {$N_3$}; 
		%%%%%%%%%%% Lines %%%%%%%%%%%
		\draw (0)  to node {$X_{14}^{p}$} (1) [<<-, thick];
		\draw (1)  to node {$X_{12}^{p'}$} (2) [->>, thick];
		\draw (2)  to node {$X_{23}^{p}$} (3) [->>, thick];
		\draw (3)  to node {$X_{43}^{p'}$} (0) [<<-, thick];
		\draw (3)  to node [pos=0.4] {$(X_{31})^{pp'}$} (1) [->>>>, thick];
		\end{tikzpicture}}
	\end{subfigure}
	\begin{subfigure}{0.4\textwidth}
		\centerline{\includegraphics[scale=0.35, trim={5.5cm 4.5cm 4.5cm 15.5cm}, clip]{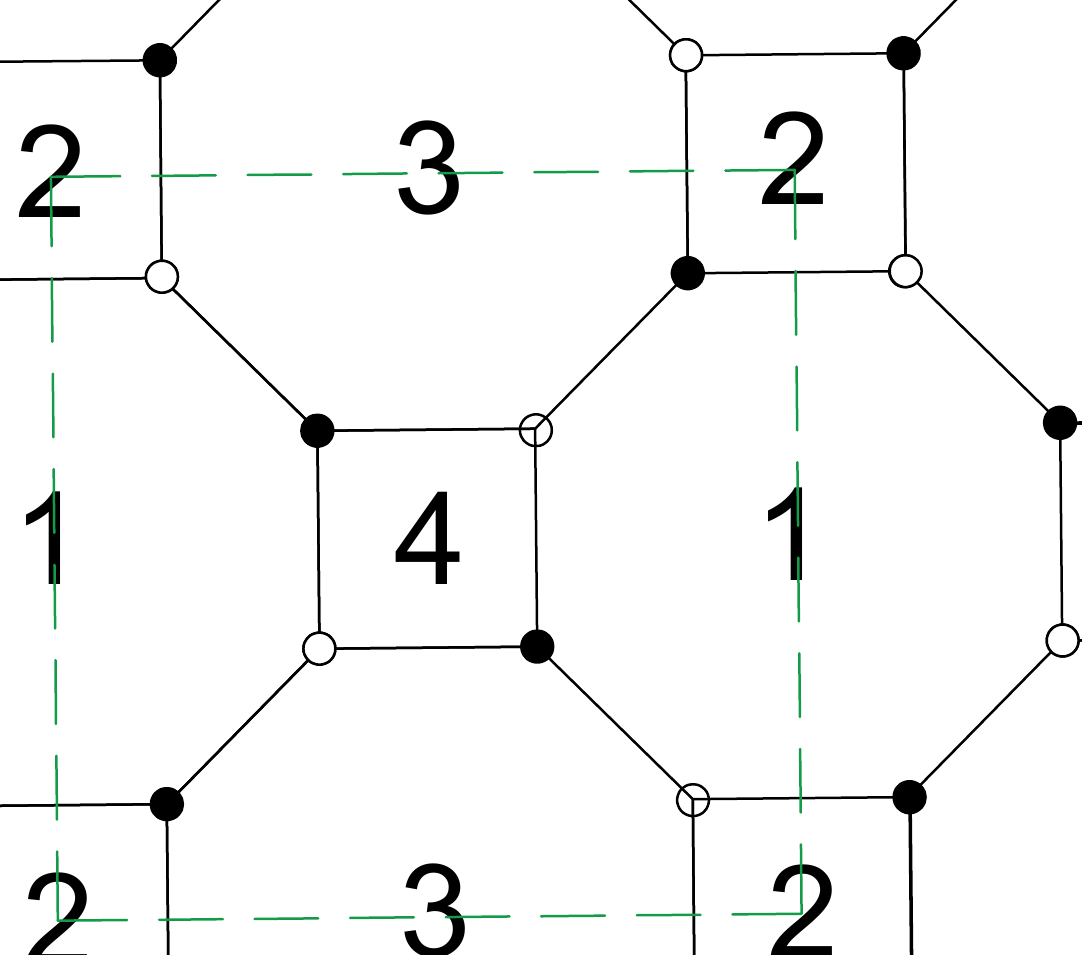}}
	\end{subfigure}
    \caption{The Seiberg dual, or magnetic, phase $\Cc/\ZZ'_2$ with dualization on node 0. The quiver is shown on the left, while the corresponding dimer is on the right.}\label{fig:CZ2mag}
\end{figure}
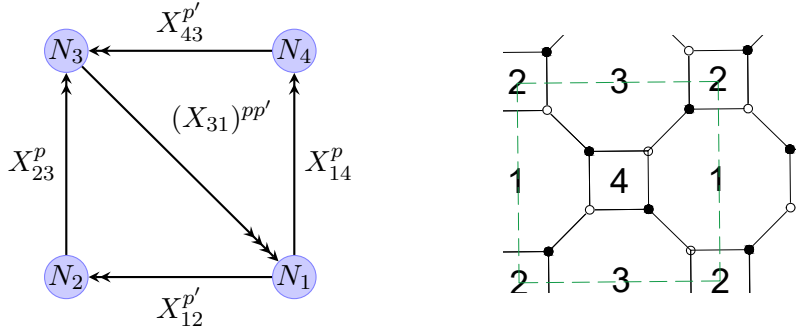

This theory is the magnetic dual of $\Cc/\ZZ'_2$. The node 4 is the dual of the node previously called node 0. There are four additional mesons, which are the fields $X_{31}^{pp'}$, $p,p'=1,2$, in Fig.\,\eqref{fig:CZ2mag}. 
The super-potential reads
\begin{align}
W = {\epsilon}_{pq}{\epsilon}_{p'q'} X_{31}^{pp'} \left( X_{14}^q X_{43}^{q'} -X_{12}^{q'} X_{23}^q  \right) 
\end{align}
and the theory still enjoys an $SU(2) \times SU(2)'$ mesonic symmetry. Again, there are two inequivalent unoriented projections: the first denoted by $\Omega$ which passes trough the nodes 4 and 2, while the second is denoted by $\widehat{\Omega}$ and crosses only fields. Since the toric diagram is the same of the electric phase, the involution on the toric diagram will be the same. Note that the $R$-charges and then the anomalous dimensions at the conformal point are such that $\gamma_{14}=\gamma_{43}=\gamma_{23}=\gamma_{12} = - 1/2$ and $\gamma_{31} = 1$.

\subsubsection*{Orientifold $\Omega$ of the Magnetic Phase of $ \Cc/\ZZ'_2$}

\begin{figure}[H]
\centering
   \begin{subfigure}{0.3\textwidth} 
		\centerline{\begin{tikzpicture}[auto, scale= 0.5]
		%%%%%%%%%%%% Nodes %%%%%%%%%
		\node [circle, draw=blue!50, fill=blue!20, inner sep=0pt, minimum size=5mm] (0) at (3,3) {$N_4$}; 
		\node [circle, draw=blue!50, fill=blue!20, inner sep=0pt, minimum size=5mm] (1) at (3,-3) {$N_1$}; 			
		\node [circle, draw=blue!50, fill=blue!20, inner sep=0pt, minimum size=5mm] (2) at (-3,-3) {$N_2$};
		\node [circle, draw=blue!50, fill=blue!20, inner sep=0pt, minimum size=5mm] (3) at (-3,3) {$N_3$}; 
		%%%%%%%%%%% Lines %%%%%%%%%%%
		\draw (0)  to node {$X_{14}^{p}$} (1) [<<-, thick];
		\draw (1)  to node {$X_{12}^{p}$} (2) [->>, thick];
		\draw (2)  to node {$X_{23}^{p}$} (3) [->>, thick];
		\draw (3)  to node {$X_{34}^{p}$} (0) [<<-, thick];
		\draw (3)  to node [pos=0.25] {$(X_{31})^{pq}$} (1) [->>>>, thick];
		\draw [thick, dashed, gray] (4, 4) to node [pos=0.3] {$\Omega$} (-4,-4) ;
		\end{tikzpicture}}
\end{subfigure}
	\begin{subfigure}{0.3\textwidth}
	\centerline{\includegraphics[scale=0.3, trim={5.5cm 3.5cm 6.5cm 15.5cm}, clip]{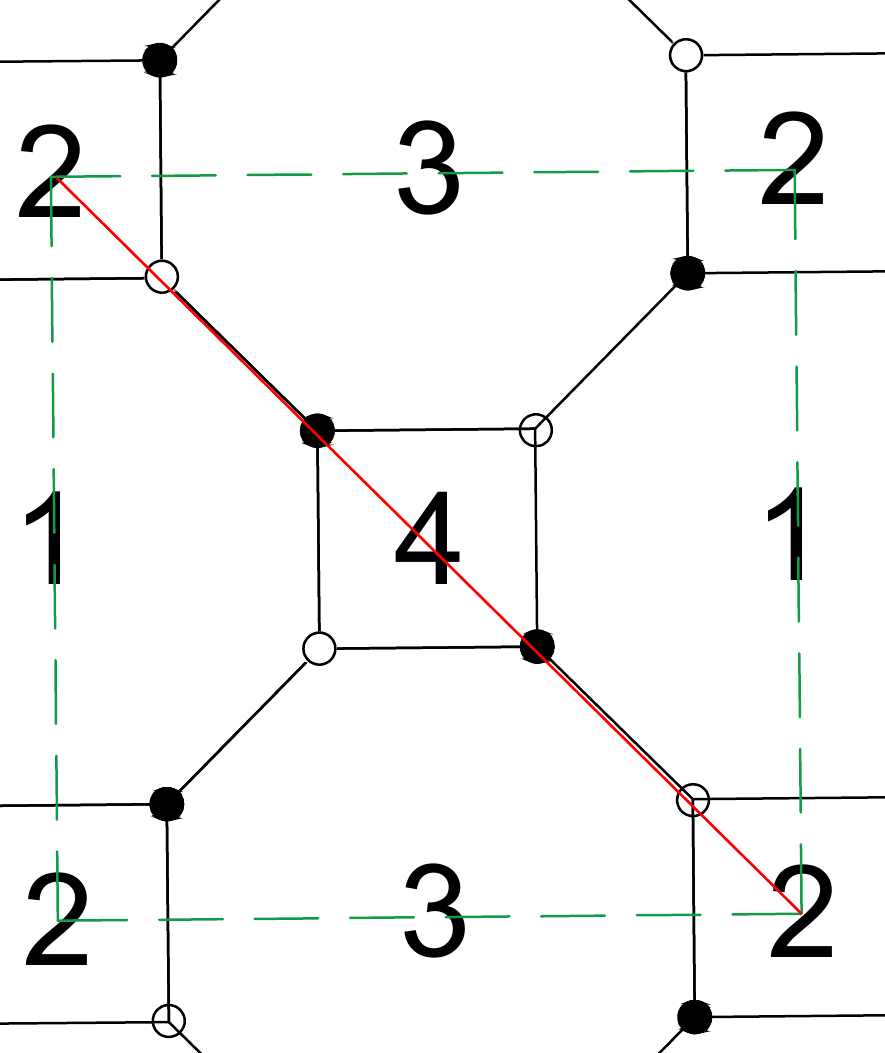}}
	\end{subfigure}
   \begin{subfigure}{0.3\textwidth}
   \centering
        \centerline{\begin{tikzpicture}[auto, scale= 0.5]
		%%%%%%%%%%%% Nodes %%%%%%%%%%
        \node [circle, fill=black, inner sep=0pt, minimum size=1mm] (1) at (-3,3) {};		
		\node [circle, fill=black, inner sep=0pt, minimum size=1.3mm] (2) at (0,3) {}; 
		\node [circle, fill=black, inner sep=0pt, minimum size=1mm] (3) at (3,3) {};
		\node [circle, fill=black, inner sep=0pt, minimum size=1.3mm] (4) at (-3,0) {};
		\node [circle, fill=black, inner sep=0pt, minimum size=1mm] (5) at (0,0) {}; 
		\node [circle, fill=black, inner sep=0pt, minimum size=1.3mm] (6) at (3,0) {}; 			
		\node [circle, fill=black, inner sep=0pt, minimum size=1mm] (7) at (-3,-3) {};
		\node [circle, fill=black, inner sep=0pt, minimum size=1.3mm] (8) at (0,-3) {}; 
		\node [circle, fill=black, inner sep=0pt, minimum size=1mm] (9) at (3,-3) {};
		%%%%%%%%%%% Lines %%%%%%%%%%%
        \draw (2) edge (6) [thick];
		\draw (6) edge (8) [thick];
		\draw (8) edge (4) [thick];
		\draw (4) edge (2) [thick];
		\draw (2) edge (8) [thick, red];
		\end{tikzpicture}}
   \end{subfigure}
\caption{The orientifold involution $\Omega$ of the magnetic $\Cc/\ZZ'_2$, whose quiver is drawn on the left and the dimer on the center, while the toric diagram on the right.}		
\label{fig:CZ2magOmega}
\end{figure}
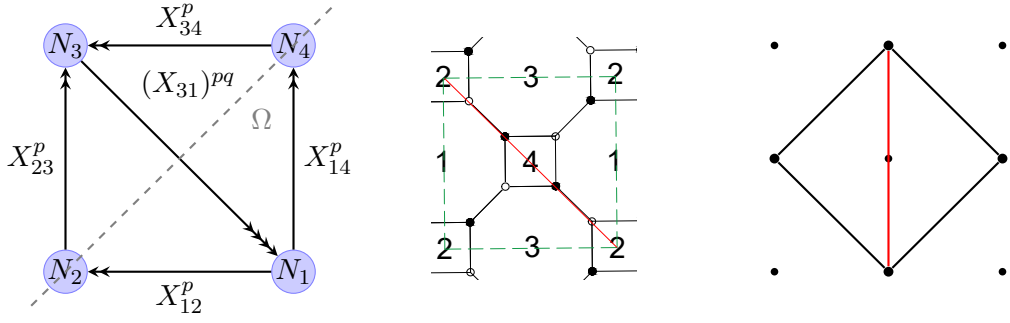

This orientifold acts as
\begin{align}
\overline{{\bf{N}}}_3 = {\bf{N}}_1 \; , \qquad
U(N_4) \rightarrow Sp/SO(N_4) \; , \qquad
U(N_2) \rightarrow Sp/SO(N_2) \; \nonumber \\ 
\end{align}
and the super-potential becomes
\begin{align}
W' = {\epsilon}_{pq}{\epsilon}_{lm} \left( X_{1'1}^{mp}X_{12}^l X_{21'}^q + X_{1'1}^{pl} X_{14}^m X_{41'}^q \right) \; .
\end{align}
The field identifications $X_{14}^p \leftrightarrow X_{41'}^{p'}$ and $X_{12}^{p'} \leftrightarrow X_{21'}^{p}$ leaves only one $SU(2)$ unbroken. The symmetry is thus reduced to $SU(2) \times U(1)_{R}$ and hence toricity is broken. We can see this also from the corresponding dimer with a fixed line shown in Fig.\eqref{fig:CZ2magOmega}. We denote with $\epsilon$ the sign of the fixed line and with $\epsilon_{1'1}^{(mp)}$, $m,p=1,2$ the orientifold sign for the four fields $X_{1'1}^{(mp)}$. Two of the four fields are on top of the fixed line and are projected onto a symmetric or antisymmetric representation, while the other two fields are identified with each other yielding one symmetric and one antisymmetric field. We can see this from the superpotential as follows.

%where an improper fundamental cell is taken on a Klein bottle rather than on a torus. In fact, it is an improper fundamental cell because the corners are identified rotating the rectangle, although each face appears once. The right fundamental cell is deformed and then the dimer is not reliable. Because of this arguments, for the $\Omega$ projection of this phase we focus on the quiver diagram, where there are two nodes projected by $\epsilon_4$ and $\epsilon_2$ and four fields (the mesons), by $\epsilon_{1'1}^{(I)}$, $I=1,2,3,4$.

%Another interesting feature of this model is that from this counting the apparent number of charges is six rather than four. This seems unnatural in view of the relation between the signs (or T-parities) and the orientifold charges, as discussed in the Setup. However, a constraint coming from the super-potential actually shows that the “true” charges are indeed four, as we now show. 

Consider the first super-potential term and momentarily restore the gauge group indices as
\begin{equation}
W_{1} = \epsilon_{pq} \epsilon_{lm} \left( X_{1'1}^{mp} \right)^{i_{1} \, j_{1}} \left( X_{12}^{l} \right)_{j_{1}}^{ l_{2}} {\cal I}_{(2)}^{[l_{2} m_{2}]_{\pm}} \left( X_{21'}^{q} \right)_{m_{2} \, i_{1}} \; ,
\end{equation}
where ${\cal I}_{(2)}^{[l_{2} m_{2}]_{\pm}} $ is the two-index invariant tensor of the gauge group at the node $2$ and $\pm$ indicates whether it is symmetric ($SO(N_0)$) or antisymmetric ($Sp(N_0)$) in the indices $  l_{2}, m_{2} $.
Due to the presence of two $\epsilon_{pq}$ of $SU(2)$, we can only have $SO(N_2)$ with a symmetric $\left( X_{11'}^{mp} \right)^{i_{1} \, j_{1}}$ or $Sp(N_2)$ with an antisymmetric $\left( X_{11'}^{mp} \right)^{i_{1} \, j_{1}}$. But a symmetric $X_{11'}$ means that we have 3 symmetric combinations $(mp)(i_{1} j_{1})$ and 1 antisymmetric $[mp][i_{1} j_{1}]$, thus $\sum_{m,p}\epsilon_{1'1}^{(mp)}=+2$ with $SO(N_2)$, otherwise the super-potential term vanishes. On the other hand, $\sum_{m,p}\epsilon_{1'1}^{(mp)}=-2$ with $Sp(N_2)$. %This conditions on the sum of the charges mean that we can replace the four charges by the sum of two “effective” charges (with the same sign). Furthermore, by looking at the super-potential, one of the two charges can be associated to the meson while the second to the product of the two remaining fields. 
If we let the terms to vanish, there are no F-terms for $X_{11'}$ yielding a no longer singular mesonic moduli space. Since it is unlikely that the theory becomes free after the addition of the orientifold plane, this scenario is implausible. The same line of reasoning holds for the second super-potential term. Moreover, both super-potential terms contain $X_{11'}^{mp}$, then groups at node 4 and 2 must be projected in the same way, i.e. by the single sign $\epsilon$ of the fixed line.

The anomaly cancellation condition requires that 
\begin{equation}
N_4 + N_2 = 2 \left( N_1 + \sum_{m,p=1,2}\epsilon_{1'1}^{(mp)}\right)  \; ,
\end{equation} 
with the constraint $\sum_{m,p}\epsilon_{1'1}^{(mp)}=\pm 2$. The results are summarized in Tab.\,\eqref{tab:CZ2mag}.\\[10pt] 

\begin{center}
\begin{tabular*}{0.85\textwidth}{@{\extracolsep{\fill}}ccc}
\toprule
Gauge groups & Anomaly condition & $ (X_{31}^{11}, X_{31}^{12}, X_{31}^{21}, X_{31}^{22}) $ \\
\midrule
$SO(N_4) \times U(N_1) \times SO(N_2)$ & $N_4 + N_2 = 2 N_1 + 4 $ & $ (S, S, S, A)  $\\
$Sp(N_4) \times U(N_1) \times Sp(N_2)$ & $N_4 + N_2 = 2 N_1 - 4 $ & $ (S, A, A, A) $ \\
\bottomrule
\end{tabular*}
\end{center}  
\captionof{table}{The orientifold involution $\Omega$ of the magnetic phase of $ \Cc/\ZZ'_2$ without flavour branes. “A” stands for “Antisymmetric representation”, while “S” for “Symmetric representation”.\\}\label{tab:CZ2mag}
%\vspace{5pt}

With the anomaly cancellation condition, the beta functions take the form
\begin{align}
2 \beta_4 &= 2N_1\left( 2 + \gamma_{14} \right) - 3 N_2  + 6\epsilon +6 \sum_{m,p=1,2}\epsilon_{1'1}^{(mp)}  \; , \nonumber\\
\beta_1 &=N_1 \left( - 1 + 2 \gamma_{14} + 2 \gamma_{1'1} \right) + N_2 \left( \gamma_{12} - \gamma_{14} \right) + \sum_{m,p=1,2}\epsilon_{1'1}^{(mp)} \left( -3 + \gamma_{1'1} + 2 \gamma_{14} \right) \; , \nonumber \\
2 \beta_2 &= 2N_1(-1 + \gamma_{12}) + 3N_2 + 6 \epsilon   \; .
\end{align}
%where ${\epsilon}_4$ and ${\epsilon}_2$ are the signs of the orientifold involution which project the groups at nodes 4 and 2, respectively. 
%This unoriented theory does not have a conformal point. Besides, 
If we set $N_a=N$ for all $a$, i.e. the condition needed at the conformal point of the parent theory, the unoriented projection $\Omega$ of the magnetic phase of $\Cc/{\ZZ}'_2$ is anomalous.
 
The sum of the beta functions yields
\begin{equation} 
N_1 (3 \gamma_{14} + 2 \gamma_{1'1} + \gamma_{12}) + N_2(\gamma_{12}-\gamma_{14}) + 3 (\epsilon_2 + \epsilon_4) + \sum_{m,p=1,2}\epsilon_{1'1}^{(mp)} (\gamma_{1'1} + 2 \gamma_{14}) = 0
\end{equation}

\subsubsection*{Orientifold $\widehat{\Omega}$ of the Magnetic Phase of $ \Cc/\ZZ'_2$}

\begin{figure}[H]
\centering
   \begin{subfigure}{0.4\textwidth} 
	\centerline{\begin{tikzpicture}[auto, scale= 0.5]
		%%%%%%%%%%%% Nodes %%%%%%%%%
		\node [circle, draw=blue!50, fill=blue!20, inner sep=0pt, minimum size=5mm] (0) at (4,3) {$N_4$}; 
		\node [circle, draw=blue!50, fill=blue!20, inner sep=0pt, minimum size=5mm] (1) at (4,-3) {$N_1$}; 			
		\node [circle, draw=blue!50, fill=blue!20, inner sep=0pt, minimum size=5mm] (2) at (-4,3) {$N_2$};
		\node [circle, draw=blue!50, fill=blue!20, inner sep=0pt, minimum size=5mm] (3) at (-4,-3) {$N_3$}; 
		%%%%%%%%%%% Lines %%%%%%%%%%%
		\draw (0)  to node [pos=0.8] {$X_{14}^p$} (1) [<<-, thick];
		\draw (1)  to node [pos=0.8] {$X_{12}^p$} (2) [->>, thick];
		\draw (2)  to node [swap, pos=0.8] {$X_{23}^p$} (3) [->>, thick];
		\draw (3)  to node [pos=0.8, swap] {$X_{43}^p$} (0) [-<<, thick];
		\draw (3)  to node [swap, pos=0.75] {$X_{31}^{pq}$} (1) [->>>>, thick];
		\draw [thick, dashed, gray] (0, -4) to node [pos=0.05] {$\widehat{\Omega}$} (0,4) ;
		\end{tikzpicture}}
		\end{subfigure}
	\begin{subfigure}{0.4\textwidth}
	\centerline{\includegraphics[scale=0.4, trim={7cm 9cm 6cm 12cm}, clip]{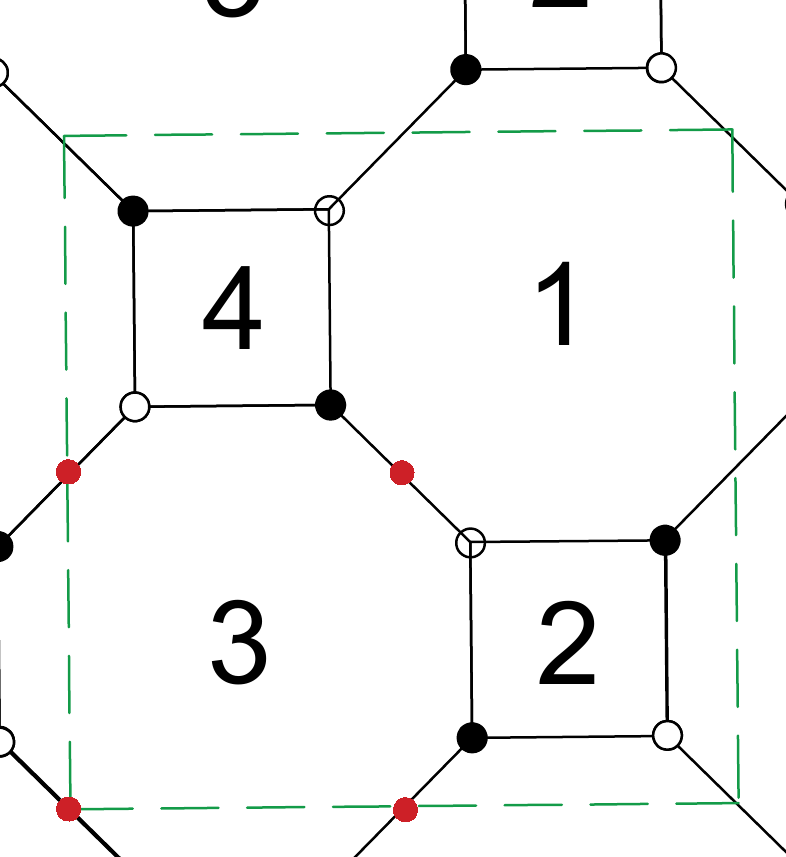}}
	\end{subfigure}
\caption{The orientifold projection $\widehat{\Omega}$ of $ \Cc/\ZZ'_2$, whose quiver is drawn on the left and the dimer on the right.}		
\label{fig:CZ2MagOmegaHat}
\end{figure}

This orientifold acts as
\begin{align}
\bf{\overline{N}_2} =\bf{ N_4 } \; , \qquad
\bf{\overline{N}_3} =\bf{N_1} \; 
\end{align}
and the super-potential reads
\begin{align}
W' = {\epsilon}_{pq}{\epsilon}_{p'q'} X_{1'1}^{pp'} \left( X_{14}^q X_{41'}^{q'} -X_{14'}^{q'} X_{4'1'}^q  \right) 
\end{align}
This unoriented involution is obtained by four fixed points in the dimer as in Fig.\,\eqref{fig:CZ2MagOmegaHat} and it preserves the mesonic symmetries. The four T-parities $({\tau}_{1'1}^1, {\tau}_{1'1}^2, {\tau}_{1'1}^1, {\tau}_{1'1}^2,) $ project fields $X_{1'1}^{pp'}$ onto the symmetric (+) representation and antisymmetric (-) representantion. Their product is constrained by Eq.\,\eqref{eq:dimersgn} and must be positive, thus also the choices for the spectrum are constrained. This reflects the choices for the $\epsilon_{1'1}^{(I)}$, since the four of them project fields as the T-parities. 

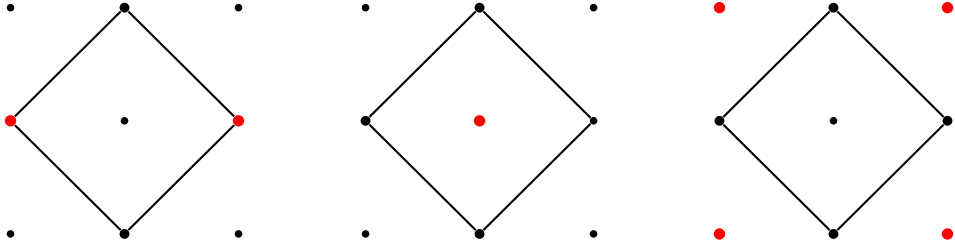
\begin{figure}[H]
\centering
   \begin{subfigure}{0.3\textwidth}
   \centering
        \centerline{\begin{tikzpicture}[auto, scale= 0.5]
		%%%%%%%%%%%% Nodes %%%%%%%%%%
        \node [circle, fill=black, inner sep=0pt, minimum size=1mm] (1) at (-3,3) {};		
		\node [circle, fill=black, inner sep=0pt, minimum size=1.3mm] (2) at (0,3) {}; 
		\node [circle, fill=black, inner sep=0pt, minimum size=1mm] (3) at (3,3) {};
		\node [circle, fill=red, inner sep=0pt, minimum size=1.5mm] (4) at (-3,0) {};
		\node [circle, fill=black, inner sep=0pt, minimum size=1mm] (5) at (0,0) {}; 
		\node [circle, fill=red, inner sep=0pt, minimum size=1.5mm] (6) at (3,0) {}; 			
		\node [circle, fill=black, inner sep=0pt, minimum size=1mm] (7) at (-3,-3) {};
		\node [circle, fill=black, inner sep=0pt, minimum size=1.3mm] (8) at (0,-3) {}; 
		\node [circle, fill=black, inner sep=0pt, minimum size=1mm] (9) at (3,-3) {};
		%%%%%%%%%%% Lines %%%%%%%%%%%
        \draw (2) edge (6) [thick];
		\draw (6) edge (8) [thick];
		\draw (8) edge (4) [thick];
		\draw (4) edge (2) [thick];
		\end{tikzpicture}}
   \end{subfigure}
\begin{subfigure}{0.3\textwidth}
   \centering
        \centerline{\begin{tikzpicture}[auto, scale= 0.5]
		%%%%%%%%%%%% Nodes %%%%%%%%%%
        \node [circle, fill=black, inner sep=0pt, minimum size=1mm] (1) at (-3,3) {};		
		\node [circle, fill=black, inner sep=0pt, minimum size=1.3mm] (2) at (0,3) {}; 
		\node [circle, fill=black, inner sep=0pt, minimum size=1mm] (3) at (3,3) {};
		\node [circle, fill=black, inner sep=0pt, minimum size=1.3mm] (4) at (-3,0) {};
		\node [circle, fill=red, inner sep=0pt, minimum size=1.5mm] (5) at (0,0) {}; 
		\node [circle, fill=black, inner sep=0pt, minimum size=1mm] (6) at (3,0) {}; 			
		\node [circle, fill=black, inner sep=0pt, minimum size=1mm] (7) at (-3,-3) {};
		\node [circle, fill=black, inner sep=0pt, minimum size=1.3mm] (8) at (0,-3) {}; 
		\node [circle, fill=black, inner sep=0pt, minimum size=1mm] (9) at (3,-3) {};
		%%%%%%%%%%% Lines %%%%%%%%%%%
        \draw (2) edge (6) [thick];
		\draw (6) edge (8) [thick];
		\draw (8) edge (4) [thick];
		\draw (4) edge (2) [thick];
		\end{tikzpicture}}
   \end{subfigure}
\begin{subfigure}{0.3\textwidth}
   \centering
        \centerline{\begin{tikzpicture}[auto, scale= 0.5]
		%%%%%%%%%%%% Nodes %%%%%%%%%%
        \node [circle, fill=red, inner sep=0pt, minimum size=1.5mm] (1) at (-3,3) {};		
		\node [circle, fill=black, inner sep=0pt, minimum size=1.3mm] (2) at (0,3) {}; 
		\node [circle, fill=red, inner sep=0pt, minimum size=1.5mm] (3) at (3,3) {};
		\node [circle, fill=black, inner sep=0pt, minimum size=1.3mm] (4) at (-3,0) {};
		\node [circle, fill=black, inner sep=0pt, minimum size=1mm] (5) at (0,0) {}; 
		\node [circle, fill=black, inner sep=0pt, minimum size=1.3mm] (6) at (3,0) {}; 			
		\node [circle, fill=red, inner sep=0pt, minimum size=1.5mm] (7) at (-3,-3) {};
		\node [circle, fill=black, inner sep=0pt, minimum size=1.3mm] (8) at (0,-3) {}; 
		\node [circle, fill=red, inner sep=0pt, minimum size=1.5mm] (9) at (3,-3) {};
		%%%%%%%%%%% Lines %%%%%%%%%%%
        \draw (2) edge (6) [thick];
		\draw (6) edge (8) [thick];
		\draw (8) edge (4) [thick];
		\draw (4) edge (2) [thick];
		\end{tikzpicture}}
   \end{subfigure}
   \caption{The various toric involutions $\widehat{\Omega}$ of the magnetic phase of $ \Cc/\ZZ'_2$. Since Seiberg duality does not change the toric diagram, the possible involutions are the same as for the electric phase. The left figure shows the toric involution with a non-compact $\Omega$7, the center toric diagram represents a toric involution with a compact $\Omega$7, the right one a toric involution with an $\Omega$3.}\label{ToricCZ2OmegaHatM}
\end{figure}

The anomaly cancellation condition reads
\begin{align}\label{eq:AnomalyMagHat}
N_4 = N_1 + \sum_{I=1}^4 {\epsilon}_{1'1}^{(I)} \; , 
\end{align}
The various unoriented theories with gauge groups $U(N_2) \times U(N_1)$ are summarized in Tab.\,\eqref{tab:CZ2MagOmegaHat}.

\begin{center}
	\begin{tabular*}{0.55\textwidth}{@{\extracolsep{\fill}}cc}
\toprule
Anomaly condition & $(X_{1'1}^{11},X_{1'1}^{12}, X_{1'1}^{21}, X_{1'1}^{22})$\\
\midrule
$N_2 = N_1 + 4$ & $ (S, S, S, S) $ \\
$N_2 = N_1 $ & $ (A,A,S,S) $\\
$N_2 = N_1 - 4$ & $ (A,A,A,A) $\\
\bottomrule
	\end{tabular*}
\end{center}
\captionof{table}{The various unoriented projections $\widehat{\Omega}$ of the magnetic $ \Cc/\ZZ'_2$, all of them with gauge groups $U(N_2) \times U(N_1)$. “A” stands for “Antisymmetric representation”, while “S” for “Symmetric representation”.\\[10pt]}\label{tab:CZ2MagOmegaHat}
\vspace{10pt}

The beta functions of this anomaly-free model read
\begin{align}
\beta_4 &= N_1 \left( 1 + \gamma_{41} + \gamma_{14}  \right) + 3 \sum_{I=1}^4 {\epsilon}_{1'1}^{(I)} \; , \nonumber \\
\beta_1 &= N_1 \left[ -1 +  \gamma_{14} + \gamma_{14'} + 2 \gamma_{1'1}\right]+ \sum_{I=1}^4 {\epsilon}_{1'1}^{(I)} \left( -3 + \gamma_{1'1} + \gamma_{14} + \gamma_{14'} \right)\; . 
\end{align}
Note that at the conformal point of the parent theory, i.e. $N_a=N$ for all $a$, this unoriented theory may be anomalous depending on the spectrum. %Also, the anomalous dimensions tend to the parent theory values in the large $N$ limit for both $N_1 = N_2$ and $N_2 = N_1 \pm 4$.
\subsection{Orientifold of $\Nn=1$ Orbifold $ \CC^3/\ZZ_4$, $(1,1,2)$ and its Mass Deformation} \label{Sec:C3Z4}

We study orientifold actions on the chiral orbifold $\CC^3/\ZZ_4$ \cite{Franco_2006, Bianchi:2013gka} and its mass deformation to the unoriented $\Cc/\ZZ'_2$ \cite{Bianchi:2014qma}. We see that the conjugacy classes listed in Tab.\,\eqref{tab:C3Z4classes} include a senior class. This corresponds to a compact 4-cycle around which, in the smooth resolved space, an $\Omega$7 plane can wrap. The (crepant) resolution of this model and its relation with the Generalized Kronheimer Construction can be found in \cite{Bruzzo:2017fwj,Bruzzo:2019asa}.

\begin{center}
\begin{tabular*}{0.55\textwidth}{@{\extracolsep{\fill}}ccc}
\toprule
$(k_1,\, k_2,\, k_3)$ & Age$=\frac{1}{4} \sum_I k_I$ & Conjugacy class\\
\midrule
$(0,\, 0,\, 0)$ & 0 & Baby \\
$(1,\, 1,\, 2)$ & 1 & Junior \\
$(2,\, 2,\, 0)$ & 1 & Junior \\
$(3,\, 3,\, 2)$ & 2 & Senior \\
\bottomrule
\end{tabular*}
\end{center}  
\captionof{table}{The conjugacy classes of the chiral orbifold model $\CC^3/\ZZ_4$.}\label{tab:C3Z4classes}
\vspace{15pt}

\begin{figure}[H]
\centering
	\begin{subfigure}{0.3\textwidth}
			\centerline{\begin{tikzpicture}[auto, scale= 0.5]
		%%%%%%%%%%%% Nodes %%%%%%%%%
		\node [circle, draw=blue!50, fill=blue!20, inner sep=0pt, minimum size=5mm] (0) at (3,3) {$N_0$}; 
		\node [circle, draw=blue!50, fill=blue!20, inner sep=0pt, minimum size=5mm] (1) at (3,-3) {$N_1$}; 			
		\node [circle, draw=blue!50, fill=blue!20, inner sep=0pt, minimum size=5mm] (2) at (-3,-3) {$N_2$};
		\node [circle, draw=blue!50, fill=blue!20, inner sep=0pt, minimum size=5mm] (3) at (-3,3) {$N_3$}; 
		%%%%%%%%%%% Lines %%%%%%%%%%%
		\draw (0)  to node [pos=0.2] {$X_{01}^{p}$} (1) [->>, thick];
		\draw (1)  to node [pos=0.2] {$X_{12}^{p}$} (2) [->>, thick];
		\draw (2)  to node [pos=0.2] {$X_{23}^{p}$} (3) [->>, thick];
		\draw (3)  to node [pos=0.2] {$X_{30}^{p}$} (0) [->>, thick];
		\draw (2)  to node [pos=0.8] {$X_{20}$} (0) [<->, thick];
		\draw (0)  to node [pos=0.8] {$X_{02}$} (2) [<->, thick];
		\draw (1)  to node [pos=0.75] {$X_{13}$} (3) [<->, thick];
		\draw (3)  to node [pos=0.75] {$X_{31}$} (1) [<->, thick];
		\end{tikzpicture}}
	\end{subfigure}
	\begin{subfigure}{0.3\textwidth}
		\centerline{\includegraphics[scale=0.25, trim={3cm 5cm 1.3cm 12cm}, clip]{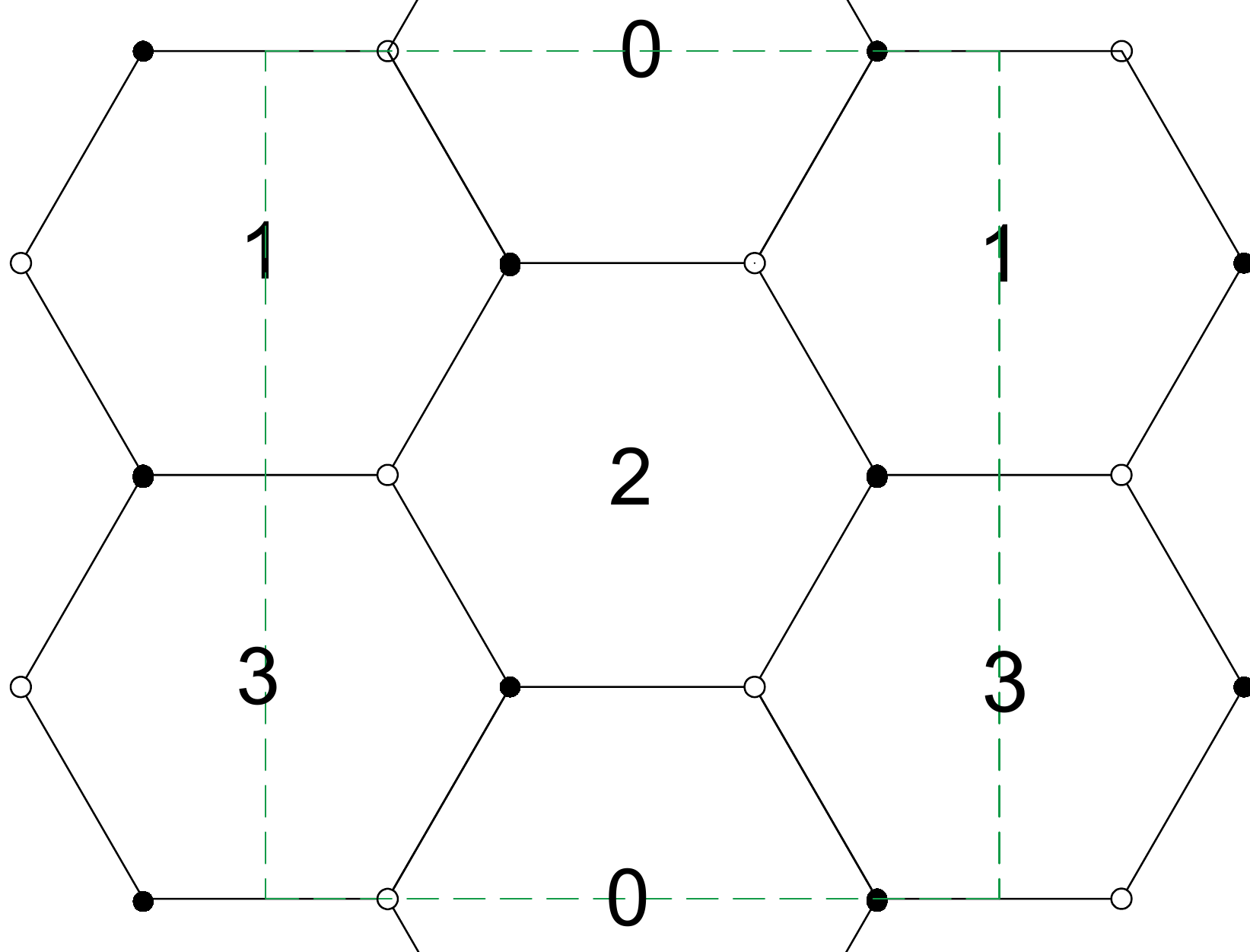}}
	\end{subfigure}
\begin{subfigure}{0.3\textwidth}
        \centerline{\begin{tikzpicture}[auto, scale= 0.5]
		%%%%%%%%%%%% Nodes %%%%%%%%%%
        \node [circle, fill=black, inner sep=0pt, minimum size=1mm] (1) at (-3,3) {};		
		\node [circle, fill=black, inner sep=0pt, minimum size=1mm] (2) at (0,3) {}; 
		\node [circle, fill=black, inner sep=0pt, minimum size=1mm] (3) at (3,3) {};
		\node [circle, fill=black, inner sep=0pt, minimum size=1mm] (4) at (-3,0) {};
		\node [circle, fill=black, inner sep=0pt, minimum size=1mm] (5) at (0,0) {}; 
		\node [circle, fill=black, inner sep=0pt, minimum size=1mm] (6) at (3,0) {}; 			
		\node [circle, fill=black, inner sep=0pt, minimum size=1mm] (7) at (-3,-3) {};
		\node [circle, fill=black, inner sep=0pt, minimum size=1mm] (8) at (0,-3) {}; 
		\node [circle, fill=black, inner sep=0pt, minimum size=1mm] (9) at (3,-3) {};
		%%%%%%%%%%% Lines %%%%%%%%%%%
        \draw (1) edge (6) [thick];
		\draw (6) edge (7) [thick];
		\draw (7) edge (4) [thick];
		\draw (4) edge (1) [thick];
		\end{tikzpicture}}
\end{subfigure}
    \caption{The quiver (left), the dimer (center) and the toric diagram (right) of $\CC^3/\ZZ_4$.}\label{fig:C3Z4}
\end{figure}

The associated field theory is described by the diagrams (quiver, dimer, toric) drawn in Fig.\,\eqref{fig:C3Z4} and the super-potential reads
\begin{equation}\label{eq:C3Z4W}
W = {\epsilon}_{pq} \left( X_{20} X_{01}^q X_{12}^p + X_{02} X_{23}^q X_{30}^p + X_{13} X_{30}^q X_{01}^p + X_{31} X_{12}^q X_{23}^p \right) \;  
\end{equation}
with mesonic symmetries $SU(2) \times U(1) \times U(1)_R$.

The spectrum contains two vector-like fields, denoted by $X_{13}$, $X_{31}$ and $X_{02}$, $X_{20}$. As discussed in \cite{Bianchi:2014qma}, a pair of vector-like fields can be integrated out with a mass deformation of the theory (see Sec.\,\eqref{MassDefo}). In general, in the low energy theory toricity is lost but, in some cases, a suitable redefinition of the fields can restore the toric symmetry. Performing this procedure for $\CC^3/\ZZ_4$ the resulting low energy theory is the chiral orbifold of the conifold $\Cc/\ZZ_2'$. It is very simple to see this from the quivers, since Fig.\,\eqref{fig:C3Z4} without vector-like fields it is exactly the quiver of (the electric phase of) $ \Cc/\ZZ'_2$ in Fig.\,\eqref{fig:CZ2El}. It is easy to see that their super-potential, after the deformation, are the same: one starts with the super-potential of $\CC^3/\ZZ_4$ Eq.\,\eqref{eq:C3Z4W} and adds a mass deformation
\begin{equation}
\Delta W = m \left( X_{31}X_{13} - X_{20} X_{02} \right) \; .
\end{equation}
The F-terms of $W_D = W + \Delta W$ for the massive fields give
\begin{align}
X_{02} & = \frac{1}{m}  {\epsilon}_{pq} X_{01}^q X_{12}^p  \; , \qquad \quad
X_{20} = \frac{1}{m}  {\epsilon}_{pq} X_{23}^q X_{30}^p  \; , \nonumber \\
X_{13} & = \frac{1}{m} {\epsilon}_{pq} X_{12}^p X_{23}^q   \; ,  \qquad \quad
X_{31} = \frac{1}{m}  {\epsilon}_{pq} X_{30}^p X_{01}^q  \; .
\end{align}
Plugging them back in $W_D$ the super-potential read
\begin{equation}
W_D = \frac{1}{m} {\epsilon}_{pq} {\epsilon}_{lm} X_{23}^q X_{30}^m X_{01}^p X_{12}^l \; ,
\end{equation}
which is indeed the super-potential of $ \Cc/\ZZ'_2$. Note that the mesonic symmetries along the flow have been enhanced from $SU(2) \times U(1) \times U(1)$ to $SU(2) \times SU(2) \times U(1)$, after integrating out the massive fields. Furthermore, the presence of a mass scale changes the dimension of the fields. Note that $\CC^3/\ZZ_4$ does not trigger a cascade, while after mass deformation we reach $ \Cc/\ZZ'_2$, which does flow along a cascade. \\

\begin{figure}[H]
\centering
\centerline{\begin{tikzpicture}[auto, scale= 0.5]
		%%%%%%%%%%%% Nodes C3Z4 %%%%%%%%%%
        \node [circle, fill=black, inner sep=0pt, minimum size=1mm] (1) at (-10,3) {};		
		\node [circle, fill=black, inner sep=0pt, minimum size=1mm] (2) at (-7,3) {}; 
		\node [circle, fill=black, inner sep=0pt, minimum size=1mm] (3) at (-4,3) {};
		\node [circle, fill=black, inner sep=0pt, minimum size=1mm] (4) at (-10,0) {};
		\node [circle, fill=black, inner sep=0pt, minimum size=1mm] (5) at (-7,0) {}; 
		\node [circle, fill=black, inner sep=0pt, minimum size=1mm] (6) at (-4,0) {}; 			
		\node [circle, fill=black, inner sep=0pt, minimum size=1mm] (7) at (-10,-3) {};
		\node [circle, fill=black, inner sep=0pt, minimum size=1mm] (8) at (-7,-3) {}; 
		\node [circle, fill=black, inner sep=0pt, minimum size=1mm] (9) at (-4,-3) {};
		%%%%%%%%%%% Lines C3Z4 %%%%%%%%%%%
        \draw (1) edge (6) [thick];
		\draw (6) edge (7) [thick];
		\draw (7) edge (4) [thick];
		\draw (4) edge (1) [thick];
		%%%%%%%%%%% Nodes CZ2 %%%%%%%%%%%
        \node [circle, fill=black, inner sep=0pt, minimum size=1mm] (11) at (4,3) {};		
		\node [circle, fill=black, inner sep=0pt, minimum size=1mm] (22) at (7,3) {}; 
		\node [circle, fill=black, inner sep=0pt, minimum size=1mm] (33) at (10,3) {};
		\node [circle, fill=black, inner sep=0pt, minimum size=1mm] (44) at (4,0) {};
		\node [circle, fill=black, inner sep=0pt, minimum size=1mm] (55) at (7,0) {}; 
		\node [circle, fill=black, inner sep=0pt, minimum size=1mm] (66) at (10,0) {}; 			
		\node [circle, fill=black, inner sep=0pt, minimum size=1mm] (77) at (4,-3) {};
		\node [circle, fill=black, inner sep=0pt, minimum size=1mm] (88) at (7,-3) {}; 
		\node [circle, fill=black, inner sep=0pt, minimum size=1mm] (99) at (10,-3) {};
		%%%%%%%%%%% Lines CZ2 %%%%%%%%%%%
        \draw (22) edge (66) [thick];
		\draw (66) edge (88) [thick];
		\draw (88) edge (44) [thick];
		\draw (44) edge (22) [thick];
		\draw (1) to (2) [->, gray];
		\draw (7) to (8) [->, gray];
        \draw[vecArrow] (-1,0) to (1,0);
		\end{tikzpicture}}
		\caption{Mass deformation on $\CC^3/\ZZ_4$ (left) moves external nodes in the toric diagram, yielding $ \Cc/\ZZ'_2$ (right).}\label{fig:ToricMassDeform}
\end{figure}
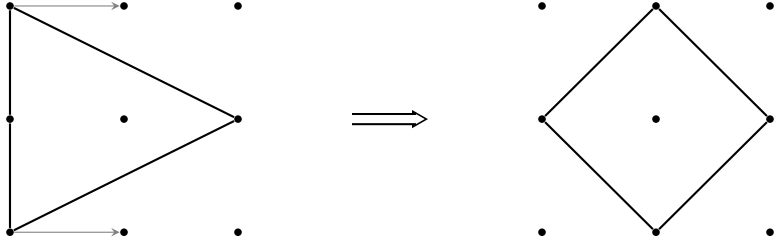

On the toric diagram, the effect of the mass deformation corresponds to moving external nodes, as drawn in Fig.\,\eqref{fig:ToricMassDeform}.

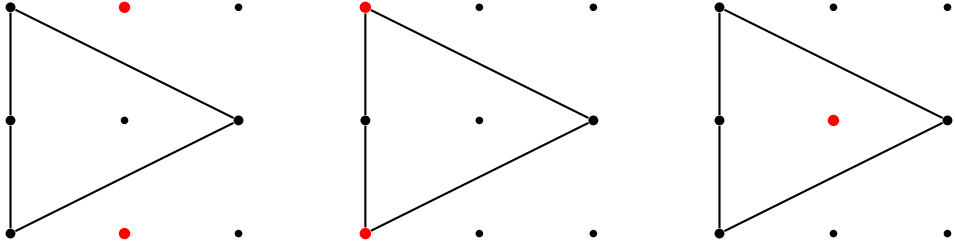
\begin{figure}[H]
\centering
\begin{subfigure}{0.3\textwidth}
        \centerline{\begin{tikzpicture}[auto, scale= 0.5]
		%%%%%%%%%%%% Nodes %%%%%%%%%%
        \node [circle, fill=black, inner sep=0pt, minimum size=1.3mm] (1) at (-3,3) {};		
		\node [circle, fill=red, inner sep=0pt, minimum size=1.5mm] (2) at (0,3) {}; 
		\node [circle, fill=black, inner sep=0pt, minimum size=1mm] (3) at (3,3) {};
		\node [circle, fill=black, inner sep=0pt, minimum size=1.3mm] (4) at (-3,0) {};
		\node [circle, fill=black, inner sep=0pt, minimum size=1mm] (5) at (0,0) {}; 
		\node [circle, fill=black, inner sep=0pt, minimum size=1.3mm] (6) at (3,0) {}; 			
		\node [circle, fill=black, inner sep=0pt, minimum size=1.3mm] (7) at (-3,-3) {};
		\node [circle, fill=red, inner sep=0pt, minimum size=1.5mm] (8) at (0,-3) {}; 
		\node [circle, fill=black, inner sep=0pt, minimum size=1mm] (9) at (3,-3) {};
		%%%%%%%%%%% Lines %%%%%%%%%%%
        \draw (1) edge (6) [thick];
		\draw (6) edge (7) [thick];
		\draw (7) edge (4) [thick];
		\draw (4) edge (1) [thick];
		\end{tikzpicture}}
\end{subfigure}
\centering
	\begin{subfigure}{0.3\textwidth}
			\centerline{\begin{tikzpicture}[auto, scale= 0.5]
		%%%%%%%%%%%% Nodes %%%%%%%%%%
        \node [circle, fill=red, inner sep=0pt, minimum size=1.5mm] (1) at (-3,3) {};		
		\node [circle, fill=black, inner sep=0pt, minimum size=1mm] (2) at (0,3) {}; 
		\node [circle, fill=black, inner sep=0pt, minimum size=1mm] (3) at (3,3) {};
		\node [circle, fill=black, inner sep=0pt, minimum size=1.3mm] (4) at (-3,0) {};
		\node [circle, fill=black, inner sep=0pt, minimum size=1mm] (5) at (0,0) {}; 
		\node [circle, fill=black, inner sep=0pt, minimum size=1.3mm] (6) at (3,0) {}; 			
		\node [circle, fill=red, inner sep=0pt, minimum size=1.5mm] (7) at (-3,-3) {};
		\node [circle, fill=black, inner sep=0pt, minimum size=1mm] (8) at (0,-3) {}; 
		\node [circle, fill=black, inner sep=0pt, minimum size=1mm] (9) at (3,-3) {};
		%%%%%%%%%%% Lines %%%%%%%%%%%
        \draw (1) edge (6) [thick];
		\draw (6) edge (7) [thick];
		\draw (7) edge (4) [thick];
		\draw (4) edge (1) [thick];
		\end{tikzpicture}}
	\end{subfigure}
	\begin{subfigure}{0.3\textwidth}
		\centerline{\begin{tikzpicture}[auto, scale= 0.5]
		%%%%%%%%%%%% Nodes %%%%%%%%%%
        \node [circle, fill=black, inner sep=0pt, minimum size=1.3mm] (1) at (-3,3) {};		
		\node [circle, fill=black, inner sep=0pt, minimum size=1mm] (2) at (0,3) {}; 
		\node [circle, fill=black, inner sep=0pt, minimum size=1mm] (3) at (3,3) {};
		\node [circle, fill=black, inner sep=0pt, minimum size=1.3mm] (4) at (-3,0) {};
		\node [circle, fill=red, inner sep=0pt, minimum size=1.5mm] (5) at (0,0) {}; 
		\node [circle, fill=black, inner sep=0pt, minimum size=1.3mm] (6) at (3,0) {}; 			
		\node [circle, fill=black, inner sep=0pt, minimum size=1.3mm] (7) at (-3,-3) {};
		\node [circle, fill=black, inner sep=0pt, minimum size=1mm] (8) at (0,-3) {}; 
		\node [circle, fill=black, inner sep=0pt, minimum size=1mm] (9) at (3,-3) {};
		%%%%%%%%%%% Lines %%%%%%%%%%%
        \draw (1) edge (6) [thick];
		\draw (6) edge (7) [thick];
		\draw (7) edge (4) [thick];
		\draw (4) edge (1) [thick];
		\end{tikzpicture}}
	\end{subfigure}
    \caption{The three toric orientifold projections on the toric diagram. The right one is performed by an $\Omega$3 plane, while the one on the left a non-compact $\Omega$7, on the right a compact $\Omega$7.}\label{fig:ToricZ4Orientifold}
\end{figure}

In studying the unoriented projections of these two models, it is interesting to analyze the relation between them. From the toric diagram of $\CC^3/\ZZ_4$ shown in Fig.\,\eqref{fig:ToricZ4Orientifold}, we see that in the resolved space there are three different types of orientifold: one with an $\Omega$3, one with a compact $\Omega$7 (which wraps the compact 4-cycle) and one with a non-compact $\Omega$7. On the other hand, from the quiver one can note the existence of only two orientifolds: $\Omega$ crossing two nodes, and $\widehat{\Omega}$ crossing fields, only. We study both cases in that order.

\subsubsection*{Orientifold $\Omega$ of $ \CC^3/\ZZ_4$}

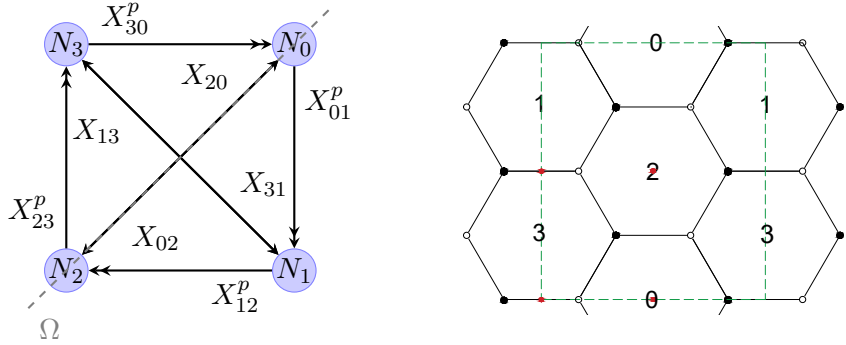
\begin{figure}[H]
\centering
\begin{subfigure}{0.4\textwidth}
	\centerline{\begin{tikzpicture}[auto, scale= 0.5]
		%%%%%%%%%%%% Nodes %%%%%%%%%
		\node [circle, draw=blue!50, fill=blue!20, inner sep=0pt, minimum size=5mm] (0) at (3,3) {$N_0$}; 
		\node [circle, draw=blue!50, fill=blue!20, inner sep=0pt, minimum size=5mm] (1) at (3,-3) {$N_1$}; 			
		\node [circle, draw=blue!50, fill=blue!20, inner sep=0pt, minimum size=5mm] (2) at (-3,-3) {$N_2$};
		\node [circle, draw=blue!50, fill=blue!20, inner sep=0pt, minimum size=5mm] (3) at (-3,3) {$N_3$}; 
		%%%%%%%%%%% Lines %%%%%%%%%%%
		\draw (0)  to node [pos=0.2] {$X_{01}^{p}$} (1) [->>, thick];
		\draw (1)  to node [pos=0.2] {$X_{12}^{p}$} (2) [->>, thick];
		\draw (2)  to node [pos=0.2] {$X_{23}^{p}$} (3) [->>, thick];
		\draw (3)  to node [pos=0.2] {$X_{30}^{p}$} (0) [->>, thick];
		\draw (2)  to node [pos=0.8] {$X_{20}$} (0) [<->, thick];
		\draw (0)  to node [pos=0.8] {$X_{02}$} (2) [<->, thick];
		\draw (1)  to node [pos=0.75] {$X_{13}$} (3) [<->, thick];
		\draw (3)  to node [pos=0.75] {$X_{31}$} (1) [<->, thick];
        \draw [thick, dashed, gray] (-4, -4) to node [pos=0, swap] {$\Omega$} (4,4) ; ;
		\end{tikzpicture}}
\end{subfigure}
\begin{subfigure}{0.4\textwidth} 
            \centerline{\includegraphics[scale=0.3, trim={3cm 5cm 1cm 12cm}, clip]{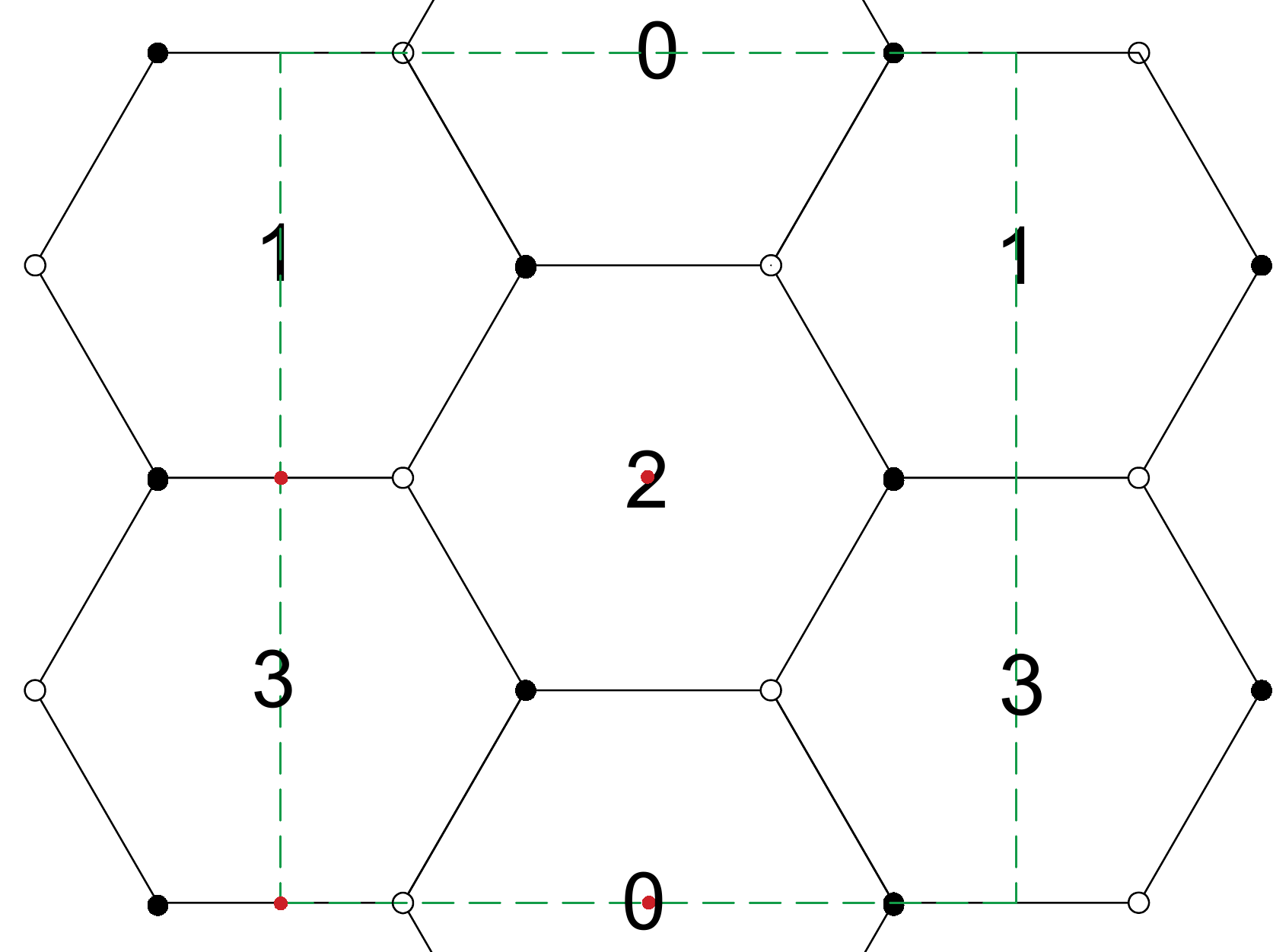}}
\end{subfigure}
\caption{The quiver and the corresponding dimer for the orientifold projection $\Omega$ of $\CC^3/\ZZ_4$.}\label{fig:C3Z4Omega}
\end{figure}

As it is clear from the quiver in Fig.\,\eqref{fig:C3Z4Omega}, the unoriented projection $\Omega$ acts as
\begin{align}
{\bf{\overline{N}}}_3 = {\bf{N}}_1 \; , \qquad
U(N_0) \rightarrow Sp/SO(N_0) \; ,\qquad
U(N_2) \rightarrow Sp/SO(N_2) \; ,
\end{align}
The super-potential reads
\begin{equation}
W' = {\epsilon}_{pq} \left( X_{20} X_{01}^q X_{12}^p + X_{02} X_{21'}^q X_{1'0}^p + X_{11'} X_{1'0}^q X_{01}^p + X_{1'1} X_{12}^q X_{21'}^p \right) \; .
\end{equation}

The anomaly cancellation condition gives
\begin{equation}
N_0 = N_2 + 2 \left( {\epsilon}_{11'}^{(3)} - {\epsilon}_{1'1}^{(3)} \right) \; .
\end{equation}
When ${\epsilon}_{11'}^{(3)} = {\epsilon}_{1'1}^{(3)}$, the fields $X_{11'}$ and $X_{1'1}$ are projected in the same symmetric or antisymmetric representation, with $N_0 = N_2$. When ${\epsilon}_{11'}^{(3)} = - {\epsilon}_{1'1}^{(3)}$ they are projected in opposite ways and the anomaly cancellation condition becomes $N_0 = N_2 + 4 {\epsilon}_{11'}^{(3)}$. However, the various possible choices are constrained from the dimer and from the super-potential. Indeed, by looking at the dimer (where this unoriented projection is obtained by fixed point involution) one can note that according to Eq.\,\eqref{eq:dimersgn}, the product of the T-parities must be positive. Hence, this limits the possible cases to four choices of the unoriented theory: the gauge groups can be $Sp/SO(N_0) \times U(N_1) \times Sp/SO(N_2)$ while the fields $X_{11'}$, $X_{1'1}$ can transform in the  $(S,S)$ or $(A,A)$ representation, only. The second constraint comes from the orientifold action on the super-potential, which imposes further conditions on the spectrum: by momentarily restoring the gauge group indices and considering the super-potential term with, for instance, the field $X_{11'}$:
\begin{equation}
W_{11'} = \epsilon_{pq} \left( X_{11'} \right)_{[i_{1} j_{1}]_{\pm}} \left( X_{1'0}^{q} \right)^{j_{1}}_{l_{0}} {\cal I}_{(0)}^{[l_{0} m_{0}]_{\pm}} \left( X_{01}^{p} \right)_{m_{0}}^{\phantom{m_0}i_1} \; ,
\end{equation}
where ${\cal I}_{(0)}^{[l_{0} m_{0}]_{\pm}} $ is the invariant tensor of the gauge group at the node $0$ and $\pm$ indicates whether it is symmetric ($SO(N_0)$) or antisymmetric ($Sp(N_0)$) in the indices $  l_{0}, m_{0} $. The whole super-potential term must be symmetric for the identification of groups $ 1 $ and $ 3=1' $, thus, we can only have $X_{11'}^S$ with $Sp(N_0)$ or $X_{11'}^A$ with $SO(N_0)$, otherwise the term vanishes. The same line of reasoning holds for the super-potential term with $X_{1'1}$ and the invariant tensor of the group at node $2$. The results are summarized in Tab.\,\eqref{tab:C3Z4Omega}.

\begin{center}
\begin{tabular*}{0.75\textwidth}{@{\extracolsep{\fill}}ccc}
\toprule
Gauge groups & Anomaly condition & $(X_{11'}, \: X_{1'1})$ \\
\midrule
$SO(N_0) \times U(N_1) \times SO(N_2)$ & $N_0 = N_2$ & $ (A,A) $ \\
$Sp(N_0) \times U(N_1) \times Sp(N_2)$ & $N_0 = N_2$ & $ (S,S) $ \\
$Sp(N_0) \times U(N_1) \times SO(N_2)$ & $N_0 = N_2 + 4$ & $ (S,A) $ \\
$SO(N_0) \times U(N_1) \times Sp(N_2)$ & $N_0 = N_2 - 4$ & $ (A,S) $ \\
\bottomrule
\end{tabular*}
\end{center}  
\captionof{table}{The various orientifold projections $\Omega$ of $\CC^3/\ZZ_4$. “A” stands for “Antisymmetric representation”, while “S” for “Symmetric representation”.}\label{tab:C3Z4Omega}
\vspace{15pt}

At this point it is natural to wonder if the unoriented involution $\Omega$ of $\CC^3/\ZZ_4$ can be mass deformed to the unoriented projection $\Omega$ of $ \Cc/\ZZ'_2$. For the $SO/Sp(N_0) \times U(N_1) \times Sp/SO(N_2)$ it is not possible to add a mass deformation term as $X_{11'}X_{1'1}$ since the two fields transform under different representations, one symmetric and the other antisymmetric: the product of the two fields vanishes, and the mass term is trivial. Besides, in this case the anomaly condition requires $N_0 = N_2 \pm 4$, in contrast to $N_0 = N_2$ for the case of $\Omega$-plane for $ \Cc/\ZZ'_2$. 

On the other hand, the case with $SO/Sp(N_0) \times U(N_1) \times SO/Sp(N_2)$ has $N_0 = N_2$ and admits a mass deformation. Integrating out massive fields one obtains 
\begin{equation}
W' = \frac{1}{m} \left( X_{12}^1 X_{21'}^2 X_{1'0}^2 X_{01}^1 + X_{12}^2 X_{21'}^1 X_{1'0}^1 X_{01}^2 - X_{12}^2 X_{21'}^2 X_{1'0}^1 X_{01}^1 - X_{12}^1 X_{21'}^1 X_{1'0}^2 X_{01}^2 \right) \; 
\end{equation}
and the first two terms are identified, since they are the transpose of each other. This is the super-potential in Eq.\,\eqref{eq:CZ2W} of the unoriented projection $\Omega$ for $ \Cc/\ZZ'_2$, which is obtained by a fixed line on the dimer and hence it is not toric, in agreement with the result of the mass deformation. 

Let us discuss conformal invariance. Plugging in the anomaly cancellation condition, the beta functions read
\begin{align}
2\beta_0 &=   2N_1\left( -1 + \gamma_{01} \right) + N_2 \left( 2 + \gamma_{02} \right) + 6\epsilon_0 + 6 \left( \epsilon_{11'}^{(3)} - \epsilon_{1'1}^{(3)} \right)  \; , \nonumber \\
\beta_1 &= N_1 \left( 2 + \frac{\gamma_{11'} + \gamma_{1'1}}{2} \right) + N_2 \left( -2 + \gamma_{12} + \gamma_{01} \right) + \epsilon_{11'}^{(3)} \left( -3 + 2 \gamma_{01} + \gamma_{11'} \right) \nonumber \\ 
&+ \epsilon_{1'1}^{(3)} \left( 1 - 2 \gamma_{01} + \gamma_{1'1} \right) \; , \nonumber \\
2 \beta_2 &=  2N_1\left( -1 + \gamma_{12} \right)+ N_2 \left( 2 + \gamma_{02} \right)  + 6\epsilon_2 +  2\epsilon_{11'}^{(3)}\left( -1 + \gamma_{02} \right) + 2 \epsilon_{1'1}^{(3)}\left( 1 - \gamma_{02} \right)   \; .
\end{align}
Summing the above beta functions we get
\begin{align}
\sum_{i = 0}^2 \beta_i &=  N_1 \left( \gamma_{01} + \gamma_{12} + \frac{ \gamma_{11'} + \gamma_{1'1}}{2} \right) +  N_2  \left( \gamma_{02} + \gamma_{01} + \gamma_{12} \right) +3 \left( \epsilon_0 + \epsilon_2 \right)  \nonumber \\ 
& + \epsilon_{11'}^{(3)}\left( -1 + 2 \gamma_{01} + \gamma_{02} + \gamma_{11'} \right) + \epsilon_{1'1}^{(3)}\left( -1 - 2 \gamma_{01} - \gamma_{02} + \gamma_{1'1} \right)
\end{align}

The unoriented theory is globally conformal (i.e. the sum of the above beta functions vanish) in the large N-limit, with non-zero anomalous dimensions for $\epsilon_0=-\epsilon_2$ and $\epsilon_{11'}^{(3)} = -\epsilon_{1'1}^{(3)}$.

%In the large $N$ limit, with $\epsilon_{11'}^{(3)} = \epsilon_{1'1}^{(3)}$ and $\epsilon_0 = \epsilon_2$, the beta functions do not vanish, not even globally (i.e. the sum of the above beta functions). Hence, the theory is not conformal and we associate this unoriented involution to a non-compact $\Omega7$ in the smooth space. On the other hand, when $\epsilon_{11'}^{(3)} = - \epsilon_{1'1}^{(3)}$ and $\epsilon_0 = - \epsilon_2$ the beta functions do not vanish separately, but their sum does. This is the case of an $\Omega3$ or a compact $\Omega7$ (wrapping the compact 4-cycle) in the resolved space.

\subsubsection*{Orientifold $\widehat{\Omega}$ of $ \CC^3/\ZZ_4$}

The unoriented involution $\widehat{\Omega}$ identifies
\begin{align}
{\bf{\overline{N}_3}}  = {\bf{N_0}} \; , \qquad
{\bf{\overline{N}_2}}  = {\bf{N_1}} \; 
\end{align}
and the super-potential reads
\begin{equation}
W' = {\epsilon}_{pq} \left( X_{1'0} X_{01}^q X_{11'}^p + X_{01'} X_{20'}^q X_{0'0}^p + X_{10'} X_{0'0}^q X_{01}^p + X_{0'1} X_{11'}^q X_{1'0'}^p \right) \; ,
\end{equation}
while the anomaly-free condition is
\begin{align}
N_0 = N_1 + 2 &\left( {\epsilon}_{11'}^{(1)} + {\epsilon}_{11'}^{(2)} \right) \; , \nonumber \\
\left( {\epsilon}_{0'0}^{(1)} + {\epsilon}_{0'0 }^{(2)} \right) = - &\left( {\epsilon}_{11'}^{(1)} + {\epsilon}_{11'}^{(2)} \right) \; .
\end{align}

\begin{figure}[H]
\centering
\begin{subfigure}{0.4\textwidth}
	\centerline{\begin{tikzpicture}[auto, scale= 0.5]
		%%%%%%%%%%%% Nodes %%%%%%%%%
		\node [circle, draw=blue!50, fill=blue!20, inner sep=0pt, minimum size=5mm] (0) at (3,3) {$N_0$}; 
		\node [circle, draw=blue!50, fill=blue!20, inner sep=0pt, minimum size=5mm] (1) at (3,-3) {$N_1$}; 			
		\node [circle, draw=blue!50, fill=blue!20, inner sep=0pt, minimum size=5mm] (2) at (-3,-3) {$N_2$};
		\node [circle, draw=blue!50, fill=blue!20, inner sep=0pt, minimum size=5mm] (3) at (-3,3) {$N_3$}; 
		%%%%%%%%%%% Lines %%%%%%%%%%%
		\draw (0)  to node [pos=0.8] {$X_{01}^{p}$} (1) [->>, thick];
		\draw (1)  to node [pos=0.8] {$X_{12}^{p}$} (2) [->>, thick];
		\draw (2)  to node [pos=0.8] {$X_{23}^{p}$} (3) [->>, thick];
		\draw (3)  to node [pos=0.75] {$X_{30}^{p}$} (0) [->>, thick];
		\draw (2)  to node [pos=0.75, swap] {$X_{20}$} (0) [<->, thick];
		\draw (0)  to node [pos=0.75, swap] {$X_{02}$} (2) [<->, thick];
		\draw (1)  to node [pos=0.8, swap] {$X_{13}$} (3) [<->, thick];
		\draw (3)  to node [pos=0.8, swap] {$X_{31}$} (1) [<->, thick];
		\draw [thick, dashed, gray] (0, -5) to node [pos=0, swap] {$\widehat{\Omega}$} (0,5) ;
		\end{tikzpicture}}
\end{subfigure}
\begin{subfigure}{0.4\textwidth} 
            \centerline{\includegraphics[scale=0.3, trim={7.5cm 5cm 0 12cm}, clip]{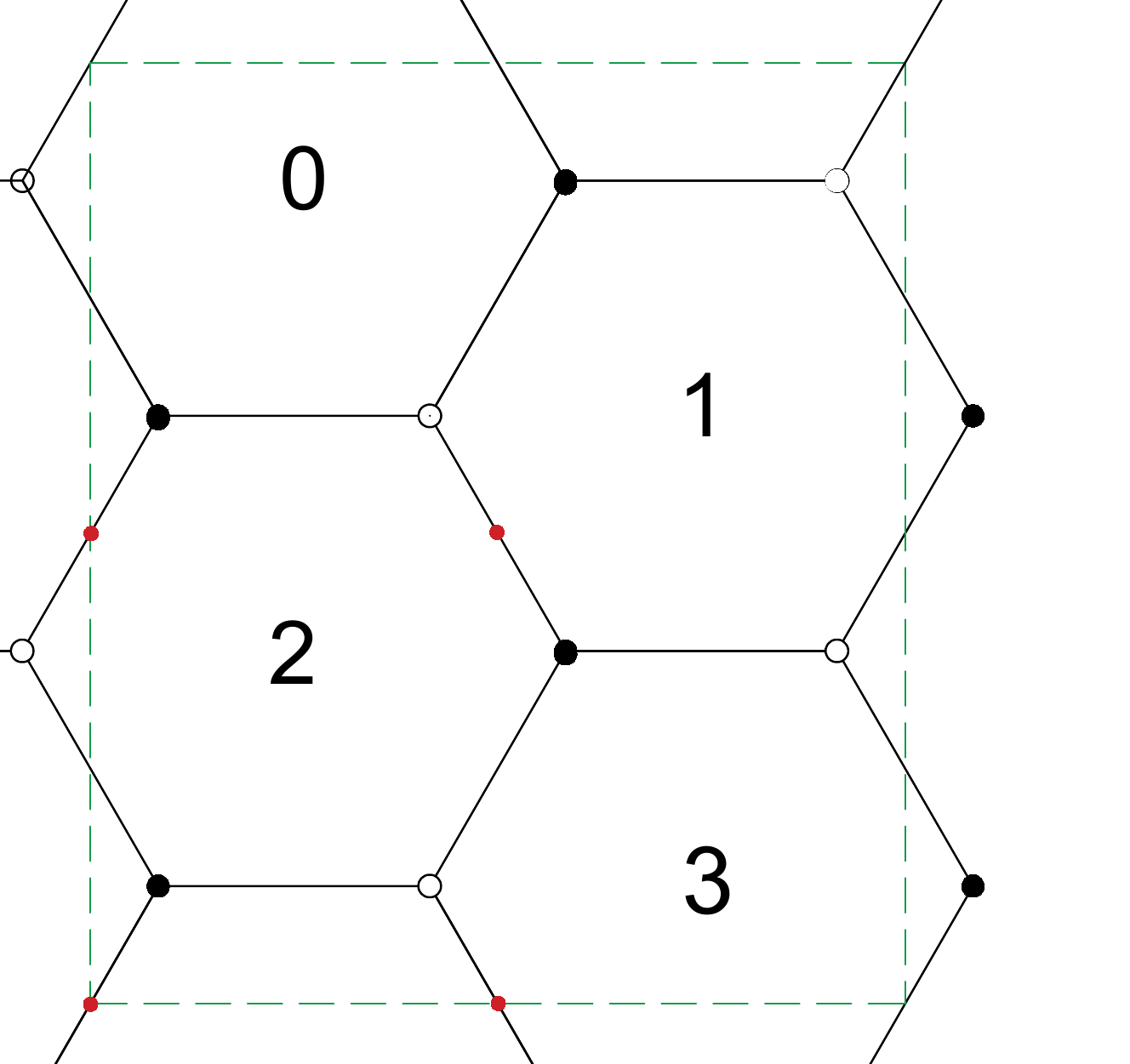}}
\end{subfigure}
\caption{The quiver and the corresponding dimer for the orientifold involution $\hat{\Omega}$ of $\CC^3/\ZZ_4$.}\label{fig:C3Z4OmegaHat}
\end{figure}

There are no constraints on the spectrum other than the anomaly cancellation condition. The different choices are summarized in Tab.\,\eqref{tab:C3Z4OmegaHat} and both lines show the same possibilities as for the unoriented projection $\widehat{\Omega}$ of $\Cc/{\ZZ}'_2$. Thus, both choices can be mass deformed with a mass term for $X_{01'}$, $X_{1'0}$ and $X_{0'1}$, $X_{10'}$. Integrating them out gives the toric super-potential
\begin{equation}
W' = \frac{1 }{m} \epsilon_{pq} \epsilon_{lm}  X_{01}^{p}X_{11'}^{l}X_{1'0'}^{m}X_{0'0}^{q}  \; ,
\end{equation}
which is also the super-potential of $(\Cc/{\ZZ}'_2)/\widehat{\Omega}$, obtained with a toric involution.

\begin{center}
	\begin{tabular*}{0.7\textwidth}{@{\extracolsep{\fill}}ccc}
		\toprule
		Anomaly condition & $\left(X_{11'}^1,X_{11'}^2 \right)$ & $\left(X_{0'0}^1,X_{0'0}^2 \right)$\\
		\midrule
		$N_1 = N_0$ & $ (S,A) $ or $ (A,S) $ & $ (A,S) $ or $ (S,A) $ \\
		$N_1 = N_0 \pm 4$ & $ (S,S) $ or $ (A,A) $ & $ (A,A) $ or $ (S,S) $ \\
		\bottomrule
	\end{tabular*}
\end{center} 
\captionof{table}{The various unoriented projections $\widehat{\Omega}$ of $\CC^3/\ZZ_4$ with gauge groups $U(N_0) \times U(N_1)$. “A” stands for “Antisymmetric representation”, while “S” for “Symmetric representation”.}\label{tab:C3Z4OmegaHat}
\vspace{15pt}

Computing the beta-functions with the anomaly-free condition we have (with $\gamma_{01'}=\gamma_{1'0}=\gamma_{0'1}=\gamma_{10'}$)
\begin{align}
\beta_0 &=  N_1 \left( \gamma_{01} + \gamma_{01'} + \frac{\gamma_{0'0}^{(1)} + \gamma_{0'0}^{(2)}}{2} \right) + \epsilon_{0'0}^{(1)} \left( -5 - \gamma_{0'0}^{(2)} \right) + \epsilon_{0'0}^{2} \left( -5 - \gamma_{0'0}^{(1)} \right) \; , \nonumber \\
\beta_1 &=  N_1 \left( \gamma_{01} + \gamma_{01'} + \frac{\gamma_{11'}^{(1)} + \gamma_{11'}^{(2)}}{2} \right) + \sum_{I=1}^2 \epsilon_{11'}^{(I)} \left( - 5 + 2 \gamma_{01} + 2 \gamma_{01'} \right) + \sum_{I=1}^2 \epsilon_{11'}^{(I)} \gamma_{11'}^{(I)}  \; 
\end{align}
\begin{align}
\beta_0 + \beta_1 &= N_1 \left( 2 \gamma_{01} + 2 \gamma_{01'} + \frac{\gamma_{0'0}^{(1)} + \gamma_{0'0}^{(2)} + \gamma_{11'}^{(1)} + \gamma_{11'}^{(2)}}{2} \right)  \nonumber \\
& + \sum_{I = 1}^{2} \epsilon_{11'}^{(I)} \left( \gamma_{11'}^{(1)} + \gamma_{11'}^{(2)} + 2 \gamma_{01} + 2 \gamma_{01'} \right) - \epsilon_{0'0 }^{(1)} \gamma_{0'0}^{(2)} - \epsilon_{0'0 }^{(2)} \gamma_{0'0}^{(1)} \ .
\end{align}
When the anomalous dimensions are trivial, both beta functions  vanish if ${\epsilon}_{11'}^{(1)} = - {\epsilon}_{11'}^{(2)}$, corresponding to an $\Omega3$ or a compact $\Omega7$ in the smooth space, while for ${\epsilon}_{11'}^{(1)} = {\epsilon}_{11'}^{(2)}$ the theory is not conformal and the unoriented projection is given by a non-compact $\Omega7$ in the resolved space.

\subsection{Orientifold Projection of Non-chiral Orbifolds}

All the examples we have discussed so far involve unoriented projection of chiral orbifolds. We are going to study also non-chiral examples \cite{Bianchi:1995xd, Bianchi:1996zj, Lawrence_1998, Hanany_1999, Yamazaki_2008}, related via mass  deformation to well known theories as the Suspended Pinch Point (SPP), as well as the Conifold and its non-chiral orbifold. The analysis follows closely what is done for chiral examples, thus it will be less detailed.

\subsubsection{Orientifold of $\Nn=2$ Orbifold $\CC^3/{\ZZ}_3'$, $(1,2,0)$}\label{sec:nonchiralC3Z3}

\begin{figure}[H]
\centering
	\begin{subfigure}{0.4\textwidth}
		\centerline{\begin{tikzpicture}[auto, scale= 0.4]
		%%%%%%%%%%%% Nodes %%%%%%%%%%
        \node [circle, fill=black, inner sep=0pt, minimum size=1.3mm] (1) at (-3,-3) {};		
		\node [circle, fill=black, inner sep=0pt, minimum size=1.3mm] (2) at (-3,0) {}; 
		\node [circle, fill=black, inner sep=0pt, minimum size=1.3mm] (3) at (-3,3) {};
		\node [circle, fill=black, inner sep=0pt, minimum size=1.3mm] (4) at (-3,6) {};
		\node [circle, fill=black, inner sep=0pt, minimum size=1.3mm] (5) at (0,-3) {}; 
		\node [circle, fill=black, inner sep=0pt, minimum size=1mm] (6) at (0,0) {}; 			
		\node [circle, fill=black, inner sep=0pt, minimum size=1mm] (7) at (0,3) {};
		\node [circle, fill=black, inner sep=0pt, minimum size=1mm] (8) at (0,6) {}; 
		\node [circle, fill=red, inner sep=0pt, minimum size=1.5mm] (a) at (-6,0) {}; 
		\node [circle, fill=red, inner sep=0pt, minimum size=1.5mm] (b) at (-6,6) {};
		\node [circle, fill=red, inner sep=0pt, minimum size=1.5mm] (c) at (0,0) {};
		\node [circle, fill=red, inner sep=0pt, minimum size=1.5mm] (d) at (0,6) {};
        \node [circle, fill=black, inner sep=0pt, minimum size=1mm] (e) at (-6,-3) {}; 
		\node [circle, fill=black, inner sep=0pt, minimum size=1mm] (f) at (-6,3) {}; 
		%%%%%%%%%%% Lines %%%%%%%%%%%
        \draw (1) edge (2) [thick];
		\draw (2) edge (3) [thick];
		\draw (3) edge (4) [thick];
		\draw (4) edge (5) [thick, gray];
		\draw (5) edge (1) [thick, gray];
		\end{tikzpicture}}
	\end{subfigure}
\begin{subfigure}{0.4\textwidth}
\centerline{\begin{tikzpicture}[auto, scale= 0.4]
		%%%%%%%%%%%% Nodes %%%%%%%%%%
        \node [circle, fill=red, inner sep=0pt, minimum size=1.5mm] (1) at (-3,-3) {};		
		\node [circle, fill=black, inner sep=0pt, minimum size=1.3mm] (2) at (-3,0) {}; 
		\node [circle, fill=red, inner sep=0pt, minimum size=1.5mm] (3) at (-3,3) {};
		\node [circle, fill=black, inner sep=0pt, minimum size=1.3mm] (4) at (-3,6) {};
		\node [circle, fill=black, inner sep=0pt, minimum size=1.3mm] (5) at (0,-3) {}; 
		\node [circle, fill=black, inner sep=0pt, minimum size=1mm] (6) at (0,0) {}; 			
		\node [circle, fill=black, inner sep=0pt, minimum size=1mm] (7) at (0,3) {};
		\node [circle, fill=black, inner sep=0pt, minimum size=1mm] (8) at (0,6) {}; 
		\node [circle, fill=red, inner sep=0pt, minimum size=1.5mm] (a) at (3,-3) {};
		\node [circle, fill=red, inner sep=0pt, minimum size=1.5mm] (b) at (3,3) {};
		\node [circle, fill=black, inner sep=0pt, minimum size=1mm] (c) at (3,0) {}; 
		\node [circle, fill=black, inner sep=0pt, minimum size=1mm] (d) at (3,6) {}; 
		%%%%%%%%%%% Lines %%%%%%%%%%%
        \draw (1) edge (2) [thick];
		\draw (2) edge (3) [thick];
		\draw (3) edge (4) [thick];
		\draw (4) edge (5) [thick, gray];
		\draw (5) edge (1) [thick, gray];
		\end{tikzpicture}}
\end{subfigure}	
	 \\[20pt]
	 \begin{subfigure}{0.4 \textwidth}
		\centerline{\begin{tikzpicture}[auto, scale= 0.5]
		%%%%%%%%%%%% Nodes %%%%%%%%%%
		\node [circle, draw=blue!50, fill=blue!20, inner sep=0pt, minimum size=5mm] (0) at (0,5) {$N_0$}; 
		\node [circle, draw=blue!50, fill=blue!20, inner sep=0pt, minimum size=5mm] (1) at (3,0) {$N_1$}; 			
		\node [circle, draw=blue!50, fill=blue!20, inner sep=0pt, minimum size=5mm] (2) at (-3,0) {$N_2$};
		%%%%%%%%%%% Lines %%%%%%%%%%%
		\draw (0) to [out=120, in=60, looseness=10, thick](0);
		\draw (1) to [out=-60, in=0, looseness=10, thick] (1);
        \draw (2) to [out=180, in=-120, looseness=10, thick] (2);
        \draw (0) to node [pos=0.8]{$X_{01}$} (1) [->, thick];
		\draw (1) to node [pos=0.8]{$X_{12}$} (2) [->, thick];
		\draw (2) to node [pos=0.8]{$X_{20}$} (0) [->, thick];
		\draw (1) to node [pos=0.8, swap]{$X_{10}$} (0) [->, thick];
		\draw (2) to node [pos=0.8, swap]{$X_{21}$} (1) [->, thick];
		\draw (0) to node [pos=0.8, swap]{$X_{02}$} (2) [->, thick];
		\draw [thick, dashed, gray] (0, -2) to node [pos=0.05]{$\Omega$}(0,7.5);
		\end{tikzpicture}}
	\end{subfigure}
	\begin{subfigure}{0.4\textwidth}
		\centerline{\includegraphics[scale=0.3, trim={5cm 5cm 2.5cm 9cm}, clip]{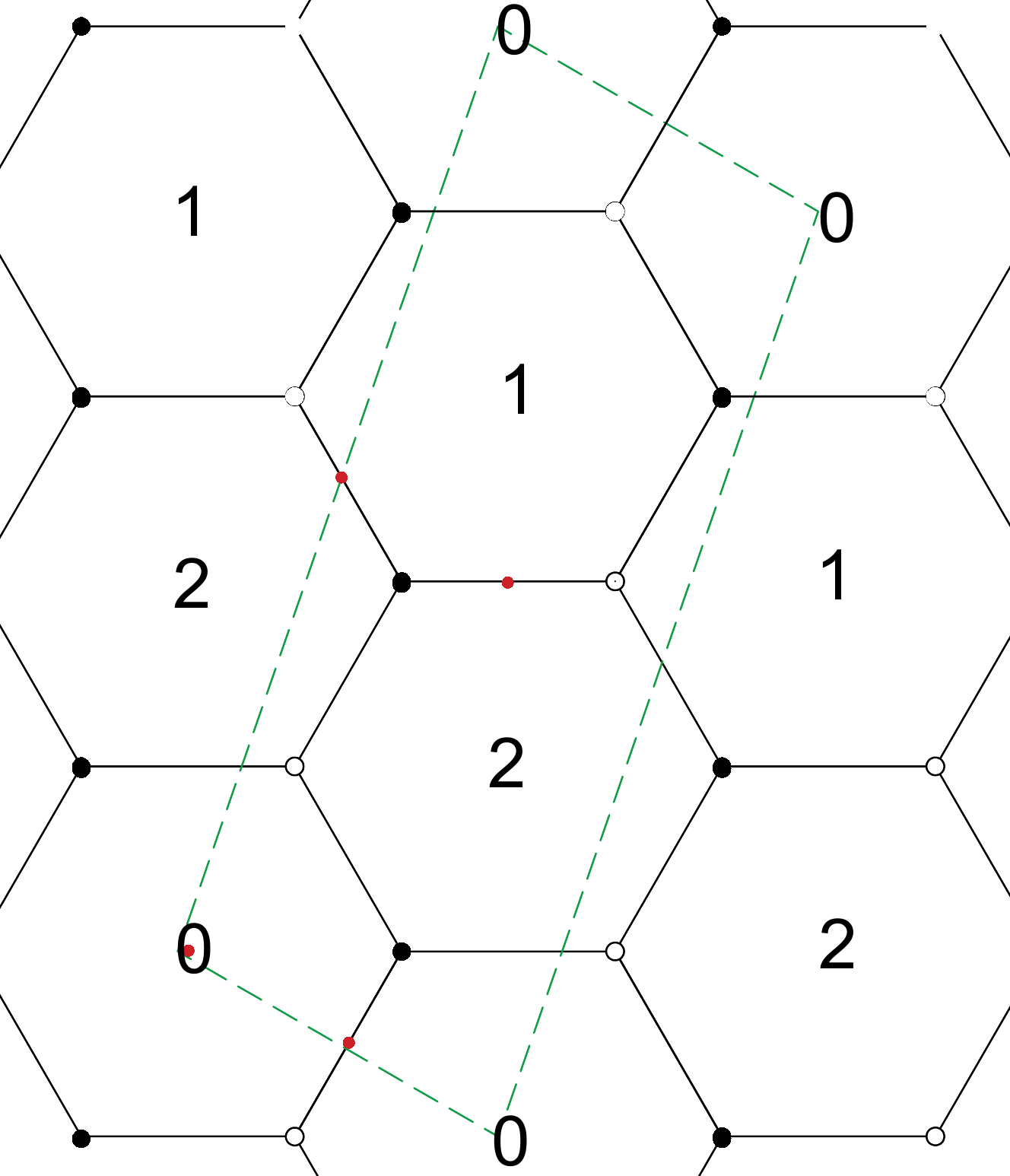}}
	\end{subfigure}
    \caption{The various unoriented descriptions of $ \CC^3/{\ZZ'}_3 $. The upper figure shows the toric diagram and the toric involution with a compact $\Omega$3 (left) and a non-compact $\Omega7$ (right). In the lower row: the left side show the quiver and the $\Omega$-line, whereas on the right side the dimer and the four fixed points in red.}\label{fig:C3Z3pr}
\end{figure}
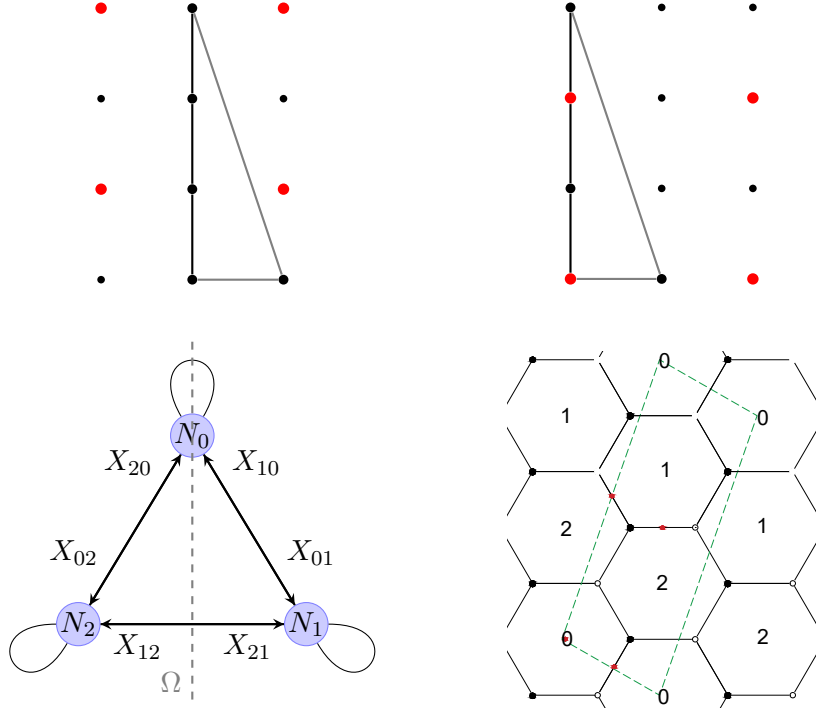

The non-chiral orbifold $\CC^3/{\ZZ}'_3$ with $k_I=(1,2,0)$ has only a junior class from the age classification, hence there are no compact 4-cycles. In fact, the toric diagram has no internal points and the unoriented projection is only given by $\Omega3$ and non-compact $\Omega7$ in the resolved space, see Fig.\,\eqref{fig:C3Z3pr}.

The unoriented projection $\Omega$ acts as 
\begin{align}
{\bf{\overline{N}_2}} = {\bf{N_1}} \;, \qquad
U(N_0) \rightarrow Sp/SO(N_0) \; ,
\end{align}
and the super-potential reads
\begin{align}
W' = {\phi'}_0 \left( X_{01'}X_{1'0} - X_{01} X_{10} \right) + \phi_1& \left( X_{10}X_{01} - X_{11'} X_{1'1} \right) + \phi_{1'} \left( X_{1'1}X_{11'} - X_{1'0} X_{01'} \right) \; ,   
\end{align}
where $\phi_a$ are the adjoint fields at node $a$, ${\phi'}_0$ is projected down to a symmetric or an antisymmetric representation. The anomaly cancellation condition reads
\begin{equation}
\epsilon_{11'} = \epsilon_{1'1} \; .
\end{equation}

From the dimer, $\Omega$ is obtained by fixed point involution and the product of T-parities is negative.Together with the anomaly-free condition, this means that a symmetric ${\phi'}_0$ requires an $Sp(N_0)$ group and an antisymmetric ${\phi'}_0$ requires an $SO(N_0)$ group. The beta functions with a non-trivial anomalous dimension for the adjoint fields are
\begin{align}
\beta_0 &= N_0 \left(1 + \half \gamma_{00}\right) - N_1 + 3 \epsilon_0 - \epsilon_{00}^{(3)}\left(1 -\gamma_{00}\right) \; , \nonumber \\
\beta_1 &= N_1 \left(1 + \gamma_{11}\right) - N_0 - 2 \epsilon_{11'}^{(1)} \; 
\end{align}
and if we suppose $ \gamma_{00} = \gamma_{11} = 0 $ we get
\begin{align}
\beta_0 &= N_0 - N_1 + 3 \epsilon_0 - \epsilon_{00}^{(3)} \; , \nonumber \\
\beta_1 &= N_1 - N_0 - 2 \epsilon_{11'}^{(1)} \; 
\end{align}
and the beta functions can vanish simultaneously when all the charges have the same sign $ (\epsilon_{0},\epsilon_{00},\epsilon_{11'}, \epsilon_{1'1}) = (\pm,\pm,\pm,\pm) $: this corresponds to an $ \Omega $3-plane. Furthermore one notes that the sum of the beta functions with the same condition, namely
\begin{equation}\label{eq:betasumZ3pr}
3\epsilon_0 = \epsilon_{00}^{(3)} + 2 \epsilon_{11'}^{(1)} \; ,
\end{equation} 
In the following table we show the possible cases compatible with the anomaly cancellation condition.

\begin{center}
\begin{tabular*}{0.5\textwidth}{@{\extracolsep{\fill}}ccc}
\toprule
Gauge groups & ${\phi'}_0$ & $(X_{11'}, X_{1'1'})$\\
\midrule
$Sp(N_0)\times U(N_1) $ & $ S $ & $ (S,S) $ or $ (A,A) $ \\
$SO(N_0)\times U(N_1) $ & $ A $ & $ (S,S) $ or $ (A,A) $ \\
\bottomrule
\end{tabular*}
\end{center}  
\captionof{table}{The orientifold involution $\Omega$ of the non-chiral orbifold $\CC^3/{\ZZ'}_3$. “A” stands for “Antisymmetric representation”, while “S” for “Symmetric representation”.}\label{tab:C3Z3pr}

\subsubsection{Orientifold of the Suspended Pinch Point (SPP)}

In \cite{Park:1999ep} it is showed that the SPP theory and its unoriented projections \cite{Franco:2007ii} may be obtained via Higgsing of the orbifold $\CC^3/\left( \ZZ_2 \times \ZZ_2 \right)$, and in \cite{Bianchi:2014qma} it is showed the mass deformed $\CC^3/{\ZZ}'_3$ model flows to the SPP. In the previous section, the same happens with mass deformation of the orientifold involution. The final super-potential reads
\begin{equation}
W' = {\phi'}_0 \left( X_{01'}X_{1'0} - X_{01}X_{01} \right) + X_{11'}X_{1'1}X_{10}X_{01} - X_{1'1}X_{11'}X_{1'0}X_{01'} \; .
\end{equation}

\begin{figure}[H]
\centering
	 \begin{subfigure}{0.3 \textwidth}
		\centerline{\begin{tikzpicture}[auto, scale= 0.5]
		%%%%%%%%%%%% Nodes %%%%%%%%%%
		\node [circle, draw=blue!50, fill=blue!20, inner sep=0pt, minimum size=5mm] (0) at (0,5) {$N_0$}; 
		\node [circle, draw=blue!50, fill=blue!20, inner sep=0pt, minimum size=5mm] (1) at (3,0) {$N_1$}; 			
		\node [circle, draw=blue!50, fill=blue!20, inner sep=0pt, minimum size=5mm] (2) at (-3,0) {$N_2$};
		%%%%%%%%%%% Lines %%%%%%%%%%%
		\draw (0) to [out=120, in=60, looseness=10, thick](0);
        \draw (0) to node [pos=0.8]{$X_{01}$} (1) [->, thick];
		\draw (1) to node [pos=0.8]{$X_{12}$} (2) [->, thick];
		\draw (2) to node [pos=0.8]{$X_{20}$} (0) [->, thick];
		\draw (1) to node [pos=0.8, swap]{$X_{10}$} (0) [->, thick];
		\draw (2) to node [pos=0.8, swap]{$X_{21}$} (1) [->, thick];
		\draw (0) to node [pos=0.8, swap]{$X_{02}$} (2) [->, thick];
		\draw [thick, dashed, gray] (0, -1.5) to node [pos=0.01]{$\Omega$}(0,7.5);
		\end{tikzpicture}}
	\end{subfigure}
	\begin{subfigure}{0.3\textwidth}
		\centerline{\includegraphics[scale=0.23, trim={4cm 5cm 1cm 7cm}, clip]{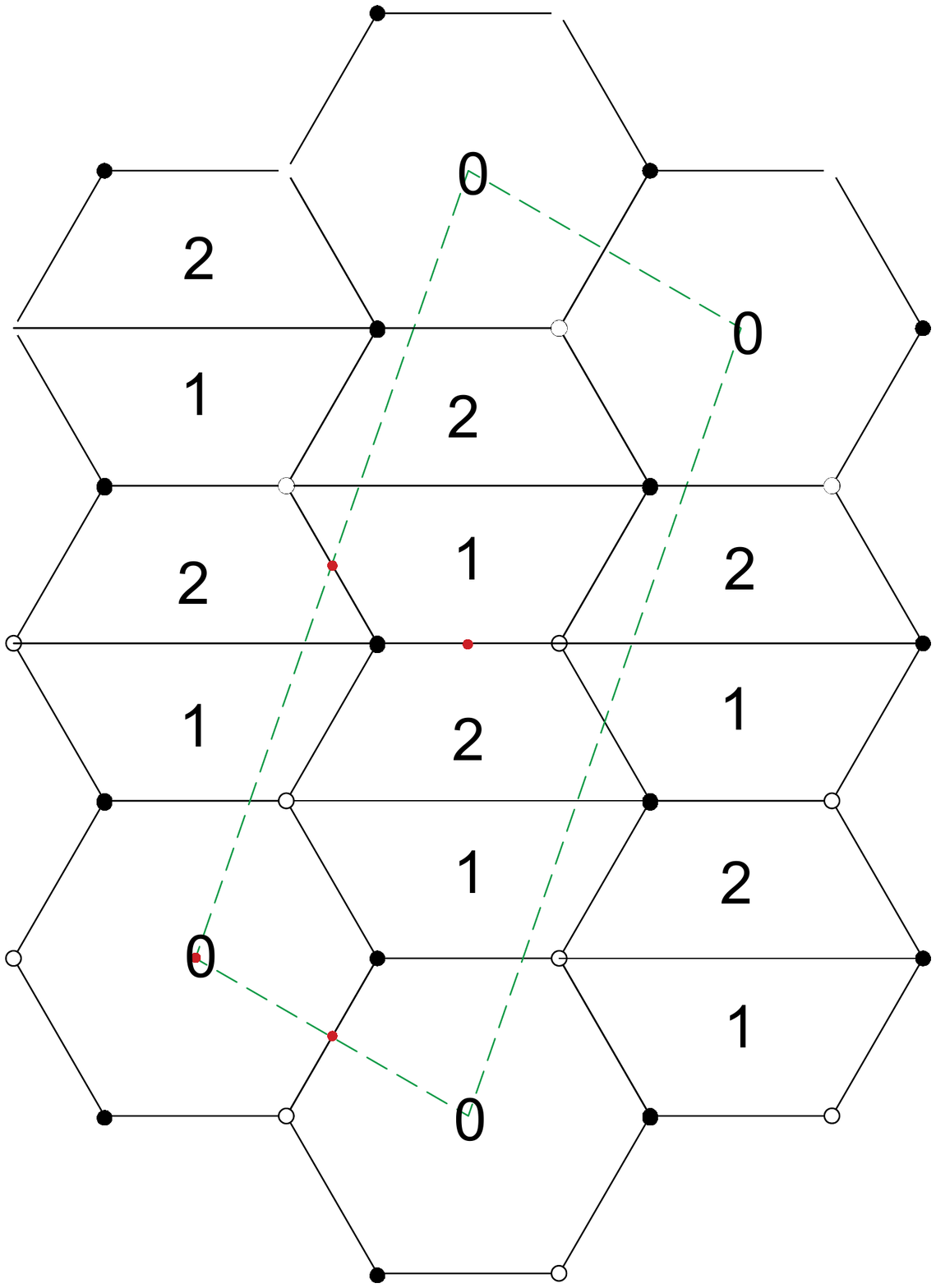}}
	\end{subfigure}
	\begin{subfigure}{0.3\textwidth}
		\centerline{\begin{tikzpicture}[auto, scale= 0.4]
		%%%%%%%%%%%% Nodes %%%%%%%%%%
        \node [circle, fill=red, inner sep=0pt, minimum size=1.5mm] (1) at (-3,3) {};		
		\node [circle, fill=black, inner sep=0pt, minimum size=1mm] (2) at (0,3) {}; 
		\node [circle, fill=red, inner sep=0pt, minimum size=1.5mm] (3) at (3,3) {};
		\node [circle, fill=black, inner sep=0pt, minimum size=1.3mm] (4) at (-3,0) {};
		\node [circle, fill=black, inner sep=0pt, minimum size=1.3mm] (5) at (0,0) {}; 
		\node [circle, fill=black, inner sep=0pt, minimum size=1mm] (6) at (3,0) {}; 			
		\node [circle, fill=red, inner sep=0pt, minimum size=1.5mm] (7) at (-3,-3) {};
		\node [circle, fill=black, inner sep=0pt, minimum size=1.3mm] (8) at (0,-3) {}; 
		\node [circle, fill=red, inner sep=0pt, minimum size=1.5mm] (9) at (3,-3) {};
		%%%%%%%%%%% Lines %%%%%%%%%%%
        \draw (1) edge (4) [thick];
		\draw (4) edge (7) [thick];
		\draw (7) edge (8) [thick];
		\draw (8) edge (5) [thick, gray];
		\draw (5) edge (1) [thick, gray];
		\end{tikzpicture}}
	\end{subfigure}
    \caption{The various unoriented descriptions of the SPP: the quiver, the dimer with the four fixed points in red, and the toric diagram  with toric involution corresponding to a non-compact $\Omega$7 or a non-compact $\Omega7$ and a $\Omega3$, depending on how the toric diagram is triangulated.}\label{fig:SPP}
\end{figure}

The theory is anomaly free if $\epsilon_{11'} = \epsilon_{1'1}$, which is the same condition as in the previous model. From the dimer, the product of the T-parities is positive, then $Sp(N_0)$ requires ${\phi'}_0$ to be antisymmetric, while $SO(N_0)$ requires a symmetric ${\phi'}_0$. The beta functions with a non-trivial anomalous dimension for the adjoint fields are
\begin{align}
\beta_0 &= N_0 \left(1 + \half \gamma_{00}\right) - N_1 + 3 \epsilon_0 - \epsilon_{00}^{(3)} \left(1 -\gamma_{00}\right) \; , \nonumber \\
\beta_1 &= 2 N_1 - N_0 - 2 \epsilon_{11'}  \; 
\end{align}
and if we assume $ \gamma_{00}= 0 $ we get
\begin{align}
\beta_0 &= N_0 - N_1 + 3 \epsilon_0 - \epsilon_{00} \; , \nonumber \\
\beta_1 &= 2 N_1 - N_0 - 2 \epsilon_{11'} \; . 
\end{align}
The beta functions can vanish separately if $N_1 = \epsilon_{00} + 2 \epsilon_{11'} - 3 \epsilon_0$ and $N_0 = 2 \left( \epsilon_{00'} + \epsilon_{11'} - 3 \epsilon_0 \right) $. The sum vanishes if $N_1 = \epsilon_{00} + 2 \epsilon_{11'} - 3 \epsilon_0$. The following table summarizes the possibilities compatible with conformal invariance. When all the fields transform in the same representation, the orientifold involution is given by an $ \Omega3 $. 

\begin{center}
\begin{tabular*}{0.55\textwidth}{@{\extracolsep{\fill}}ccc}
\toprule
Gauge groups & ${\phi'}_0$ & $(X_{11'}, X_{1'1'})$\\
\midrule
$Sp(N_0)\times U(N_1) $ & $ A $ & $ (S,S) $ or $ (A,A) $ \\
$SO(N_0)\times U(N_1) $ & $ S $ & $ (S,S) $ or $ (A,A) $ \\
\bottomrule
\end{tabular*}
\captionof{table}{The orientifold projection $\Omega$ of the SPP. “A” stands for “Antisymmetric representation”, while “S” for “Symmetric representation”.}\label{tab:SPP}
\end{center}

\subsubsection{Orientifold of $\Nn=2$ Orbifold $\CC^3/{\ZZ}'_4 $, $(1,3,0)$}

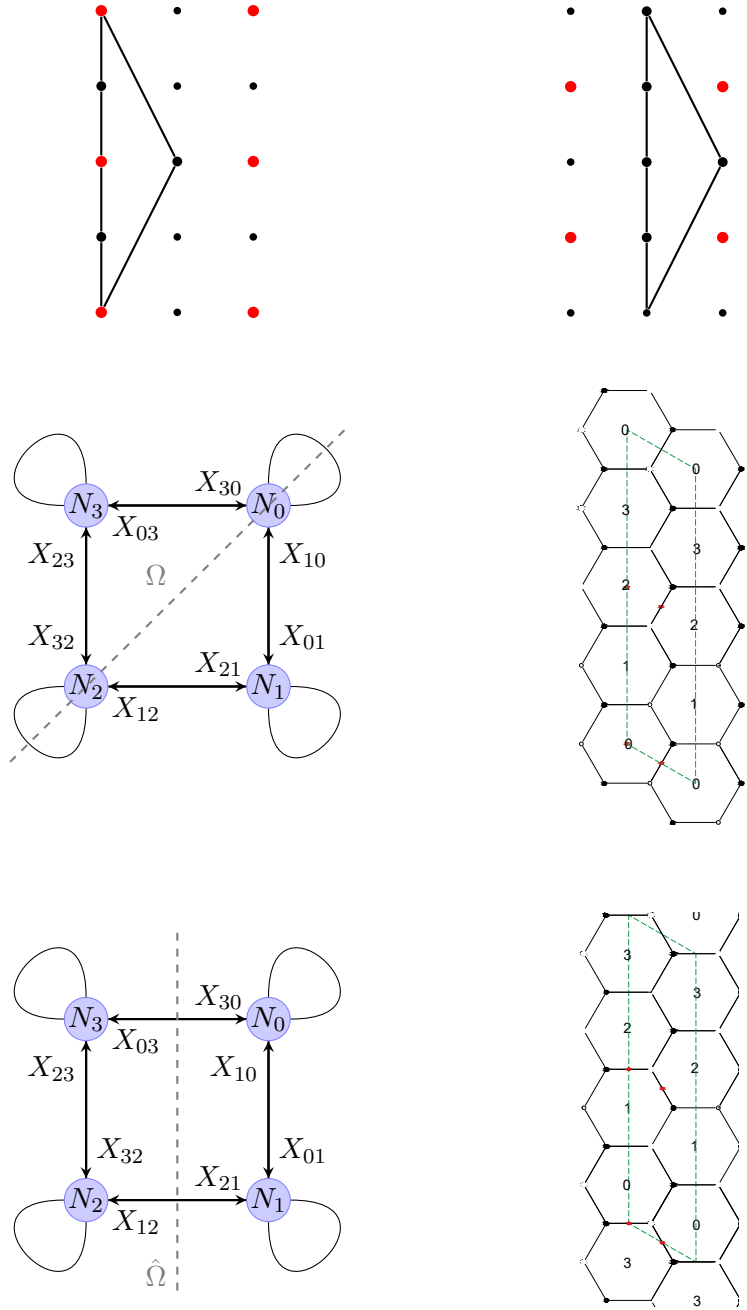
\begin{figure}[H]
\centering
	\begin{subfigure}{0.4\textwidth}
		\centerline{\begin{tikzpicture}[auto, scale= 0.4]
		%%%%%%%%%%%% Nodes %%%%%%%%%%
        \node [circle, fill=red, inner sep=0pt, minimum size=1.5mm] (1) at (-2.5,-2.5) {};		
		\node [circle, fill=black, inner sep=0pt, minimum size=1.3mm] (2) at (-2.5,0) {}; 
		\node [circle, fill=red, inner sep=0pt, minimum size=1.5mm] (3) at (-2.5,2.5) {};
		\node [circle, fill=black, inner sep=0pt, minimum size=1.3mm] (4) at (-2.5,5) {};
		\node [circle, fill=red, inner sep=0pt, minimum size=1.5mm] (5) at (-2.5,7.5) {}; 
		\node [circle, fill=black, inner sep=0pt, minimum size=1mm] (6) at (0,-2.5) {}; 			
		\node [circle, fill=black, inner sep=0pt, minimum size=1mm] (7) at (0,0) {};
		\node [circle, fill=black, inner sep=0pt, minimum size=1.3mm] (8) at (0,2.5) {}; 
		\node [circle, fill=black, inner sep=0pt, minimum size=1mm] (9) at (0,5) {}; 
		\node [circle, fill=black, inner sep=0pt, minimum size=1mm] (10) at (0,7.5) {};
		\node [circle, fill=red, inner sep=0pt, minimum size=1.5mm] (11) at (2.5,-2.5) {};
        \node [circle, fill=black, inner sep=0pt, minimum size=1mm] (12) at (2.5,0) {}; 
		\node [circle, fill=red, inner sep=0pt, minimum size=1.5mm] (13) at (2.5,2.5) {}; 
		\node [circle, fill=black, inner sep=0pt, minimum size=1mm] (14) at (2.5,5) {};
        \node [circle, fill=red, inner sep=0pt, minimum size=1.5mm] (15) at (2.5,7.5) {};
		%%%%%%%%%%% Lines %%%%%%%%%%%
        \draw (1) edge (2) [thick];
		\draw (2) edge (3) [thick];
		\draw (3) edge (4) [thick];
		\draw (4) edge (5) [thick];
		\draw (5) edge (8) [thick];
		\draw (8) edge (1) [thick];
		\end{tikzpicture}}
	\end{subfigure}
\begin{subfigure}{0.4\textwidth}
\centerline{\begin{tikzpicture}[auto, scale= 0.4]
		%%%%%%%%%%%% Nodes %%%%%%%%%%
        \node [circle, fill=black, inner sep=0pt, minimum size=1mm] (1) at (-2.5,-2.5) {};		
		\node [circle, fill=red, inner sep=0pt, minimum size=1.5mm] (2) at (-2.5,0) {}; 
		\node [circle, fill=black, inner sep=0pt, minimum size=1mm] (3) at (-2.5,2.5) {};
		\node [circle, fill=red, inner sep=0pt, minimum size=1.5mm] (4) at (-2.5,5) {};
		\node [circle, fill=black, inner sep=0pt, minimum size=1mm] (5) at (-2.5,7.5) {}; 
		\node [circle, fill=black, inner sep=0pt, minimum size=1mm] (6) at (0,-2.5) {}; 			
		\node [circle, fill=black, inner sep=0pt, minimum size=1.3mm] (7) at (0,0) {};
		\node [circle, fill=black, inner sep=0pt, minimum size=1.3mm] (8) at (0,2.5) {}; 
		\node [circle, fill=black, inner sep=0pt, minimum size=1.3mm] (9) at (0,5) {}; 
		\node [circle, fill=black, inner sep=0pt, minimum size=1.3mm] (10) at (0,7.5) {};
		\node [circle, fill=black, inner sep=0pt, minimum size=1mm] (11) at (2.5,-2.5) {};
        \node [circle, fill=red, inner sep=0pt, minimum size=1.5mm] (12) at (2.5,0) {}; 
		\node [circle, fill=black, inner sep=0pt, minimum size=1.3mm] (13) at (2.5,2.5) {}; 
		\node [circle, fill=red, inner sep=0pt, minimum size=1.5mm] (14) at (2.5,5) {};
        \node [circle, fill=black, inner sep=0pt, minimum size=1mm] (15) at (2.5,7.5) {};
		%%%%%%%%%%% Lines %%%%%%%%%%%
        \draw (6) edge (7) [thick];
		\draw (7) edge (8) [thick];
		\draw (8) edge (9) [thick];
		\draw (9) edge (10) [thick];
		\draw (10) edge (13) [thick];
		\draw (13) edge (6) [thick];
		\end{tikzpicture}}
\end{subfigure}	
	 \\[15pt]
	 \begin{subfigure}{0.4 \textwidth}
		\centerline{\begin{tikzpicture}[auto, scale= 0.4]
		%%%%%%%%%%%% Nodes %%%%%%%%%%
		\node [circle, draw=blue!50, fill=blue!20, inner sep=0pt, minimum size=5mm] (0) at (3,3) {$N_0$}; 
		\node [circle, draw=blue!50, fill=blue!20, inner sep=0pt, minimum size=5mm] (1) at (3,-3) {$N_1$}; 			
		\node [circle, draw=blue!50, fill=blue!20, inner sep=0pt, minimum size=5mm] (2) at (-3,-3) {$N_2$};
		\node [circle, draw=blue!50, fill=blue!20, inner sep=0pt, minimum size=5mm] (3) at (-3,3) {$N_3$}; 
		%%%%%%%%%%% Lines %%%%%%%%%%%
		\draw (0) to [out=0, in=90, looseness=10, thick](0);
		\draw (1) to [out=-90, in=0, looseness=10, thick] (1);
        \draw (2) to [out=180, in=-90, looseness=10, thick] (2);
        \draw (3) to [out=90, in=180, looseness=10, thick] (3);
        \draw (0) to node [pos=0.8]{$X_{01}$} (1) [->, thick];
		\draw (1) to node [pos=0.8]{$X_{12}$} (2) [->, thick];
		\draw (2) to node [pos=0.8]{$X_{23}$} (3) [->, thick];
		\draw (3) to node [pos=0.8]{$X_{30}$} (0) [->, thick];
        \draw (1) to node [pos=0.8, swap]{$X_{10}$} (0) [->, thick];
		\draw (0) to node [pos=0.8]{$X_{03}$} (3) [->, thick];
		\draw (3) to node [pos=0.8, swap]{$X_{32}$} (2) [->, thick];
		\draw (2) to node [pos=0.8]{$X_{21}$} (1) [->, thick];
		\draw [thick, dashed, gray] (-5.5, -5.5) to node [pos=0.5]{$\Omega$}(5.5,5.5);
		\end{tikzpicture}}
	\end{subfigure}
	\begin{subfigure}{0.4\textwidth}
		\centerline{\includegraphics[scale=0.25, trim={2.4cm 2.5cm 8.3cm 3cm}, clip]{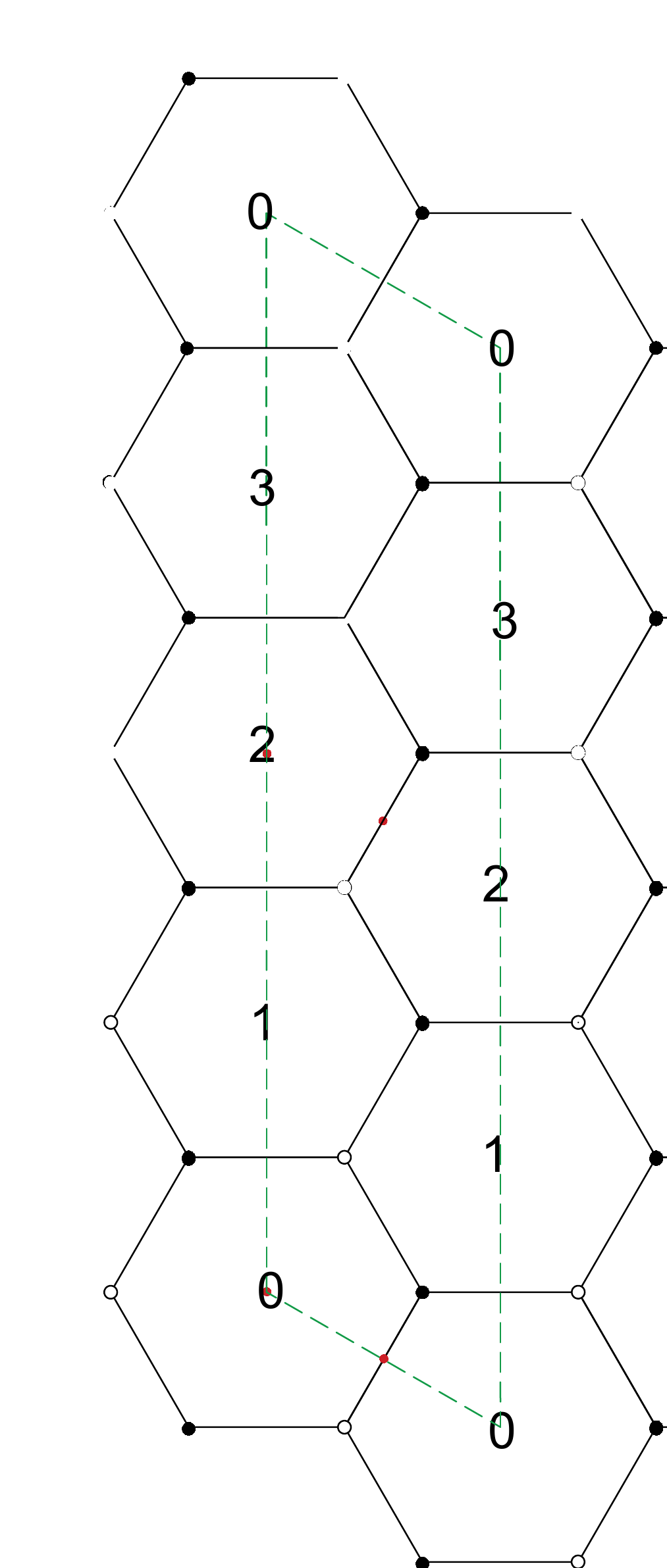}}
	\end{subfigure} \\[15pt]
	\begin{subfigure}{0.4 \textwidth}
		\centerline{\begin{tikzpicture}[auto, scale= 0.4]
		%%%%%%%%%%%% Nodes %%%%%%%%%%
		\node [circle, draw=blue!50, fill=blue!20, inner sep=0pt, minimum size=5mm] (0) at (3,3) {$N_0$}; 
		\node [circle, draw=blue!50, fill=blue!20, inner sep=0pt, minimum size=5mm] (1) at (3,-3) {$N_1$}; 			
		\node [circle, draw=blue!50, fill=blue!20, inner sep=0pt, minimum size=5mm] (2) at (-3,-3) {$N_2$};
		\node [circle, draw=blue!50, fill=blue!20, inner sep=0pt, minimum size=5mm] (3) at (-3,3) {$N_3$}; 
		%%%%%%%%%%% Lines %%%%%%%%%%%
		\draw (0) to [out=0, in=90, looseness=10, thick](0);
		\draw (1) to [out=-90, in=0, looseness=10, thick] (1);
        \draw (2) to [out=180, in=-90, looseness=10, thick] (2);
        \draw (3) to [out=90, in=180, looseness=10, thick] (3);
        \draw (0) to node [pos=0.8]{$X_{01}$} (1) [->, thick];
		\draw (1) to node [pos=0.8]{$X_{12}$} (2) [->, thick];
		\draw (2) to node [pos=0.8]{$X_{23}$} (3) [->, thick];
		\draw (3) to node [pos=0.8]{$X_{30}$} (0) [->, thick];
        \draw (1) to node [pos=0.8]{$X_{10}$} (0) [->, thick];
		\draw (0) to node [pos=0.8]{$X_{03}$} (3) [->, thick];
		\draw (3) to node [pos=0.8]{$X_{32}$} (2) [->, thick];
		\draw (2) to node [pos=0.8]{$X_{21}$} (1) [->, thick];
		\draw [thick, dashed, gray] (0, -6) to node [pos=0.05]{$\hat{\Omega}$}(0,6);
		\end{tikzpicture}}
	\end{subfigure}
	\begin{subfigure}{0.4\textwidth}
		\centerline{\includegraphics[scale=0.3, trim={2.4cm 4.6cm 10.5cm 7.7cm}, clip]{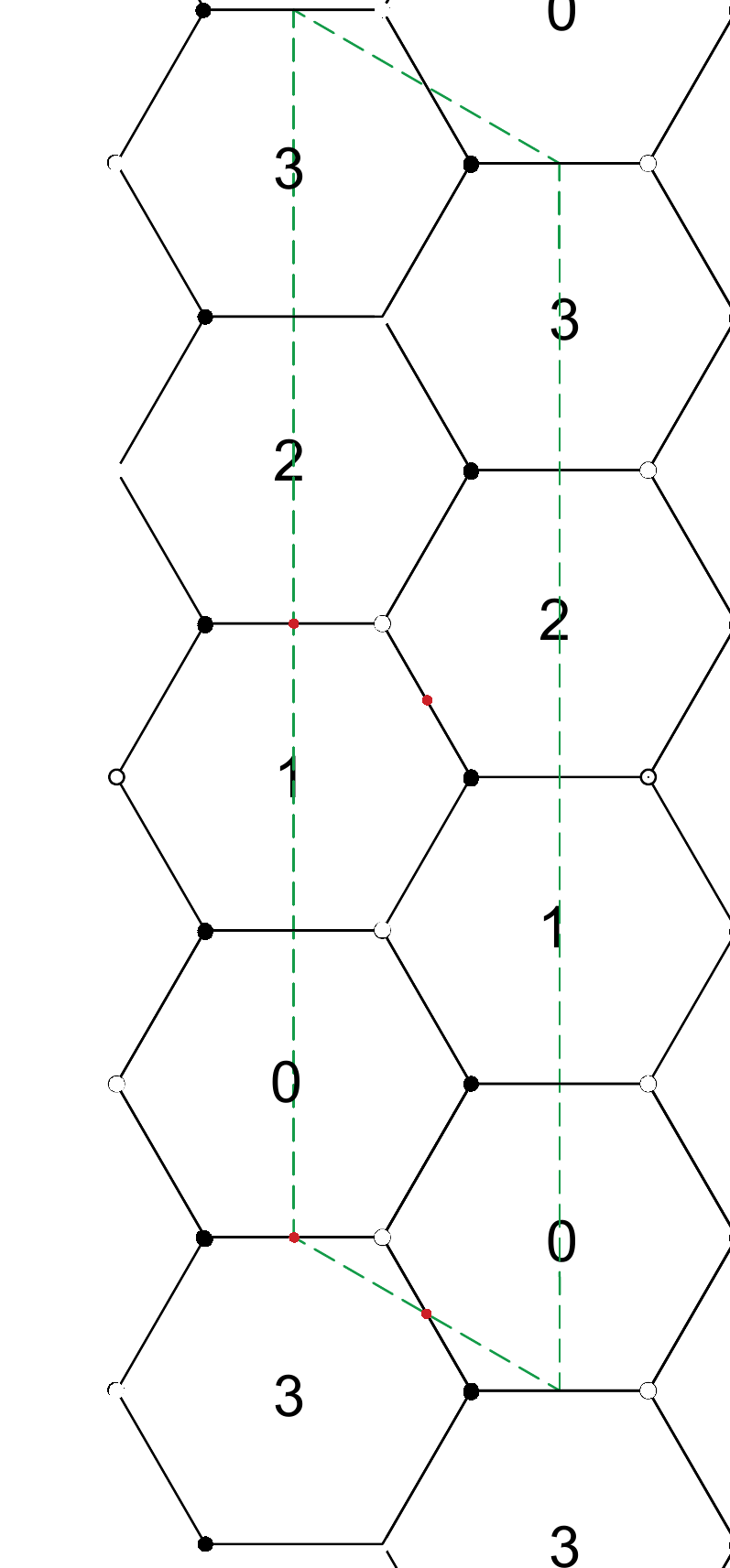}}
	\end{subfigure}
    \caption{The various unoriented descriptions of $ \CC^3/{\ZZ}'_4 $. The upper figure shows the toric diagram and the toric involution with a non-compact $\Omega$7 (left) and a $\Omega3$ (right). The middle row shows the orientifold involution $\Omega$, on quiver (left) and dimer (right), while in the lower row are drawn the quiver (left) and the dimer (right) for the $\hat{\Omega}$.}\label{fig:C3Z4pr}
\end{figure}

In this section we study the unoriented projections $\Omega$ and $\hat{\Omega}$ of the non-chiral $\CC^3/{\ZZ}'_4$ model with $k_I=(1,3,0)$, whose conjugacy classes are only junior classes, from $(1,3,0)$ and $(2,2,0)$. There are no compact 4-cycles, in agreement with the fact that the toric diagram has no internal point and hence no compact $\Omega7$ in the resolved space. The various diagrams are drawn in Fig.\,\eqref{fig:C3Z4pr}.

\subsubsection*{Orientifold $\Omega$ of $ \CC^3/{\ZZ}'_4$}

The action of the involution is
\begin{align}
{\bf{\overline{N}_3}} = {\bf{N_1}} \; , \qquad
U(N_0) \rightarrow Sp/SO(N_0) \; , \qquad
U(N_2) \rightarrow Sp/SO(N_2) \; ,
\end{align}
and the super-potential reads
\begin{align}
W' &= {\phi'}_0 \left( X_{01} X_{10} - X_{01'} X_{1'0} \right) + {\phi}_1 \left( X_{12} X_{21} - X_{10} X_{01} \right) \nonumber \\
&+ {\phi'}_{2} \left( X_{21'} X_{1'2} - X_{21} X_{12} \right) + {\phi}_{1'} \left( X_{1'0} X_{01'} - X_{1'2} X_{21'} \right) \; .
\end{align}
The theory is anomaly-free without any relevant restriction on the gauge group ranks and on the spectrum. From the dimer, this orientifold configuration is given by four T-parities whose product is positive. The choices are displayed in Tab \eqref{tab:C3Z4pr}. \\

The beta-functions read
\begin{align}
2 \beta_0 &= N_0 \left( 2 + \gamma_{00} \right) - 2 N_1 + 6 \epsilon_0 -  2 \epsilon_{00}\left( 1 - \gamma_{00} \right) \; , \nonumber \\
\beta_1 &= 2 N_1 \left( 2 + \gamma_{11} \right) - N_0 - N_2 \; , \nonumber \\
2\beta_2 &= N_2 \left( 2 + \gamma_{22} \right) - 2 N_1 + 6 \epsilon_2 - 2 \epsilon_{22} \left( 2 + \gamma_{22} \right) \; . 
\end{align}
In case of trivial anomalous dimensions, by imposing the simultaneous vanishing of the individual beta functions we get a condition on the charges
\begin{equation} 
3 (\epsilon_{0} + \epsilon_2) = \epsilon_{22} + \epsilon_{00}
\end{equation}
which can be satisfied only if $\epsilon_0 = - \epsilon_2$ and $\epsilon_{00} = - \epsilon_{22}$. This corresponds to projecting the group and the adjoint fields in opposite manner.

Their sum is $\sum_a \beta_0 = 3 \left( \epsilon_0 + \epsilon_2 \right) + \epsilon_{00} + \epsilon_{22} $, which vanishes, again, only if $\epsilon_0 = - \epsilon_2$ and $\epsilon_{00} = - \epsilon_{22}$. This corresponds to an $\Omega3$ plane, while other choices (for which $\sum_a \beta_a \neq 0$) are given by a non-compact $\Omega7$.

\begin{center}
\begin{tabular*}{0.6\textwidth}{@{\extracolsep{\fill}}ccc}
\toprule
Gauge groups & ${\phi'}_{0}$ & ${\phi'}_{2}$\\
\midrule
$Sp(N_0) \times U(N_1) \times Sp(N_2) $ & S/A & S/A \\
$Sp(N_0) \times U(N_1) \times SO(N_2) $ & S/A & A/S \\
$SO(N_0) \times U(N_1) \times Sp(N_2) $ & S/A & A/S \\
$SO(N_0) \times U(N_1) \times SO(N_2) $ & S/A & S/A \\
\bottomrule
\end{tabular*}
\end{center}  
\captionof{table}{The orientifold projection $\Omega$ of $\CC^3/{\ZZ'}_4$. “A” stands for “Antisymmetric representation”, while “S” for “Symmetric representation”.}\label{tab:C3Z4pr}

\subsubsection*{Orientifold $\hat{\Omega}$ of $\CC^3/{\ZZ'}_4$}

The action of the involution is
\begin{align}
{\bf{\overline{N}_3}}= {\bf{N_0}} \; , \qquad
{\bf{\overline{N}_2}} = {\bf{N_1}} \; , 
\end{align}
and the super-potential reads
\begin{align}
W' &= {\phi}_0 \left( X_{01} X_{10} - X_{00'} X_{0'0} \right) + {\phi}_1 \left( X_{12'} X_{2'1} - X_{10} X_{01} \right) \nonumber \\
&+ {\phi}_{1'} \left( X_{1'0'} X_{0'1'} - X_{1'1} X_{11'} \right) + {\phi}_{0'} \left( X_{0'0} X_{00'} - X_{0'1'} X_{1'0'} \right) \; .
\end{align}
The anomaly-free condition gives
\begin{align}
\epsilon_{00'} = \epsilon_{0'0} \; , \nonumber \\
\epsilon_{11'} = \epsilon_{1'1} \; ,
\end{align}
which is in agreement with the constraint from the dimer, since the product of the T-paritites must be positive. The choices are reported in Tab.\,\eqref{tab:C3Z4prhat}. 

The beta-functions read
\begin{align}
\beta_0 &= N_0 \left( 1 + \gamma_{00} \right) - N_1 - 2 \epsilon_{00'} \; , \nonumber \\
\beta_1 &= N_1 \left( 1 + \gamma_{11} \right) - N_0 - 2 \epsilon_{11'} \; ,
\end{align}
whose sum vanishes at large $ N $ only if $\epsilon_{00'} = - \epsilon_{11'}$, which corresponds to an $\Omega3$ plane, while the other choice (for which $\sum_a \beta_a \neq 0$) are given by a non-compact $\Omega7$. The same condition holds for each $\beta_a=0$, with $N_1 = N_0 - 2 \epsilon_{00'}$.

\begin{center}
\begin{tabular*}{0.6\textwidth}{@{\extracolsep{\fill}}ccc}
\toprule
Gauge groups & $(X_{00'}, X_{0'0})$ & $(X_{11'}, X_{1'1})$\\
\midrule
$U(N_0) \times U(N_1)$ & $ (S,S) $ or $ (A,A) $ & $ (S,S) $ or $ (A,A) $ \\
$U(N_0) \times U(N_1)$ & $ (S,S) $ or $ (A,A) $ & $ (A,A) $ or $ (S,S) $ \\
\bottomrule
\end{tabular*}
\end{center}  
\captionof{table}{The unoriented involution $\hat{\Omega}$ of $\CC^3/{\ZZ'}_4$. “A” stands for “Antisymmetric representation”, while “S” for “Symmetric representation”.}\label{tab:C3Z4prhat}

\subsubsection{Orientifold of the Non-chiral orbifold of the conifold $\Cc/\ZZ_2 $}

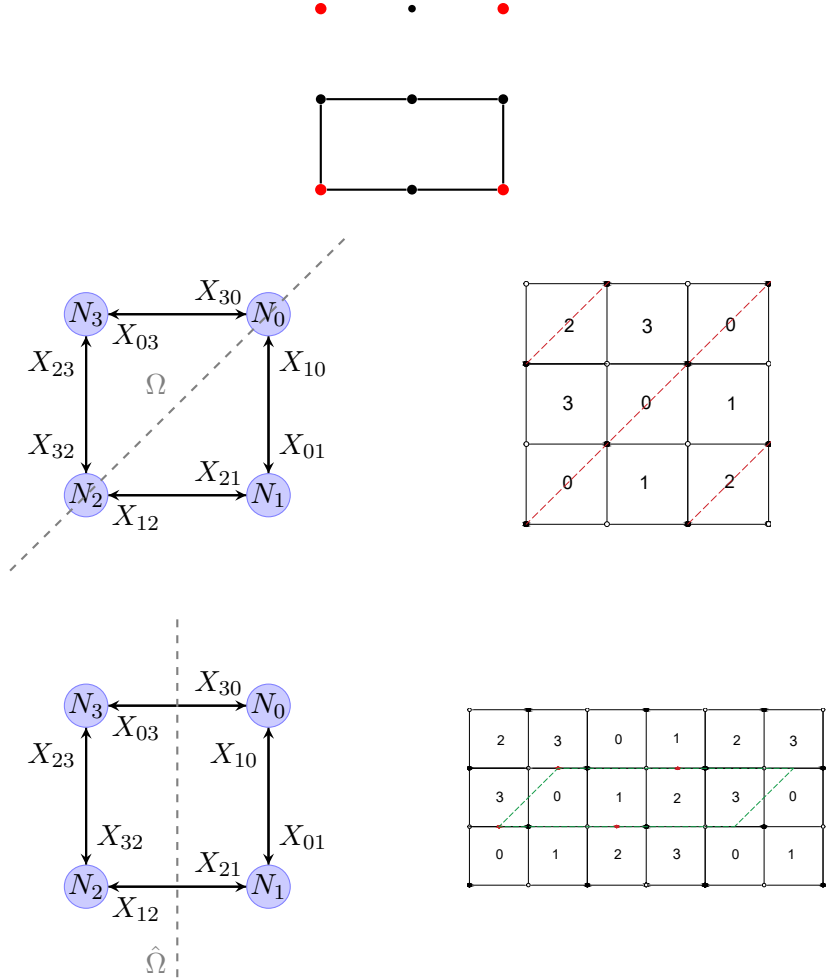
\begin{figure}[H]
\centering
	\begin{subfigure}{0.4\textwidth}
		\centerline{\begin{tikzpicture}[auto, scale= 0.4]
		%%%%%%%%%%%% Nodes %%%%%%%%%%
        \node [circle, fill=red, inner sep=0pt, minimum size=1.5mm] (1) at (-3,-3) {};		
		\node [circle, fill=black, inner sep=0pt, minimum size=1.3mm] (2) at (-3,0) {}; 
		\node [circle, fill=red, inner sep=0pt, minimum size=1.5mm] (3) at (-3,3) {};
		\node [circle, fill=black, inner sep=0pt, minimum size=1.3mm] (4) at (0,-3) {};
		\node [circle, fill=black, inner sep=0pt, minimum size=1.3mm] (5) at (0,0) {}; 
		\node [circle, fill=black, inner sep=0pt, minimum size=1mm] (6) at (0,3) {}; 			
		\node [circle, fill=red, inner sep=0pt, minimum size=1.5mm] (7) at (3,-3) {};
		\node [circle, fill=black, inner sep=0pt, minimum size=1.3mm] (8) at (3,0) {}; 
		\node [circle, fill=red, inner sep=0pt, minimum size=1.5mm] (9) at (3,3) {}; 
		%%%%%%%%%%% Lines %%%%%%%%%%%
        \draw (1) edge (2) [thick];
		\draw (2) edge (5) [thick];
		\draw (5) edge (8) [thick];
		\draw (8) edge (7) [thick];
		\draw (7) edge (4) [thick];
		\draw (4) edge (1) [thick];
		\end{tikzpicture}}
	\end{subfigure}	
	 \\[15pt]
	 \begin{subfigure}{0.4 \textwidth}
		\centerline{\begin{tikzpicture}[auto, scale= 0.4]
		%%%%%%%%%%%% Nodes %%%%%%%%%%
		\node [circle, draw=blue!50, fill=blue!20, inner sep=0pt, minimum size=5mm] (0) at (3,3) {$N_0$}; 
		\node [circle, draw=blue!50, fill=blue!20, inner sep=0pt, minimum size=5mm] (1) at (3,-3) {$N_1$}; 			
		\node [circle, draw=blue!50, fill=blue!20, inner sep=0pt, minimum size=5mm] (2) at (-3,-3) {$N_2$};
		\node [circle, draw=blue!50, fill=blue!20, inner sep=0pt, minimum size=5mm] (3) at (-3,3) {$N_3$}; 
		%%%%%%%%%%% Lines %%%%%%%%%%%
        \draw (0) to node [pos=0.8]{$X_{01}$} (1) [->, thick];
		\draw (1) to node [pos=0.8]{$X_{12}$} (2) [->, thick];
		\draw (2) to node [pos=0.8]{$X_{23}$} (3) [->, thick];
		\draw (3) to node [pos=0.8]{$X_{30}$} (0) [->, thick];
        \draw (1) to node [pos=0.8, swap]{$X_{10}$} (0) [->, thick];
		\draw (0) to node [pos=0.8]{$X_{03}$} (3) [->, thick];
		\draw (3) to node [pos=0.8, swap]{$X_{32}$} (2) [->, thick];
		\draw (2) to node [pos=0.8]{$X_{21}$} (1) [->, thick];
		\draw [thick, dashed, gray] (-5.5, -5.5) to node [pos=0.5]{$\Omega$}(5.5,5.5);
		\end{tikzpicture}}
	\end{subfigure}
	\begin{subfigure}{0.4\textwidth}
		\centerline{\includegraphics[scale=0.3, trim={4.3cm 2.8cm 5.8cm 16cm}, clip]{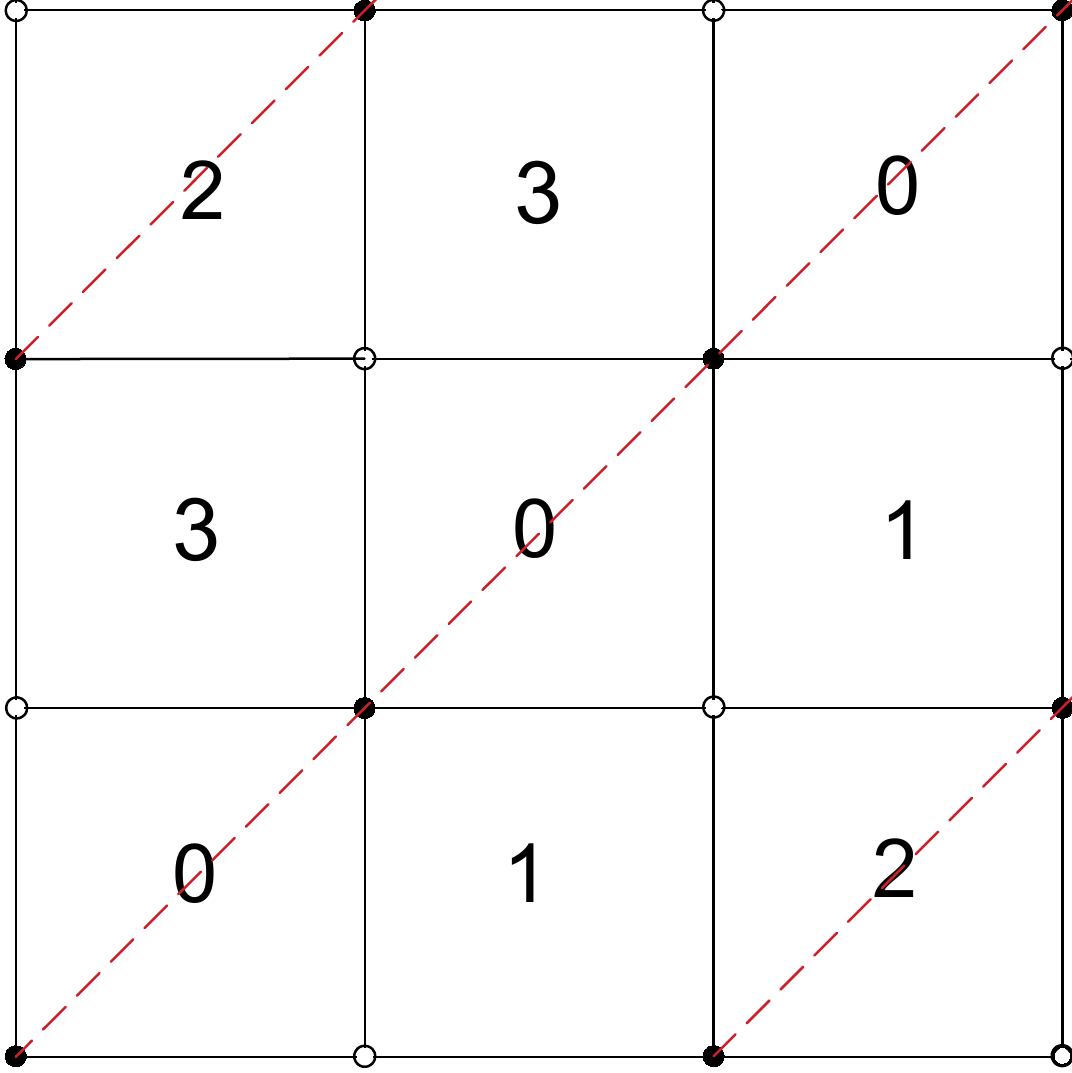}}
	\end{subfigure} \\[15pt]
	\begin{subfigure}{0.4 \textwidth}
		\centerline{\begin{tikzpicture}[auto, scale= 0.4]
		%%%%%%%%%%%% Nodes %%%%%%%%%%
		\node [circle, draw=blue!50, fill=blue!20, inner sep=0pt, minimum size=5mm] (0) at (3,3) {$N_0$}; 
		\node [circle, draw=blue!50, fill=blue!20, inner sep=0pt, minimum size=5mm] (1) at (3,-3) {$N_1$}; 			
		\node [circle, draw=blue!50, fill=blue!20, inner sep=0pt, minimum size=5mm] (2) at (-3,-3) {$N_2$};
		\node [circle, draw=blue!50, fill=blue!20, inner sep=0pt, minimum size=5mm] (3) at (-3,3) {$N_3$}; 
		%%%%%%%%%%% Lines %%%%%%%%%%%
        \draw (0) to node [pos=0.8]{$X_{01}$} (1) [->, thick];
		\draw (1) to node [pos=0.8]{$X_{12}$} (2) [->, thick];
		\draw (2) to node [pos=0.8]{$X_{23}$} (3) [->, thick];
		\draw (3) to node [pos=0.8]{$X_{30}$} (0) [->, thick];
        \draw (1) to node [pos=0.8]{$X_{10}$} (0) [->, thick];
		\draw (0) to node [pos=0.8]{$X_{03}$} (3) [->, thick];
		\draw (3) to node [pos=0.8]{$X_{32}$} (2) [->, thick];
		\draw (2) to node [pos=0.8]{$X_{21}$} (1) [->, thick];
		\draw [thick, dashed, gray] (0, -6) to node [pos=0.05]{$\hat{\Omega}$}(0,6);
		\end{tikzpicture}}
	\end{subfigure}
	\begin{subfigure}{0.4\textwidth}
		\centerline{\includegraphics[scale=0.3, trim={3.5cm 2.1cm 1.8cm 19.7cm}, clip]{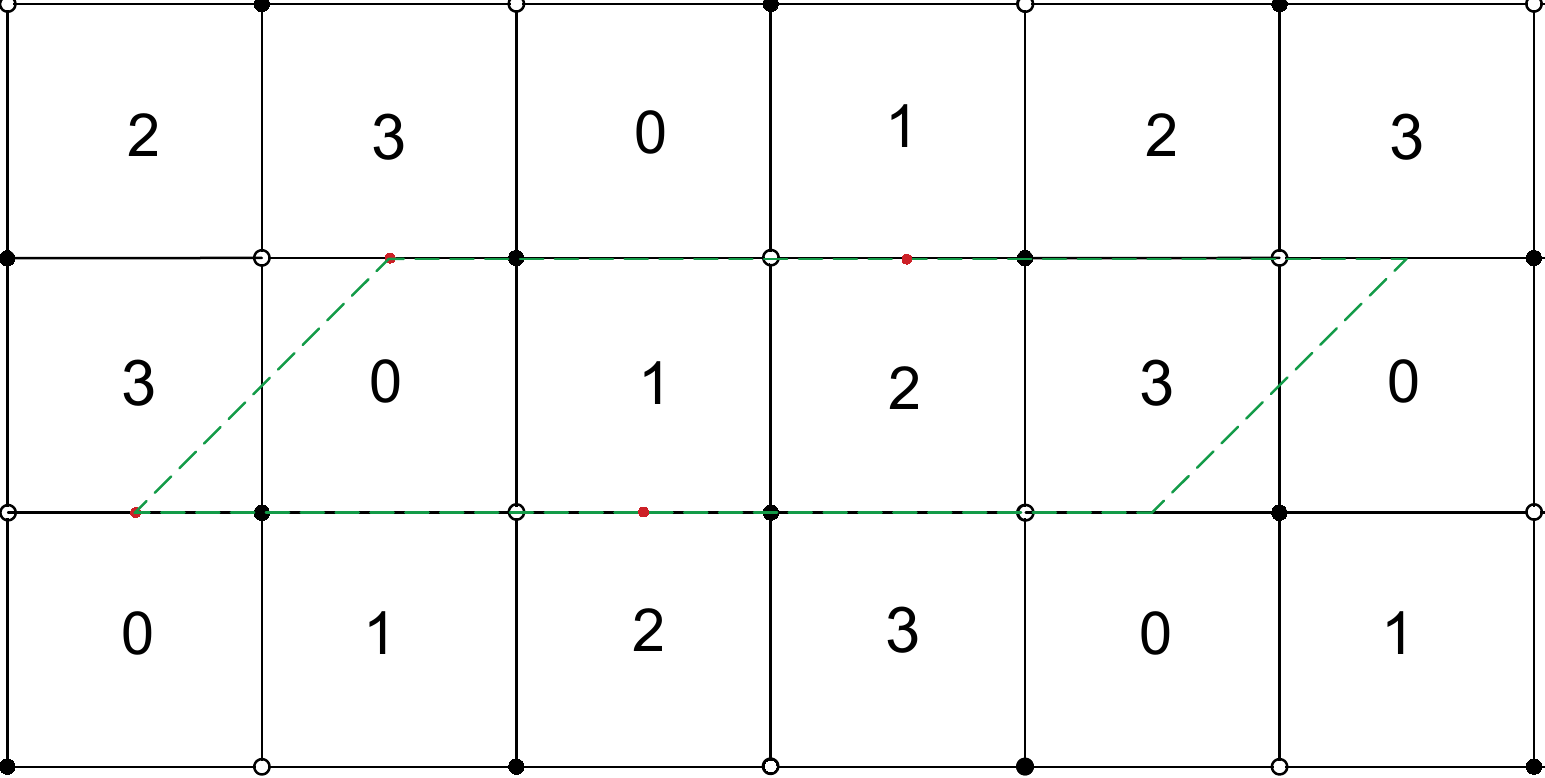}}
	\end{subfigure}
    \caption{The various unoriented descriptions of the non-chiral $ \Cc/\ZZ_2 $. The upper figure shows a possible toric diagram from which one can notice the presence of a non-compact $\Omega$7 and a $\Omega3$ or only a non-compact $\Omega7$, depending on how the toric diagram is triangulated. The middle row shows the orientifold involution $\Omega$, on quiver (left) and dimer (right), while in the lower row are drawn the quiver (left) and the dimer (right) for the involution $\hat{\Omega}$.}\label{fig:CZ2pr}
\end{figure}

In \cite{Bianchi:2014qma} it is showed that the mass deformation of the non-chiral orbifold $\CC^3/\left( \ZZ_2 \times \ZZ_2 \right)$ flows to the non-chiral orbifold of the conifold $\Cc/\ZZ_2$ \cite{Forcella_2009, Franco:2007ii}. Also, the mass deformation of the adjoint fields in the non-chiral orbifold $\CC^3/{\ZZ}'_4$ flows to $\Cc/\ZZ_2$, as well as the orientifolds $\Omega$ and $\hat{\Omega}$, whose various diagrams are drawn in Fig.\,\eqref{fig:CZ2pr}. We now study them.

\subsubsection*{Orientifold $\Omega$ of $\Cc/\ZZ_2 $}

The action of the (non-toric) involution is
\begin{align}
{\bf{\overline{N}_3}} = {\bf{N_1}}  \; , \qquad
U(N_0) \rightarrow Sp/SO(N_0) \; ,\qquad
U(N_2) \rightarrow Sp/SO(N_2) \; , 
\end{align}
and the super-potential reads
\begin{align}
W' &= X_{21} X_{12} X_{21'} X_{1'2} - X_{1'2} X_{21'} X_{1'0} X_{01'} + X_{01'} X_{1'0} X_{01} X_{10} - X_{10} X_{01} X_{12} X_{21} \; ,
\end{align}
where fields factors $X_{12}X_{21}$ and $X_{01'}X_{1'0}$ absorb the $(1/m)$ coming from the mass deformation. Being non-chiral, the theory is anomaly-free. From the dimer, this orientifold configuration is obtained by a fixed line involution, then the groups at nodes 0 and 2 are projected in the same way, i.e.  $\epsilon_0 = \epsilon_2$. The only choices are $Sp(N_0) \times U(N_1) \times Sp(N_2)$ and $SO(N_0) \times U(N_1) \times SO(N_2) $. The beta-functions read
\begin{align}
2\beta_0 &= 3N_0 - 2N_1 + 6 \epsilon_0 \; , \nonumber \\
\beta_1 &= 3 N_1 - N_0 - N_2 \; , \nonumber \\
2\beta_2 &= 3N_2 - 2N_1 + 6 \epsilon_0 \; . 
\end{align}
Their sum vanishes only if $N_1 = - 6 \epsilon_0 - \half (N_0 + N_2) $ and $N_0 + N_2 < - 12 \epsilon_{0}$, allowed only for $SO$ groups. Individually, the beta functions do not vanish simultaneously. 

\subsubsection*{Orientifold $\hat{\Omega}$ of $\Cc/\ZZ_2 $}

The action of the involution is
\begin{align}
{\bf{\overline{N}_3}} = {\bf{N_0}} \; ,  \qquad
{\bf{\overline{N}_2}} = {\bf{N_1}} \; , 
\end{align}
and the super-potential reads
\begin{align}
W' &= X_{1'1} X_{11'} X_{1'0'} X_{0'1'} - X_{0'1'} X_{1'0'} X_{0'0} X_{00'} + X_{00'} X_{0'0} X_{01} X_{10} - X_{10} X_{01} X_{11'} X_{1'1} \; ,
\end{align}
As for the previous model, the anomaly-free condition gives
\begin{align}
\epsilon_{00'} = \epsilon_{0'0} \; , \nonumber \\
\epsilon_{11'} = \epsilon_{1'1} \; ,
\end{align}
which is again in agreement with the constraint from the dimer, since the product of the four T-paritites must be positive. The choices are reported in Tab \eqref{tab:C3Z4prhat}. 

The beta-functions read
\begin{align}
\beta_0 &= 2 N_0 - N_1 - 2 \epsilon_{00'}\; , \nonumber \\
\beta_1 &= 2 N_1 - N_0 - 2 \epsilon_{11'} \; ,
\end{align}
whose sum vanishes only if $\epsilon_{00'}=\epsilon_{11'}=1$ from which $ N_0 + N_1 = 4 $. On the other hand, $\beta_0 = 0 = \beta_1$ is not allowed.  

\begin{center}
\begin{tabular*}{0.6\textwidth}{@{\extracolsep{\fill}}ccc}
\toprule
Gauge groups & $(X_{00'}, X_{0'0})$ & $(X_{11'}, X_{1'1})$\\
\midrule
$U(N_0) \times U(N_1)$ & $ (S,S) $ or $ (A,A) $ & $ (S,S) $ or $ (A,A) $ \\
$U(N_0) \times U(N_1)$ & $ (S,S) $ or $ (A,A) $ & $ (A,A) $ or $ (S,S) $ \\
\bottomrule
\end{tabular*}
\end{center}  
\captionof{table}{The orientifold projection $\hat{\Omega}$ of $\Cc/\ZZ_2$. “A” stands for “Antisymmetric representation”, while “S” for “Symmetric representation”.}\label{tab:CZ2prhat}

\section{Seiberg Duality and Orientifolds}
\label{Sec:Seiberg}

In the preceding sections we showed the commutativity between mass deformation and unoriented projection. In the following, we ask whether a similar relation holds between Seiberg duality and orientifold. Recall that Seiberg duality relates two theories which have the same fixed point in the IR. When a duality cascade occurs, the true IR is the end of the cascade. For this reason, it is meaningful to ask whether the unoriented projection at the beginning of a duality cascade yields the same theory as the unoriented projection at the end of the cascade. 
%In the following, we would like to study the relation between Seiberg duality and unoriented projections. In particular we would like to clarify whether the unoriented projection at the beginning of a duality cascade yields the same theory as the unoriented projection at the end of the cascade. 
Duality cascade in unoriented quiver theories have been studied in \cite{Franco:2015kfa,Argurio:2017upa}, where each node is dualized and the theory flows from the UV to the IR. In our case, nodes are dualized following the order $ (0,2,1,3) $. We start again with the $\Cc/\ZZ_2'$ theory in Fig.\,\eqref{fig:CZ2El} with $N_0 = N_2 = N + M$, $N_1= N_3 = N$, where $N$ is the number of regular branes and $M$ is the number of fractional branes and $N>M$ in order to dualize the nodes. Along the cascade, the number of fractional branes diminishes. 

Before proceeding any further, recall that Seiberg duality for a gauge group $Sp(N_c)$ with $N_f$ fundamentals yields a magnetic theory $Sp(N_f - N_c - 4)$, $N_f$ fundamentals and a singlet in the antisymmetric `meson' of the $U(N_f)$ `flavour' group  \cite{Intriligator:1995ne}, Seiberg duality for a gauge group $SU(N_c)$ with $N_f$ fundamentals and antifundamentals yields a magnetic theory $SU(N_f - N_c)$, $N_f$ fundamentals and anti-fundamentals and a singlet `meson' in the bifundamental of the $U(N_f) \times U(N_f)$ `flavour' groups \cite{Seiberg:1994pq}, whereas the magnetic dual of $SO(N_c)$ with $N_f$ quarks in the vector representation is a theory with $SO(N_f - N_c + 4)$, $N_f$ quarks and a singlet `meson' in the symmetric of the $U(N_f)$ `flavour' group \cite{Intriligator:1995id}. 

Let us denote the two ways of performing the projections as \textbf{A} and \textbf{B}. 
\begin{itemize}
\item \textbf{A}: Orientifold + duality cascade.\\
Let us perform the projection $\Omega$ in Fig.\,\eqref{fig:CZ2b} with ${\epsilon_0}= +1$, which gives $Sp(N+M)$ at the node 0, $U(N)$ at the node 1 and $Sp(N+M)$ at the node 2. We dualize all nodes in the order (0,2,1), with node 3 identified with 1 by the orientifold. First, at node 0 the gauge theory changes as $Sp(N + M) \rightarrow Sp(N - M - 4)$ and there are four additional antisymmetric mesons $M^{pq}$. The orientifold projection is $\Omega$ in Fig.\,\eqref{fig:CZ2magOmega}. It is important to note that these mesons are composite in terms of the electric quarks, namely
\begin{equation}
M^{pq} = ({\bf{N}}_{1'} , \overline{{\bf{N}}}_0)^p ({\bf{N}}_{0} , \overline{{\bf{N}}}_1)^q = (\overline{{\bf{N}}}_{1} , \overline{{\bf{N}}}_0)^p ({\bf{N}}_{0} , \overline{{\bf{N}}}_1)^q \; ,
\end{equation}
which transform under two of the groups. In order to make the combination antisymmetric we get: $[i_0 j_0](l_1 m_1) + (i_0 j_0)[l_1 m_1]$, where $i_0, j_0= 1, ... N_0$ run on the group 0 and $l_1, m_1= 1, ... N_1$ along the group 1. This gives the proper orientifold signs for the mesons, thus $\sum_{I=1}^4 {\epsilon}_{1'1}^{(I)} = -2$. Besides, in this way the theory is anomaly-free. We can proceed to dualization of the node 2, whose gauge group becomes $Sp(N - M - 4)$. Furthermore, there are other mesons with ``opposite orientation'' to the previous ones, since they transform under two conjugate representations. Whenever that happens, we integrate them out. What remains is to dualize node 1 (and 3), with gauge group $U(N -2M - 8)$. This completes the first step in the duality cascade, and the process can be repeated several times as long as the duality is allowed. After $k$ steps in the cascade the theory is $Sp(N_0^A) \times U(N_1^A) \times Sp(N_0^A)$ with
\begin{align}
N_0^A &= N -(2k - 1)M - 4k^2 \; , \nonumber \\
N_1^A &= N - 2kM - 4k(k+1) \; 
\end{align}
and it represents the bottom of the duality cascade, namely the IR theory whose quiver is showed in Fig.\,\eqref{fig:OrientCascadeA}, if 
\begin{equation}
N < M (2k + 1 ) + 4k(k+2) \;  
\end{equation}
and 
\begin{equation}
\begin{cases}
\;(2k - 1)M + 4k^2 < N \qquad \textrm{if } M < 4k \; , \nonumber \\[5pt]
\; 2k M + 4k(k+1) < N \qquad \textrm{if } M > 4k \; .
\end{cases}
\end{equation}
When these condition holds, no more dualities are allowed and the cascade stops.\\[15pt]
\begin{figure}[H]
	\centerline{\begin{tikzpicture}[auto, scale= 0.5]
		%%%%%%%%%%%% Nodes %%%%%%%%%
		\node [circle, draw=blue!50, fill=blue!20, inner sep=0pt, minimum size=5mm] (0) at (-3.3,0) {$N_0^A$}; 
		\node [circle, draw=blue!50, fill=blue!20, inner sep=0pt, minimum size=5mm] (1) at (0,0) {$N_1^A$}; 			
		\node [circle, draw=blue!50, fill=blue!20, inner sep=0pt, minimum size=5mm] (2) at (+3.3,0) {$N_0^A$};
		%%%%%%%%%%% Lines %%%%%%%%%%%
		\draw (0)  to node {} (1) [->>, thick];
		\draw (1)  to node {} (2) [->>, thick];
		\end{tikzpicture}}
\caption{The theory at the end of the duality cascade of $(\Cc/{\ZZ}'_2)/\Omega$. The orientifold projection is performed before the cascade.}
\label{fig:OrientCascadeA}
\end{figure}
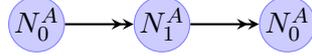
\item \textbf{B}: duality cascade + orientifold.\\
We exchange now the order and study the orientifold involution at the end of a duality cascade. We start with $M'$ fractional branes and eventually we compare this with $M$ of the previous case. The order of dualization is (0,2,1,3), with all gauge groups $U(N)$ and again integrating out fields in two conjugate representations. The cascade stops after $k'$ steps when
\begin{equation}
2k' M' < N < M' (2k' + 1 ) \; .
\end{equation}
The unoriented projection over nodes 0 and 2 yields an anomaly-free theory $Sp(N_0^B) \times U(N_1^B) \times Sp(N_0^B)$ at the IR with
\begin{align}
N_0^B &= N -(2k' - 1)M' \; , \nonumber \\
N_1^B &= N - 2k'M' \; ,
\end{align}
whose quiver is drawn in Fig.\,\eqref{fig:OrientCascadeB}.\\[15pt]
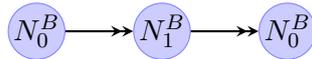
\begin{figure}[H]
	\centerline{\begin{tikzpicture}[auto, scale= 0.5]
		%%%%%%%%%%%% Nodes %%%%%%%%%
		\node [circle, draw=blue!50, fill=blue!20, inner sep=0pt, minimum size=5mm] (0) at (-3.3,0) {$N_0^B$}; 
		\node [circle, draw=blue!50, fill=blue!20, inner sep=0pt, minimum size=5mm] (1) at (0,0) {$N_1^B$}; 			
		\node [circle, draw=blue!50, fill=blue!20, inner sep=0pt, minimum size=5mm] (2) at (+3.3,0) {$N_0^B$};
		%%%%%%%%%%% Lines %%%%%%%%%%%
		\draw (0)  to node {} (1) [->>, thick];
		\draw (1)  to node {} (2) [->>, thick];
		\end{tikzpicture}}
\caption{The unoriented theory at the end of the duality cascade of $\Cc/{\ZZ}'_2$. The orientifold projection is performed after the cascade.}
\label{fig:OrientCascadeB}
\end{figure}
\end{itemize} 

Comparing the theories in $\bf{A}$ and $\bf{B}$ at the bottom of the cascades, they are equal if
\begin{align}
(2k' - 1 ) M' & = (2k - 1)M + 4k^2 \; , \nonumber \\
2k' M' & = 2k M + 4k (k+1) \; ,
\end{align}
which leads to 
\begin{align}
M' & = M + 4k \; , \nonumber \\[5pt]
k' & = \frac{ k M + 2k (k+1)}{M + 4k}  \; .
\end{align}
The solution in terms of integers $p$ and $q$ reads
\begin{align}
k' &= p \; , \nonumber \\
k &= q + p \; , \nonumber \\
M &= 2 \left[ \frac{p}{q} (p - 1) - q - 1 \right] \; , \nonumber \\
M' &= M + 4(p + q) \; , 
\end{align} 
with the condition $\frac{p}{q} (p - 1) \in \NN$ and $\frac{p}{q} (p - 1) \geq q + 1$.  

Note that $k=k'$ is allowed only if $k=k'=0$ or $k=k'=1$, where the former stands for $M=M'$ and no duality cascade is triggered and the latter describe a solution with $M' = M + 4$ and the flow stops if 
\begin{equation}
2M + 8 < N < 3M + 12
\end{equation} 
and 
\begin{equation}
\begin{cases}
\; M + 4 < N \qquad \textrm{if } M < 4 \; , \nonumber \\[5pt]
\; 2M + 8 < N \qquad \textrm{if } M > 4 \; .
\end{cases}
\end{equation}

If we perform the same process but with an unoriented projection giving $SO$ gauge groups instead of $Sp$, the path\footnote{Along the way, the mesons are symmetric and $\sum_{I=1}^4 {\epsilon}_{1'1}^{(I)}= + 2$.} $\bf{A}$ stops at ranks
\begin{align}
N_0^A &= N -(2k - 1)M + 4k^2 \; , \nonumber \\
N_1^A &= N - 2kM + 4k(k+1) \; ,
\end{align}
if 
\begin{equation}
N < M (2k + 1 ) - 4k(k+2) \;  
\end{equation}
and 
\begin{equation}
\begin{cases}
\;(2k - 1)M - 4k^2 < N \qquad \textrm{if } M < 4k \; , \nonumber \\[5pt]
\; 2k M - 4k(k+1) < N \qquad \textrm{if } M > 4k \; ,
\end{cases}
\end{equation}
while $\bf{B}$ remains the same since the orientifold projection is performed at the end of the cascade. The IR theories are the same if
\begin{align}
M' & = M - 4k \; , \nonumber \\[5pt]
k' & = \frac{ k M - 2k (k+1)}{M - 4k}  \; ,
\end{align}
and in terms of integers $p$ and $q$ it is solved as
\begin{align}
k' &= p \; , \nonumber \\
k &= p - q \; , \nonumber \\
M &= 2 \left[ \frac{p}{q} (p - 1) - q + 1 \right] \; , \nonumber \\
M' &= M + 4(p - q) \; , 
\end{align} 
with conditions $\frac{p}{q} (p - 1) \in \NN$, $\frac{p}{q} (p - 1) \geq q - 1$, $p - q > 1$.

%The order of the two operations, duality cascade and unoriented projection, over a gauge theory is then important, since the final degrees of freedom are different. However, the corresponding geometric setup of branes is not clear, in particular why the orientifold should appear at some point along the cascade. Perhaps, this problem could be better defined in the context of F-theory.
In general, cascades $\bf{A}$ and $\bf{B}$ do not end at the same step since insisting that the theories are the same in the IR gives $ k > k'$ in the $Sp$ case and $k< k'$ in the $SO$ case. Starting instead from the same theory in the UV, the IR theories are different. This is because the unoriented projection in the UV changes the degrees of freedom even before the flow along the cascade. Thus, the order of duality cascade and orientifold matters. Besides, the physical interpretation of cascade $\bf{B}$, where the orientifold projection is performed in the IR, is geometrically  unclear\footnote{We thank M. Bertolini and R. Argurio for stressing this point.}, although in the (non-perturbative) context of F-theory a certain geometric configuration could appear as an $\Omega$-plane at some distance, providing a possible physical scenario. The relation between Seiberg duality and unoriented projection will be further investigated  in an upcoming work, where several other cases will be studied.

%%%%%%%%%%%%%%%%%%%%%%%%%%%%%%%%%%%%%%%%%%%%%%%%%%%%%%%%%%%%%%
\section{Discussion and Outlook}\label{Sec:Discussion}

Let us conclude and summarise our results in order to draw some lines for future investigation.
We have discussed unoriented theories arising from the addition of $\Omega$-planes on stacks of D3-branes probing toric Calabi-Yau singularities. We focused on $\CC^3/\ZZ_3$ and $\CC^3/\ZZ_4$ (both chiral and non-chiral) and their non-orbifold descendant obtained by means of mass-deformations \cite{Bianchi:2014qma} and Higgsing/Un-Higgsing \cite{Feng:2002fv}. Examples of chiral non-orbifold theories include $ dP_1 $ \cite{Bertolini:2004xf} and the chiral $ \ZZ_2 $ quotient of the Conifold $ \Cc/\ZZ_2' $ \cite{Uranga:1998vf}, while non-chiral models include the Suspended Pinch Point (SPP) \cite{Park:1999ep} and the non-chiral $ \ZZ_2 $ quotient of the Conifold $ \Cc/\ZZ_2 $. When possible, we have simultaneously used both Quiver and Dimer descriptions in order to spell out the conditions for anomaly cancellation and super-conformal invariance, sometimes retrieved at the perturbative level after the inclusion of flavour branes \cite{Bianchi:2013gka}. For the unoriented projection of $\CC^3$ and $\CC^3/\ZZ_3$ we have found the relation between the orientifold charges $(\epsilon_0, \epsilon_I)$, which have a direct geometric interpretation, and the T-parities $\tau$ of the Dimer \cite{Franco:2007ii, Imamura:2008fd}. Orientifold charges are given by the action of T-parities on basic mesonic operators but a general relation was not evident before.

Moreover, by exploiting the combination of Toric diagram and the Ito-Reid Theorem \cite{Ito:1994zx}, we have addressed the problem of the distinction between $ \Omega $3-planes and compact/non-compact $ \Omega $7-planes for orbifold singularities, in the resolved geometry. Although theories with flavour branes admit a description in terms of bipartite graph on bordered Riemann surfaces \cite{Franco:2006es, Franco:2012mm, Franco:2013ana}, in general the resulting super-potential does not satisfy the toric condition and it is not obvious to us how far one can go with the use of Toric and Dimer diagrams in the context of unoriented projection\footnote{We thank I\~naki  Garcia-Etxebarria for clarifying comments on this issue.}. This is one of the reasons why it has been important for us to recover a satisfactory Quiver description of unoriented CY singularities: it allows the inclusion of non-compact D7-branes. The Quiver description can be used, even in the presence of both flavour branes and Orientifold planes, to easily compute RR-tadpole cancellation conditions \cite{Bianchi:2000de, Aldazabal:1999nu} and the vanishing of beta functions, needed in order to obtain an anomaly-free super-conformal field theory at the perturbative level. However, it should be noted that the superpotential can be unequivocally determined from the dimer diagram.  We stress that the anomaly cancellation condition, initially derived for orbifold theories in \cite{Bianchi:2013gka}, was used also for non-orbifold models and we justified the procedure by means of Mass Deformation and Higgsing. Hence, Mass Deformation and Higgsing are crucial for the validity of the anomaly-free condition in non-orbifold theories.

We have illustrated how, in general, each Quiver model admits more than one possible Orientifold projection (some of which preserve toricity). We have not explored non-perturbative phases that can be reached using S-duality \cite{Franco:2007ii, GarciaEtxebarria:2012qx, Garcia-Etxebarria:2013tba, Garcia-Etxebarria:2015hua, Garcia-Etxebarria:2016bpb}. Yet, by generalizing the anomaly cancellation condition derived in \cite{Bianchi:2013gka}, we have recovered results already present in the literature. Furthermore, we have exploited the symmetries of the invariant tensors of $SO(N)$ and $Sp(N)$ and the symmetries induced by the action of the unoriented projection on the fields present in super-potential in order to further constraint the spectrum and interactions. Our analysis has also shown that some particular unoriented projections, combined with the requirement of vanishing RR tadpoles, do not admit the existence of anomaly-free super-conformal theories, barring non-perturbative sectors that may emerge at strong coupling in the IR \cite{Garcia-Etxebarria:2015hua, Garcia-Etxebarria:2016bpb}.

Finally, we have studied the interplay between duality cascade and unoriented projections following similar analyses \cite{Bertolini:2005di, Argurio:2017upa}. A first analysis shows that performing the unoriented projection in the UV or in the IR yields similar theories, in the sense that the matter content are the same, but different in the ranks of gauge groups, i.e. the degrees of freedom. It would be interesting to further explore Seiberg (toric) duality and duality cascades in the context of unoriented theories with flavour. We plan to elaborate on this point in an upcoming work.

We have almost not touched the issue of non-perturbative corrections induced by stringy instantons \cite{Billo:2002hm, Billo:2006jm, Bianchi:2007fx, Bianchi:2007wy, Bianchi:2009bg, Blumenhagen:2009qh, Bianchi:2012ud, Argurio:2007}. They may play an important role in correcting the geometry, as already observed in some cases in \cite{Franco:2018vqd, Tenreiro:2017fon, Franco:2015kfa}. Extending these analyses to the unoriented case with flavour should be possible along the lines of \cite{Bianchi:2013gka}.
In the present work, we have not considered at all the issue of dynamical supersymmetry breaking in unoriented theories, which was recently addressed in \cite{Argurio:2019} and represents an interesting line of research.

%%%%%%%%%%%%%%%%%%%%%%%%%%%%%%%%%%%%%%%%%%%%%%%%%%%%%%%%%%%%%%

\section*{Acknowledgments}

We would like to thank Alice Aldi, Andrea Antinucci, Riccardo Argurio, Stephanie Baines, Sergio Benvenuti, Matteo Bertolini, Ugo Bruzzo, Dario Consoli, Pietro Fr\'e, Giorgio Di Russo, Alfredo Grillo, Maurizio Firrotta, Roberto Frezzotti, Francesco Fucito, Amihay Hanany, Juan Maldacena, Dario Martelli, Francisco Morales, Sami Rawash, Michele Santagata, Raffaele Savelli, David Turton for useful discussions and above all I\~naki  Garcia-Etxebarria for valuable comments and clarifying suggestions on an earlier version of the manuscript.\\
M.~B. was partially supported by the MIUR-PRIN contract 2015MP2CX4002 {\it ``Non-perturbative aspects of gauge theories and strings''} and by the grant {\it ``Strong Interactions: from Lattice QCD to Strings, Branes and Holography''} (CUP E84I19002260005, CUN Area 02), within the scheme {\it ``Beyond Borders''} of the University of Roma ``Tor Vergata". The work of D.B. was supported by the Royal Society Grant RGF$\backslash$R1$\backslash$181019. D.B. and S.M. would like to thank the Galileo Galilei Institute for hospitality and support while part of this work was developed.

%%%%%%%%%%%%%%%%%%%%%%%%%%%%%%%%%%%%%%%%%%%%%%%%%%%%%%%%%%%%%%
\newpage
\appendix

\section{Higgsing}\label{sec:higgs}

Mass deformation is not the only tool we have to deform a theory. Consider a supersymmetric gauge theory from the toric setup we described, with super-potential $W(X_{ab})$. If we give non-zero VEV to one of the bi-fundamental fields we obtain a new theory with a different toric diagram and a different mesonic moduli space. For instance if  $\langle X_{ij}\rangle  = v$ \cite{Feng:2002fv}, we are taking out from the dimer the edge corresponding to $X_{ij}$ and consequently the two adjacent polygons $i$ and $j$ merge into one, providing only one gauge group, which we denote $i$. In the case $X_{ij}$ enters in the super-potential in a cubic term, when it takes non-zero VEV there are quadratic terms as
\begin{equation}
W(X_{ij})= \ldots + \langle X_{ij}\rangle X_{ja}X_{ai} - \langle X_{ij}\rangle X_{jb}X_{bi}  + \ldots
\end{equation}
and the fields $X_{ja}$, $X_{ai}$, $X_{jb}$ and $X_{bi}$ become massive. Their mass will set an energy scale for the new theory. After integrating them out, by computing the corresponding $F$-terms and plugging them back in $W$, one obtain the low energy theory setting $i = j$. The dimer will change accordingly, in a different way from a mass deformation \cite{Bianchi:2014qma}. In the quiver the two nodes $i$ and $j$ merge as well and the connection/fields are pulled with them, but when we need to integrate out massive fields we should draw the quiver from the final dimer. 

The reverse method is called UnHiggsing. Starting from the dimer, we unhiggs a field drawing a new edge, which splits a polygon into two. This generates a new gauge group and new terms in the super-potential, which can be read from the new dimer.

In Sec.~\eqref{Sec:dP1} the Higgsing of the $dP_1$ to $\CC^3/\ZZ_3$ has been shown in detail. The fact that a non-orbifold theory can be higgsed down to a orbifold one is extremely useful for the identification of the conditions for anomaly cancellation.

\section{Seiberg Duality}\label{SeibergDual}

Seiberg duality relates two theories, denoted as `electric' and `magnetic',  that flow to the same conformal point in the IR, even though their Lagrangians are different in the UV. In the porto-typical case, the electric theory is a $\Nn = 1$ SQCD with gauge group $SU(N_c)$ and $N_f$ fundamentals $Q_i$ and antifundamentals $\widetilde{Q}^i$. The magnetic phase is another $\Nn = 1$ gauge theory with $SU(N_f - N_c)$, where $N_f > N_c$, $N_f$ $q_i$ fundamentals and $\overline{q}^i$ antifundamentals, in addition some mesonic fields ${M_i}^j= Q_i\widetilde{Q}^j$ and an extra super-potential term as $W_M = \widetilde{q}^i M_i^j q_j$. The theories have the same moduli space, even though R-charges may differ.

\begin{figure}[h!]
            	\begin{subfigure}{0.3\textwidth}
		\centerline{\begin{tikzpicture}[auto, scale= 0.5]
		%%%%%%%%%%%% Nodes %%%%%%%%%
		\node [circle, draw=blue!50, fill=blue!20, inner sep=0pt, minimum size=5mm] (0) at (3,3) {$N_0$}; 
		\node [circle, draw=blue!50, fill=blue!20, inner sep=0pt, minimum size=5mm] (1) at (3,-3) {$N_1$}; 			
		\node [circle, draw=blue!50, fill=blue!20, inner sep=0pt, minimum size=5mm] (2) at (-3,-3) {$N_2$};
		\node [circle, draw=blue!50, fill=blue!20, inner sep=0pt, minimum size=5mm] (3) at (-3,3) {$N_3$}; 
		%%%%%%%%%%% Lines %%%%%%%%%%%
		\draw (0)  to node {$X_{01}^{r'}$} (1) [->>, thick];
		\draw (1)  to node {$X_{12}^{r}$} (2) [->>, thick];
		\draw (2)  to node {$X_{23}^{r'}$} (3) [->>, thick];
		\draw (3)  to node {$X_{30}^{r}$} (0) [->>, thick];
		\end{tikzpicture}}
	\end{subfigure}
	\begin{subfigure}{0.3\textwidth}
		\centerline{\includegraphics[scale=0.35, trim={2cm 2.2cm 6.2cm 15cm}, clip]{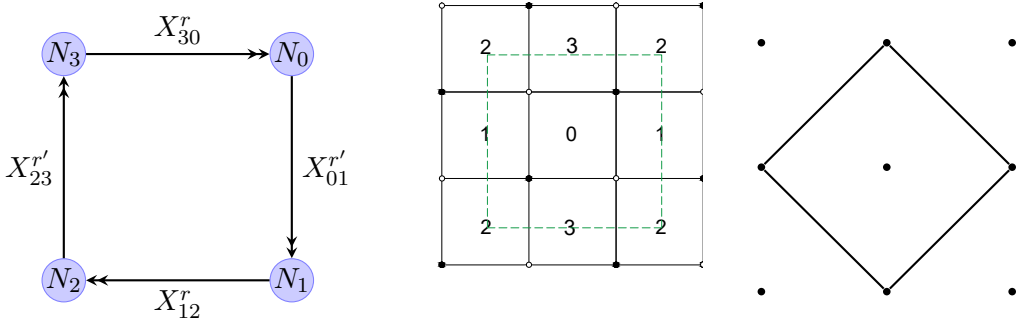}}
	\end{subfigure}
            \begin{subfigure}{0.3\textwidth} 
            \centerline{\begin{tikzpicture}[auto, scale= 0.55]
		%%%%%%%%%%%% Nodes %%%%%%%%%%
        \node [circle, fill=black, inner sep=0pt, minimum size=1mm] (1) at (-3,3) {};		
		\node [circle, fill=black, inner sep=0pt, minimum size=1mm] (2) at (0,3) {}; 
		\node [circle, fill=black, inner sep=0pt, minimum size=1mm] (3) at (3,3) {};
		\node [circle, fill=black, inner sep=0pt, minimum size=1mm] (4) at (-3,0) {};
		\node [circle, fill=black, inner sep=0pt, minimum size=1mm] (5) at (0,0) {}; 
		\node [circle, fill=black, inner sep=0pt, minimum size=1mm] (6) at (3,0) {}; 			
		\node [circle, fill=black, inner sep=0pt, minimum size=1mm] (7) at (-3,-3) {};
		\node [circle, fill=black, inner sep=0pt, minimum size=1mm] (8) at (0,-3) {}; 
		\node [circle, fill=black, inner sep=0pt, minimum size=1mm] (9) at (3,-3) {};
		%%%%%%%%%%% Lines %%%%%%%%%%%
        \draw (2) edge (6) [thick];
		\draw (6) edge (8) [thick];
		\draw (8) edge (4) [thick];
		\draw (4) edge (2) [thick];
		\end{tikzpicture}}
            \end{subfigure}\caption{The quiver, the dimer and the toric diagram of the electric phase of the $\Cc/\ZZ'_2$ model.}\label{fig:Seib1}
\end{figure}

\begin{figure}[h!]
            \begin{subfigure}{0.3\textwidth}
		\centerline{\begin{tikzpicture}[auto, scale= 0.5]
		%%%%%%%%%%%% Nodes %%%%%%%%%
		\node [circle, draw=blue!50, fill=blue!20, inner sep=0pt, minimum size=5mm] (0) at (3,3) {$N_4$}; 
		\node [circle, draw=blue!50, fill=blue!20, inner sep=0pt, minimum size=5mm] (1) at (3,-3) {$N_1$}; 			
		\node [circle, draw=blue!50, fill=blue!20, inner sep=0pt, minimum size=5mm] (2) at (-3,-3) {$N_2$};
		\node [circle, draw=blue!50, fill=blue!20, inner sep=0pt, minimum size=5mm] (3) at (-3,3) {$N_3$}; 
		%%%%%%%%%%% Lines %%%%%%%%%%%
		\draw (0)  to node {$X_{41}^{r'}$} (1) [<<-, thick];
		\draw (1)  to node {$X_{12}^{r}$} (2) [->>, thick];
		\draw (2)  to node {$X_{23}^{r'}$} (3) [->>, thick];
		\draw (3)  to node {$X_{43}^{r}$} (0) [<<-, thick];
		\draw (3)  to node [pos=0.4] {$X_{31}^{rr'}$} (1) [->>>>, thick];
		\end{tikzpicture}}
	\end{subfigure}
	\begin{subfigure}{0.3\textwidth}
		\centerline{\includegraphics[scale=0.35, trim={5.5cm 4.5cm 4.5cm 15.5cm}, clip]{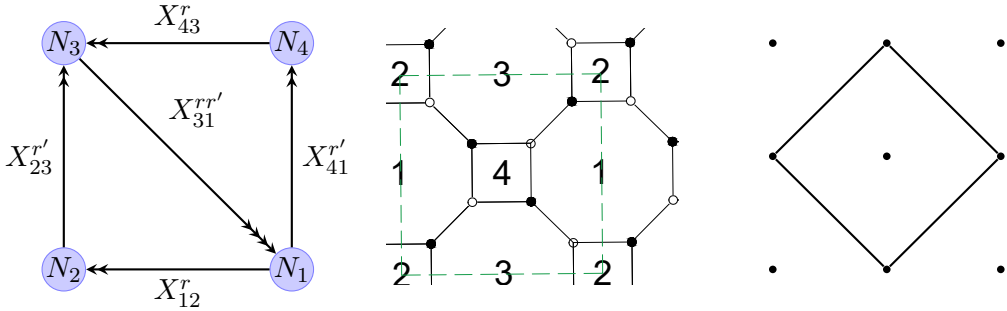}}
	\end{subfigure}
	\begin{subfigure}{0.3\textwidth} 
            \centerline{\begin{tikzpicture}[auto, scale= 0.5]
		%%%%%%%%%%%% Nodes %%%%%%%%%%
        \node [circle, fill=black, inner sep=0pt, minimum size=1mm] (1) at (-3,3) {};		
		\node [circle, fill=black, inner sep=0pt, minimum size=1mm] (2) at (0,3) {}; 
		\node [circle, fill=black, inner sep=0pt, minimum size=1mm] (3) at (3,3) {};
		\node [circle, fill=black, inner sep=0pt, minimum size=1mm] (4) at (-3,0) {};
		\node [circle, fill=black, inner sep=0pt, minimum size=1mm] (5) at (0,0) {}; 
		\node [circle, fill=black, inner sep=0pt, minimum size=1mm] (6) at (3,0) {}; 			
		\node [circle, fill=black, inner sep=0pt, minimum size=1mm] (7) at (-3,-3) {};
		\node [circle, fill=black, inner sep=0pt, minimum size=1mm] (8) at (0,-3) {}; 
		\node [circle, fill=black, inner sep=0pt, minimum size=1mm] (9) at (3,-3) {};
		%%%%%%%%%%% Lines %%%%%%%%%%%
        \draw (2) edge (6) [thick];
		\draw (6) edge (8) [thick];
		\draw (8) edge (4) [thick];
		\draw (4) edge (2) [thick];
		\end{tikzpicture}}
            \end{subfigure}\caption{The quiver, the dimer and the toric diagram of the magnetic phase of the $\Cc/\ZZ'_2$ model.}\label{fig:Seib2}
\end{figure}

An example will make this clear. Let us consider the chiral orbifold of the conifold $\Cc/\ZZ_2'$, whose resolution is the canonical bundle over the Hirzebruch $\FF_0 = \PP^1 \times \PP^1 = \SSS^2 \times \SSS^2$. This model has two Seiberg dual phases. The electric one has quiver and dimer as in Fig.\,\eqref{fig:Seib1}, where each node/gauge group in the quiver is seen as flavour from another linked node/group. Let us dualize the node 0. Denoting with 4 the dual node, its gauge group has rank $N_4 = 2N_1 - N_0$. The fields $(X_{01}^{r})_{i_0}^{i_1}$ and $(X_{30}^{s'})_{i_3}^{i_0}$, $r,s'=1,2$, change their chirality and in the magnetic theory become $(X_{14}^{r})_{i_1}^{i_4}$ and $(X_{43}^{s'})_{i_4}^{i_3}$. In addition, there are four mesonic fields 
\begin{equation}
(M^{rs'})_{i_3}^{i_1}=(X_{31}^{rs'})_{i_3}^{i_1}= (X_{30}^{r})_{i_3}^{i_0} (X_{01}^{s'})_{i_0}^{i_1} = \epsilon^{rp} \epsilon^{s'q'}(X_{31, \,pq'})_{i_3}^{i_1} \; ,
\end{equation}
being ${\epsilon}_{_{rp}}$ and ${\epsilon}_{_{s'q'}}$ the invariant tensors of the $SU(2)\times SU(2)'$ mesonic symmetry and a comma has been added to distinguish node indices and mesonic symmetry indices. The resulting magnetic phase is shown in Fig.\,\eqref{fig:Seib2}.
The quiver in the magnetic phase is easily obtained by drawing the connections corresponding to the new fields. In the dimer, there are new edges at the vertices of the dualized face/group 0. Finally, since the mesonic moduli space is the same for the two phases, the toric diagram is the same\footnote{Actually, each point on the toric diagram corresponds to a set of matter fields with the same R-charge. Then, each point has its multiplicity, from the number of fields in the corresponding set. Two Seiberg dual theories have the same toric diagram, in the sense that the shape and the area are the same. However, since R-charge may change under Seiberg duality, the multiplicity of points changes accordingly.}.
Once we have the dimer of the two phases, we can easily write the super-potential. Denoting with $W_e$ the super-potential of the electric phase, the super-potential of the magnetic one is 
\begin{equation}
 W_m = W_e + {\epsilon}_{_{rs}} {\epsilon}_{_{r's'}} (X_{43}^{s})_{i_4}^{i_3} (X_{31}^{rr'})_{i_3}^{i_1} (X_{14}^{s'})_{i_1}^{i_4} \; .
\end{equation}

Seiberg duality of $SU(N_c)$ SQCD theories with fundamental fields can be generalized to theories with adjoint fields and $SO/Sp$ gauge groups, to which we will now turn our attention.

\subsection*{Seiberg Duality with $SO(N_c)$ and $Sp(N_c)$ Gauge Group}

Unoriented projections can produce theories with gauge groups $SO$ and $Sp$. It may happen that the original theory has a Seiberg dual  description, then we expect the unoriented theory to have a magnetic description, too. 

Consider an $\Nn = 1$ gauge theory with group $SO(N_c)$ and $N_f$ quarks $Q^i$ in the vector representation. The flavour symmetry is $U(N_f)$. In \cite{Intriligator:1995id} it is argued that in the case $N_f > N_c - 2$, the theory at the origin of the space of vacua has magnetic dual with gauge group $SO(N_f - N_c +4)$, quarks $q_i$ and gauge singlet $M^{ij}=Q^i Q^j$ in the symmetric representation of the flavour group $U(N_f)$. There are various cases:
\begin{itemize}
\item for $N_c - 2 < N_f \leq \frac{3}{2} (N_c - 2) $ the magnetic theory is IR free.
\item for $ \frac{3}{2}(N_c - 2) < N_f < 3(N_c -2)$ the electric and magnetic theories flow to the same fixed point in the IR.
\end{itemize}
With $N_c \geq 4$, the magnetic theory has super-potential
\begin{equation}\label{eq:MagnetSO}
W_m = \frac{1}{2 \mu} M^{ij}q_i q_j \; ,
\end{equation}
with an additional term proportional to $\det \left( q_i q_j \right)$ if $N_f = N_c - 1$, needed so that the two dual phases have the same global symmetries. We may give mass $m$ to the $N_f$ quark $Q^{N_f}$ and the magnetic theory acquire a term as
\begin{equation}
W_{mass} = 	\frac{1}{2} m M^{N_f \, N_f} \; .
\end{equation}
Integrating out the massive quark the gauge group breaks to $SO(N_f - N_c + 3)$ and instantonic contributions are generated for $N_f \leq N_c $.

The super-potential in Eq.~\eqref{eq:MagnetSO} is similar to the super-potential in the case of $SU(N_c)$ theories, but the quarks transform in the vector representation. \\

Finally, let us focus on $\Nn = 1$ gauge theories with group $Sp(N_c)$, whose Seiberg dual is analyzed in \cite{Intriligator:1995ne}. Consider an electric $Sp(N_c)$ gauge theory with $N_f$ quarks in the fundamental `symplectic' representation, $Q_i$, $i=1,\ldots, N_f$. The flavour symmetry is $U(N_f)$.
There is a magnetic dual description if $ N_c + 3 \leq N_f \leq 3 (N_c +1) $, which is 
\begin{itemize}
\item IR free if $ (N_c + 3) \leq N_f \leq \frac{3}{2} (N_c + 1)$;
\item interacting in the IR if $ \frac{3}{2}(N_c + 1) < N_f < 3 (N_c + 1)$.
\end{itemize}
In this last regime, the magnetic theory has gauge group $Sp(N_f - N_c - 4)$ with quarks $q^i$ and an antisymmetric gauge-invariant operator $M_{ij} = Q_{i \, c} Q_{j \, d} \, {\cal I}^{cd}$, where ${\cal I} = 1 \otimes i \sigma_2$ and $c, d$ are color indices. The super-potential reads
\begin{equation}
W_m = \frac{1}{4 \mu} M_{ij} q^i_c q^j_d \; \mathcal{I}^{cd} \; .
\end{equation}
We may add masses to quarks with a relevant super-potential term 
\begin{equation}
W_{\mathrm{mass}} = \frac{1}{2} m^{ij} M_{ij} \; ,
\end{equation}
where $m^{ij}$ is the mass matrix. Integrating out the massive quarks reduces the number of flavour to $N_f - r$, being $\mathrm{rank}(m^{ij})=2 r$, and breaks the gauge group to $Sp(N_f - r - N_c - 4)$.

\begin{center}
\begin{tabular*}{0.85\textwidth}{@{\extracolsep{\fill}}ccc}
\toprule
Gauge group & Flavour group & Matter \\
\midrule
$SU(N_c)$ & $U(N_f)$ & $Q_i$, $\widetilde{Q}^i$, $i=1,\ldots, N_f$ \\
$SO(N_c)$ & $U(N_f)$ & $Q_i$, $i=1,\ldots, N_f$ \\
$Sp(N_c)$ & $U(N_f)$ & $Q_i$, $i=1,\ldots, N_f$ \\
\midrule
$SU(N_f - N_c)$ & $U(N_f)$ & $q_i$, $\widetilde{q}^i$, $M_{i}^{j} = Q_i \widetilde{Q}^j$, $i=1,\ldots, N_f$ \\
$SO(N_f - N_c + 4)$ & $U(N_f)$ & $q_i$, $M_{(ij)} = Q_i Q_j$, $i=1,\ldots, N_f$ \\
$Sp(N_f - N_c - 4)$ & $U(N_f)$ & $q_i$, $M_{[ij]} = Q_i Q_j \, \mathcal{I}$, $i=1,\ldots, N_f$ \\
\bottomrule
\end{tabular*}
\end{center}  
\captionof{table}{The various Seiberg dualities for gauge groups $SU(N_c)$, $SO(N_c)$, $Sp(N_c)$.\\[10pt]}\label{tab:Seiberg}

\bibliographystyle{JHEP}
\bibliography{Unoriented}
 
%\printbibliography
\end{document}